\documentclass{article}
\usepackage[a4paper]{geometry}
\usepackage[colorlinks, linkcolor=blue,urlcolor=blue]{hyperref}
\usepackage{graphicx}
\usepackage{amsmath,amsfonts,amssymb,bbm}
\usepackage{cases}			
\usepackage{color}
\usepackage{ulem}
\renewcommand{\emph}[1]{\textit{#1}}
\usepackage{listings}

\usepackage{tikz}
\usetikzlibrary{calc,arrows,decorations.pathmorphing,decorations.markings,positioning,fit,patterns}
\pgfdeclarelayer{background}
\pgfdeclarelayer{foreground}
\pgfsetlayers{background,main,foreground}

\tikzset{
	baseline=-.5ex,
	thck/.style={thick},
	every text node part/.style={font=\scriptsize},
	tight/.style={inner sep=1pt},
	point/.style={circle,draw,solid,tight},
	vertex/.style={circle,draw,solid,fill,tight},
	otight/.style={outer sep=0pt,inner xsep=-0.7pt,inner ysep=0pt},
	vtight/.style={outer sep=0pt,inner sep=-0.5pt},
	prop/.style={thck},
	arprop/.style={thck,postaction={decorate},
        	decoration={markings,mark=at position .55 with {\arrow{>}}}},
    	corr/.style={densely dotted,thck},
    	rectcorr/.style={dotted,thck},
    	arcorr/.style={corr,postaction={decorate},
        	decoration={markings,mark=at position .55 with {\arrow{>}}}},
	rectarcorr/.style={rectcorr,postaction={decorate},
        	decoration={markings,mark=at position .55 with {\arrow{>}}}},
	barcorr/.style={corr,postaction={decorate},
        	decoration={markings,mark=at position .55 with	{\arrow{<}}}},
        }

\newcommand{\Vq}
{\begin{tikzpicture}[baseline=-.5ex]
\node [vertex]	 (v) at (0,0) {} ;
\draw [arprop]		(v) +(210:1) -- node[above left]{$k$}(v) ; 
\draw [arprop]		(v) -- node[above right]{$k+q$}  +(330:1)	; 
\draw [arcorr]		(v) +(90:0.75) -- node[right]{$q$}(v)	;
\end{tikzpicture}
}

\newcommand{\fieldcorr}[1]
{\begin{tikzpicture}[baseline=-.5ex]
\draw[arcorr] (0,0) node {$\bullet$} -- node[above] {$#1$} (1,0) node {$\bullet$} ; 
\end{tikzpicture}
}

\newcommand{\prop}
{\begin{tikzpicture}[baseline=-.5ex]
\draw[prop] (0,0) -- 
					(0.5,0) ; 
\end{tikzpicture}
}

\newcommand{\arprop}[1]
{
\tikz{\draw[arprop] (0,0) -- node[below] {$#1$} (0.75,0);}
}

\newcommand{\Grrprime}
{\begin{tikzpicture}
\draw[arprop] (0,0) node[left]{$r$} -- (1,0) node[right]{$r'$}; 
\end{tikzpicture}
}

\newcommand{\GoVGo}
{\begin{tikzpicture}[scale=0.75]
\draw [arprop]	(0,0)	-- node[below]{$k$} (1,0)  node[vertex](v){}; 
\draw [arprop]	(v)	-- node[below]{$k'$}(2,0);
\draw [corr] (v) -- +(0,1); 
\end{tikzpicture}
}

\newcommand{\GoVGoVGo}
{\begin{tikzpicture}[scale=0.75]
\draw [arprop]	(0,0)	-- node[below]{$k$} (1,0)  node[vertex](v1){}; 
\draw [arprop]	(v1)	-- node[below]{$k''$} (2,0)  node[vertex](v2){}; 
\draw [arprop]	(v2)	-- node[below]{$k'$}(3,0);
\draw [corr] (v1) -- +(0,1); 
\draw [corr] (v2) -- +(0,1); 
\end{tikzpicture}
}

\newcommand{\vrtx}
{\begin{tikzpicture}[scale=0.75]
  \draw[arprop] (0,0) -- node[below]{$k$}  (1,0) node[vertex] (v) {};  
  \draw[arprop]	(v) --  node[below]{$k'$} +(1,0) ; 
  \draw[arcorr]	(v) -- node[right] {$q'$} +(65:1) ; 
  \draw[barcorr]	(v) -- node[left] {$q$} ++(115:1) ; 
  \end{tikzpicture}
}

\newcommand{\VPr}[2]
{\begin{tikzpicture}[scale=0.75]
\draw [rectcorr]	(0,0) node[vertex,label=below:$#1$] {} |- (0.5,0.5)
		-| (1,0)  node[vertex,label=below:$#2$] {}; 
\end{tikzpicture}
}

\newcommand{\VTr}[3]
{\begin{tikzpicture}[scale=0.75]
\draw [rectcorr]	(0,0) node[vertex,label=below:$#1$] {} 
		|- (0.5,0.5)
		-| (1,0)  node[vertex,label=below:$#2$] {}; 
\draw [rectcorr] 	(2,0) node[vertex,label=below:$#3$] {} 
		|- (1.5,0.5)
		-| (1,0)  ; 
\end{tikzpicture}
}

\newcommand{\VPk}[1]
{\begin{tikzpicture}[scale=0.75]
\draw [rectarcorr]	(0,0) node[vertex] {} 
		|- (0.5,0.5) node [above]{$#1$}
		-| (1,0)  node[vertex] {}; 
\end{tikzpicture}
}

\newcommand{\VTk}[2]
{\begin{tikzpicture}[scale=0.75]
\draw [rectarcorr]	   (0,0) node[vertex] {} 
		|- (0.5,0.5) node [above]{$#1$}
		-| (1,0)  node[vertex] (v){}; 
\draw [rectarcorr] 	   (v) 
		|- (1.5,0.5) node [above]{$#2$}
		-| (2,0)   node[vertex] {}; 
\end{tikzpicture}
}

\newcommand{\VQ}
{\begin{tikzpicture}[scale=0.75]
\draw [rectcorr]	(0,0) node[vertex] {} 
		|- (1.0,0.5) 
		(1,0)  node[vertex] {} 
		|- (2.0,0.5)
		(2,0) node[vertex] {} 
		|- (3.0,0.5) 
		-- (3,0) node[vertex] {}  ; 
\end{tikzpicture}
}

\newcommand{\Vgaussone}
{\begin{tikzpicture}[scale=0.75]
\node[vertex] (v1) at (0,0) {};  
\node[vertex] (v2) at (1,0) {};  
\node[vertex] (v3) at (2,0) {};  
\node[vertex] (v4) at (3,0) {};  
\draw [corr]	(v1) to [out=45]		(v2) ; 
\draw [corr]	(v3) to [out=45]		(v4) ; 
\end{tikzpicture}
}
\newcommand{\Vgausstwo}
{\begin{tikzpicture}[scale=0.75]
\node[vertex] (v1) at (0,0) {};  
\node[vertex] (v2) at (1,0) {};  
\node[vertex] (v3) at (2,0) {};  
\node[vertex] (v4) at (3,0) {};  
\draw [corr]	(v1) to [out=45]		(v3) ; 
\draw [corr]	(v2) to [out=45]		(v4) ; 
\end{tikzpicture}
}
\newcommand{\Vgaussthree}
{\begin{tikzpicture}[scale=0.75]
\node[vertex] (v1) at (0,0) {};  
\node[vertex] (v2) at (1,0) {};  
\node[vertex] (v3) at (2,0) {};  
\node[vertex] (v4) at (3,0) {};  
\draw [corr]	(v1) to [out=45]		(v4) ; 
\draw [corr]	(v2) to [out=45]		(v3) ; 
\end{tikzpicture}
}

\newcommand{\spcklcorr}[2]
{\begin{tikzpicture}
   \node[vtight,label=below:$#1$] (ast)  at (0,0) {$*$};	
   \node[tight,label=below:$#2$] (circ) at (1,0)   {};
   \draw[corr] 	(ast)  to [out=45] 
		(circ)  ;
   \draw (1,0) circle (2pt); 
\end{tikzpicture} 
}
\newcommand{\gammaself}
{\begin{tikzpicture}
   \node [tight](n) {$\circledast$};
   \draw[corr] 	(n)  .. controls +(65:0.5) and +(115:0.5) ..
		(n) ;
\end{tikzpicture} 
}
\newcommand{\Vspckltwo}
{\begin{tikzpicture}
   \node[vtight,label=below:$1$] (1)  at (0,0) {$\circledast$};	
   \node[vtight,label=below:$2$] (2) at (1,0)   {$\circledast$};
   \draw[corr] 	(1)  to [bend left=55] (2) 
				(1)  to [bend left=35] (2) ;
\end{tikzpicture} 
}

\newcommand{\Vspcklthree}
{\begin{tikzpicture}
   \node[vertex,label=below:$1$] (1) at (0,0) {};	
   \node[vertex,label=below:$2$] (2) at (1,0)   {};
   \node[vertex,label=below:$3$] (3) at (2,0)   {};
   \draw[corr] (1)    to [out=45] (2) ; 
   \draw[corr] (2)	to [out=45] (3) ;
   \draw[corr] (3.north)	to [out=135,in=45] (1.north) ;
\end{tikzpicture} 
}

\newcommand{\GoPGo}
{\begin{tikzpicture}[scale=0.75]
\draw [rectcorr]	(0,0) node[vertex](v1) {} |- (0.5,0.5)
		-| (1,0)  node[vertex] (v2) {}; 
\draw[prop] (-1,0) -- (v1) -- (v2) -- +(1,0);
\end{tikzpicture}
}

\newcommand{\GoTGo}
{\begin{tikzpicture}[scale=0.75]
\draw [rectcorr]	(0,0) node[vertex] (v1) {} 
		|- (0.5,0.5)
		-| (1,0)  node[vertex](v2) {}; 
\draw [rectcorr] 	(2,0) node[vertex] (v3) {} 
		|- (1.5,0.5)
		-| (1,0)  ; 
\draw[prop] (-1,0) -- (v1) -- (v2) -- (v3) -- +(1,0);
\end{tikzpicture}
}

\newcommand{\Sigmaonewithk}
{\begin{tikzpicture}[scale=0.75]
\path 	(-1,0) node [vertex] (v1) {}
		(1,0) node [vertex] (v2) {};
\draw[arprop] (v1) -- node[below] {$k'$} (v2) ;
\draw[arprop] (-1.5,0) -- node[below] {$k$} (v1) ;
\draw[arprop] (v2) -- node[below] {$k$} (1.5,0) ;
\draw[arcorr] (v1) to [bend left=65] node [above]{$k-k'$} (v2) ; 
\end{tikzpicture}
}

\newcommand{\Sigmatwo}
{\begin{tikzpicture}[scale=0.5]
\path 	(-1,0) node [vertex] (psi1) {}
		(1,0) node [vertex] (psi2) {};
\draw[prop] (psi1) -- (psi2) ;
\draw[corr] (psi1) to [bend left=55] (psi2) ; 
\draw[corr] (psi1) to [bend left=35] (psi2) ; 
\end{tikzpicture}
}

\newcommand{\Sigmathree}
{\begin{tikzpicture}[scale=0.5]
\path 	(-2,0) node [vertex] (psi1) {}
		(0,0) node [vertex] (psi2) {}
		(2,0) node [vertex] (psi3) {};
\draw[prop] (psi1) -- (psi2) -- (psi3) ;
\draw[corr] (psi1) .. controls (-1.5,0.5)  and (-0.5,0.5) .. 
			(psi2) .. controls (0.5,0.5)  and (1.5,0.5) .. 
			(psi3) .. controls (1,1)  and (-1,1) .. (psi1); 
\end{tikzpicture}
}

\newcommand{\mvGRmvGA}
{\begin{tikzpicture}[scale=0.5]
\draw[arprop] (0,1) -- node[above]{$k$} (2,1);
\draw[arprop] (2,-1) -- node[below]{$k'$} (0,-1);
\end{tikzpicture}
}

\newcommand{\Uboxwithk}
{\begin{tikzpicture}[baseline=-.5ex,scale=0.75] 
   \node [draw, minimum height=1cm,minimum width=0.5cm] (U) {};  
\draw[->] (U.north east)  -- +(0.25,0)   node [right] {$k'$}  ; 
\draw[->] (U.south west)  -- +(-0.25,0)   node [left] {$k$}  ; 
\draw[<-] (U.north west)  -- +(-0.25,0)   node [left] {$k$}  ; 
\draw[<-] (U.south east)  -- +(0.25,0)   node [right] {$k'$}  ; 
\end{tikzpicture}
}

\newcommand{\UGone}
{\begin{tikzpicture}[baseline=-.5ex,scale=0.5]
\node [vertex] (v1) at (0,1) {};
\node [vertex] (v2) at (0,-1){}; 
\draw[corr]  (v1) -- (v2)  ; 
\end{tikzpicture}
}

\newcommand{\UGtwoA}
{\begin{tikzpicture}[baseline=-.5ex,scale=0.5]
\node [vertex] (v1) at (0,1) {};
\node [vertex] (v2) at (0,-1){}; 
\node [vertex] (v3) at (1,1) {};
\node [vertex] (v4) at (1,-1){}; 
\draw[corr]  	(v1) -- (v4)  
		(v2) -- (v3) ; 
\draw[prop]  	(v1) -- (v3)  
		(v4) -- (v2) ; 
\end{tikzpicture}
}
\newcommand{\UGtwoB}
{\begin{tikzpicture}[baseline=-.5ex,scale=0.5]
\node [vertex] (v1) at (0,1) {};
\node [vertex] (v2) at (1,1){}; 
\node [vertex] (v3) at (1,-1) {};
\node [vertex] (v4) at (2,1){}; 
\draw[corr]  	(v1) [bend left=45] to (v4)  
		(v2) -- (v3) ; 
\draw[prop]  	(v1) -- (v2)  -- (v4) ; 
\end{tikzpicture}
}
\newcommand{\UGtwoC}
{\begin{tikzpicture}[baseline=-.5ex,scale=0.5]
\node [vertex] (v1) at (0,-1) {};
\node [vertex] (v2) at (1,-1){}; 
\node [vertex] (v3) at (1,1) {};
\node [vertex] (v4) at (2,-1){}; 
\draw[corr]  	(v1) [bend right=45] to (v4)  
		(v2) -- (v3) ; 
\draw[prop]  	(v1) -- (v2)  -- (v4) ; 
\end{tikzpicture}
}

\newcommand{\UBreal}
{\begin{tikzpicture}[baseline=-.5ex,node distance=0.75]
\node [vertex,label=0:$r_1$] (v) {}; 
\node [point,above left=of v,label=180:$r$] (r1) {};
\node [point,below left=of v,label=180:$r'$]  (r2)  {}; 
\draw [arprop] (r1.east) -- (v.north); 
\draw [arprop] (v.south) -- (r2.east); 
\draw [arprop,dashed] (r1.south) -- (v.west);  
\draw [arprop,dashed]  (v.west) -- (r2.north); 
\end{tikzpicture}
}
\newcommand{\Umcreal}
{\begin{tikzpicture}[baseline=-.5ex,node distance=0.75]
\node [vertex,label=0:$r_1$] (v1) at (0,0.5) {}; 
\node [vertex,label=0:$r_2$] (v2) at (0,-0.5){}; 
\node [point,above left=of v1,label=180:$r$] (r1) {};
\node [point,below left=of v2,label=180:$r'$]  (r2)  {}; 
\draw [arprop] (r1.east) -- (v1.north); 
\draw [arprop] (v1.east) -- (v2.east); 
\draw [arprop] (v2.south) -- (r2.east); 
\draw [arprop,dashed] (r1.south) -- (v2.west);  
\draw [arprop,dashed]  (v2.west) -- (v1.west); 
\draw [arprop,dashed] (v1.west) -- (r2.north);  
\end{tikzpicture}
}
\newcommand{\Umcrealclose}
{\begin{tikzpicture}[baseline=-.5ex]
\node [vertex,label=0:$r_1$] (v1) at (0,0.5) {}; 
\node [vertex,label=0:$r_2$] (v2) at (0,-0.5){}; 
\node [point,label=180:$r$] (r1) at (-1,0.1){};
\node [point,label=180:$r'$]  (r2) at (-1,-0.1) {}; 
\draw [arprop] (r1.east) -- (v1.north); 
\draw [arprop] (v1.east) -- (v2.east); 
\draw [arprop] (v2.south) -- (r2.east); 
\draw [arprop,dashed] (r1.south) -- (v2.west);  
\draw [arprop,dashed]  (v2.west) -- (v1.west); 
\draw [arprop,dashed] (v1.west) -- (r2.north);  
\end{tikzpicture}
}

\newcommand{\UGthreeA}
{\begin{tikzpicture}[baseline=-.5ex,scale=0.5]
\node [vertex] (v1) at (0,1) {};
\node [vertex] (v2) at (0,-1){}; 
\node [vertex] (v3) at (1,1) {};
\node [vertex] (v4) at (1,-1){}; 
\node [vertex] (v5) at (2,1) {};
\node [vertex] (v6) at (2,-1){}; 
\draw[corr]  	(v1) -- (v6)  
		(v2) -- (v5) 
		(v3) -- (v4); 
\draw[prop]  	(v1) -- (v3) -- (v5) 
		(v6) -- (v4) -- (v2) ; 
\end{tikzpicture}
}

\newcommand{\Ladderfour}
{\begin{tikzpicture}[baseline=-.5ex,scale=0.5]
\node [vertex] (v1) at (0,1) {};
\node [vertex] (v2) at (0,-1){}; 
\node [vertex] (v3) at (1,1) {};
\node [vertex] (v4) at (1,-1){}; 
\node [vertex] (v5) at (2,1) {};
\node [vertex] (v6) at (2,-1){}; 
\node [vertex] (v7) at (3,1) {};
\node [vertex] (v8) at (3,-1){}; 
\draw[corr]  	(v1) -- (v2)  
		(v3) -- (v4) 
		(v5) -- (v6) 
		(v7) -- (v8); 
\draw[prop]  	(v1) -- (v3) -- (v5) -- (v7) 
		(v8) -- (v6) -- (v4) -- (v2); 
\end{tikzpicture}
}
\newcommand{\Crossedfour}
{\begin{tikzpicture}[baseline=-.5ex,scale=0.5]
\node [vertex] (v1) at (0,1) {};
\node [vertex] (v2) at (0,-1){}; 
\node [vertex] (v3) at (1,1) {};
\node [vertex] (v4) at (1,-1){}; 
\node [vertex] (v5) at (2,1) {};
\node [vertex] (v6) at (2,-1){}; 
\node [vertex] (v7) at (3,1) {};
\node [vertex] (v8) at (3,-1){}; 
\draw[corr]  	(v1) -- (v8)  
		(v3) -- (v6) 
		(v5) -- (v4) 
		(v7) -- (v2); 
\draw[prop]  	(v1) -- (v3) -- (v5) -- (v7) 
		(v8) -- (v6) -- (v4) -- (v2); 
\end{tikzpicture}
}

 \definecolor{BLACK}{gray}{0}
 \definecolor{WHITE}{gray}{1}
 \definecolor{RED}{rgb}{1,0,0}
 \definecolor{GREEN}{rgb}{0,1,0}
 \definecolor{BLUE}{rgb}{0,0,1}
 \definecolor{CYAN}{cmyk}{1,0,0,0}
 \definecolor{MAGENTA}{cmyk}{0,1,0,0}
 \definecolor{YELLOW}{cmyk}{0,0,1,0}

\newcommand{\be}{\begin{equation}} 
\newcommand{\belab}[1]{\begin{equation}\label{#1}} 
\newcommand{\ee}{\end{equation}}

\renewcommand{\Re}{\mathrm{Re}}
\renewcommand{\Im}{\mathrm{Im}}
\newcommand{\bra}[1]{\langle #1 |} 
\newcommand{\ket}[1]{|#1 \rangle} 
\newcommand{\mv}[1]{{\left\langle #1 \right\rangle}}

\newcommand{\rmd}{\mathrm{d}} 		
\newcommand{\rme}{\mathrm{e}}

\newcommand{\tr}{\mathrm{tr}} 
\newcommand{\arcosh}{\mathrm{arcosh}}

\newcommand{\msf}{\mathsf} 		
\newcommand{\sfS}{{\msf{S}}}
\newcommand{\sfM}{{\msf{M}}}

\newcommand{\trsp}[1]{#1^\text{t}} 	

\newcommand{\eps}{\varepsilon}

\newcommand{\psiLout}{\psi_\text{L}^\text{out}}
\newcommand{\psiRout}{\psi_\text{R}^\text{out}}
\newcommand{\psiLin}{\psi_\text{L}^\text{in}}
\newcommand{\psiRin}{\psi_\text{R}^\text{in}}

\newcommand{\xiloc}{\xi_\text{loc}}
\newcommand{\lcor}{\zeta} 		
 		
\newcommand{\ls}{l_\text{s}} 		
\newcommand{\lB}{l_\text{B}}		
\newcommand{\DB} {D_\text{B}}		%
\newcommand{\UB} {U_\text{B}}		%
\newcommand{\UC} {U_\text{C}}		%
\newcommand{\class}[1]{\mathring{#1}}
\newcommand{\critical}[1]{#1_\text{c}} 
\newcommand{\gc}{\critical{g}} 		
\newcommand{\Wc}{\critical{W}} 		
\newcommand{\GR}{G^\mathrm{R}}
\newcommand{\GA}{G^\mathrm{A}}

\newenvironment{smallpmatrix}{\left(\begin{smallmatrix}}
        {\end{smallmatrix}\right)}

\newcounter{exo}
\newenvironment{exercise}
{\stepcounter{exo} 
\small
\medskip \hrule\nopagebreak\medskip
\noindent {\sffamily\textbf{Exercise \arabic{exo} -- }}} 
{\medskip\nopagebreak\hrule \medskip} 

\newcommand{\arxiv}[1]{\href{http://www.arxiv.org/abs/#1}{\texttt{arXiv:#1}}}
\newcommand{\journal}[4]{#1 \textbf{#2}, #3 (#4)}
\newcommand{\EPL}[3]{\journal{Europhys.\ Lett.}{#1}{#2}{#3}}
\newcommand{\JETP}[3]{\journal{JETP}{#1}{#2}{#3}}
\newcommand{\JOSA}[4]{\journal{J.\ Opt.\ Soc.\ Am.\ #1}{#2}{#3}{#4}}
\newcommand{\JPA}[3]{\journal{J.\ Phys.\ A: Math.\ Gen.}{#1}{#2}{#3}}

\newcommand{\nature}[3]{\journal{Nature}{#1}{#2}{#3}}

\newcommand{\NJP}[4]{\href{http://www.iop.org/EJ/abstract/1367-2630/#1/#2/#3}
{\journal{New J. Phys.}{#1}{#3}{#4}} }	

\newcommand{\PR}[3]{\href{http://link.aps.org/abstract/PR/v#1/e#2}
{\journal{Phys.\ Rev.}{#1}{#2}{#3}}}
\newcommand{\PRL}[3]{\href{http://link.aps.org/abstract/PRL/v#1/e#2}
{\journal{Phys.\ Rev.\ Lett.}{#1}{#2}{#3}}}
\newcommand{\PRA}[3]{\href{http://link.aps.org/abstract/PRA/v#1/e#2}
{\journal{Phys.\ Rev.\ A}{#1}{#2}{#3}}}
\newcommand{\PRB}[3]{\href{http://link.aps.org/abstract/PRB/v#1/e#2}
{\journal{Phys.\ Rev.\ B}{#1}{#2}{#3}}}
\newcommand{\PRE}[3]{\href{http://link.aps.org/abstract/PRE/v#1/e#2}
{\journal{Phys.\ Rev.\ E}{#1}{#2}{#3}}}
\newcommand{\RMP}[3]{\href{http://link.aps.org/abstract/RMP/v#1/e#2}
{\journal{Rev.\ Mod.\ Phys.}{#1}{#2}{#3}}}

\newcommand{\PRep}[3]{\journal{Phys.\ Rep.}{#1}{#2}{#3}}

\newcommand{\vthree}[1]{\textcolor{magenta}{#1}}

\title{Disorder and interference: localization phenomena} 
\author{Cord A.\ M\"uller%
\thanks{Centre for Quantum Technologies, National University of Singapore, Singapore 117543 and Department of Physics, University of Konstanz, D-78457 Konstanz, Germany}
\ and Dominique Delande\thanks{Laboratoire Kastler-Brossel, 
Universit\'e Pierre et Marie Curie, Ecole Normale Sup\'erieure, CNRS;
4 Place Jussieu, F-75005 Paris, France}}
\date{Chapter 9 in: ``Les Houches 2009 -
  Session XCI: Ultracold Gases and Quantum Information'', 
C. Miniatura, L.-C. Kwek, M. Ducloy, B. Gr\'emaud, B.-G. Englert, L.F. Cugliandolo, A. Ekert, eds.
(Oxford University Press, Oxford 2011)}

\begin{document}
\maketitle
\begin{center}
\vthree{This is v3 of \arxiv{1005.0915} with minor corrections marked in color. \\
Communications from B.~Nowak and C.~Texier are gratefully acknowledged.}
\end{center} 

\tableofcontents 
 
\section{Introduction}
Although complex systems are ubiquitous in nature, physicists tend to prefer ``simple'' systems.
The reason if of course that simple systems obey simple laws, which can be represented
by simple mathematical equations, as expressed by Goldenfeld and Kadanoff \cite{Kadanoff}:
\begin{quote}
\noindent 
\it{One of the most striking aspects of physics is
the simplicity of its laws. Maxwell's equations,
Schr\"odinger's equation, and Hamiltonian
mechanics can each be expressed in a
few lines. The ideas that form the foundation
of our worldview are also very simple indeed:
The world is lawful, and the same basic laws
hold everywhere. Everything is simple, neat,
and expressible in terms of everyday mathematics,
either partial differential or ordinary
differential equations. 

\noindent Everything is simple and neat---except, of
course, the world.}
\end{quote}
Even though the world obviously is not simple, many systems
can be split, be it only in \textit{Gedanken}, into simple components, each obeying simple laws. This is
the viewpoint of standard reductionism,
upon which modern science has been built~\cite{Anderson:More}:
\begin{quote}
\noindent
\it{The reductionist hypothesis may still be a topic for controversy
among philosophers, but among the 
great majority of active scientists I think it is accepted without
question. The workings of our 
minds and bodies, and of all the animate or inanimate matter of which
we have any detailed knowledge, are assumed to be controlled by the
same set of fundamental laws, which except under certain extreme
conditions we feel we know pretty well.} 
\end{quote}
Reductionism was a key ingredient for the development of
physics in the 19th and the first half of the 20th century, when 
(classical and quantum) mechanics, electromagnetism, relativity,
thermodynamics, etc.\  came to huge success.
But is there anything fundamental beyond the simple laws of physics, or
can one always reconstruct the properties of composed systems from the
workings of their parts? This credo of ``constructivism'' has been challenged by 
 P.~W.~Anderson, the author of the preceding quotation and one of the physicists who played
a major role in the analysis of complex physical systems:
\begin{quote}
\noindent
\it{The ability to reduce everything to simple fundamental laws does
not imply the ability to start from those laws and   
reconstruct the universe. In fact, the more the elementary particle
physicists tell us about the nature of the 
fundamental laws, the less relevance they seem to have to the very
real problems of the rest of science, much 
less to those of society.  

\noindent The constructionist hypothesis breaks down when confronted
with the twin difficulties of scale and complexity. 
The behavior of large and complex aggregates of elementary particles, it turns out, is not to be understood 
in terms of a simple extrapolation of the properties of a few particles. Instead, at each level 
of complexity, entirely new properties appear, and the understanding
of the new behaviors requires research which I think is as fundamental
in its nature as any other.} 
\end{quote}
which he summarized shortly with:
\begin{quote}
\noindent
   \textit{More Is Different.} 
\end{quote}

It is one aim of these lectures to show---in a restricted
context---how intricate the interplay between the small and the large
can be in complex systems.  
We did not yet give a precise definition of the word ``complex". It turns out that the very same word may be used
in different contexts with different meanings. In these lectures, we will address a specific example of complexity,
viz., \emph{disorder}. In physics as in everyday life, disorder is associated with some lack
of regularity. In a disordered material, 
atoms are not arranged in crystalline periodic patterns, but appear in more or less \emph{random}
positions. Randomness occurs because some agents, called ``degrees of
freedom'' by physicist, are not under control, either because we
cannot or chose not to control them. It means that we have to learn to
deal not with a single specific 
complex system---that is called a single \emph{realization} of the
disorder---but with a whole family of systems whose 
properties  are described in terms of distribution laws, correlation
functions, etc. 
The goal of the game is not to describe as accurately as possible 
a single system, but rather to predict global properties shared
by (almost) all systems, i.e.\ to acquire knowledge of universal features independent of
the precise realization of the disorder. 
These are instances of the ``new behaviors" mentioned by
Anderson. And this is also the viewpoint taken long ago by classical 
thermodynamics where one forfeits the microscopic description of a gas
in terms of all positions and momenta, 
concentrating instead on new concepts like entropy and temperature, which
prove to be the relevant, and therefore fundamental, concepts at this level 
of complexity. 

The specific problem we address in these lectures is the problem of transport and localization in disordered systems,
when \emph{interference} is present, as characteristic for waves.
A wave propagates in some medium (be it vacuum), and interference
occurs when different waves overlap, for example scattered from
different positions with various
wavevectors.  
The ``simple laws'' are the wave equation in the homogeneous medium
together with a  microscopic description of scattering by the
impurities. 
The complex behavior we want to describe is, for example,  the
propagation of the wave over long distances and for long times. 
Physical situations of this type  cover the propagation of sound in a concert hall with complicated shape, 
seismic waves multiply scattered inside the earth, electronic matter
waves in dirty semiconductor crystals, atomic matter waves in the
presence of a disordered potential, etc. In this context, it is good
to keep in mind a warning issued by W.\ Thirring \cite{Thirring1978}: 
\begin{quote}
\textit{It is notoriously difficult to obtain reliable results for quantum
mechanical scattering problems. Since they involve complicated
interference phenomena of waves, any simple uncontrolled approximation
is not worth more than the weather forecast.}
\end{quote}

\subsection{Anderson localization with atomic matter waves}

\begin{figure}
\begin{center}
\includegraphics[width=9cm]{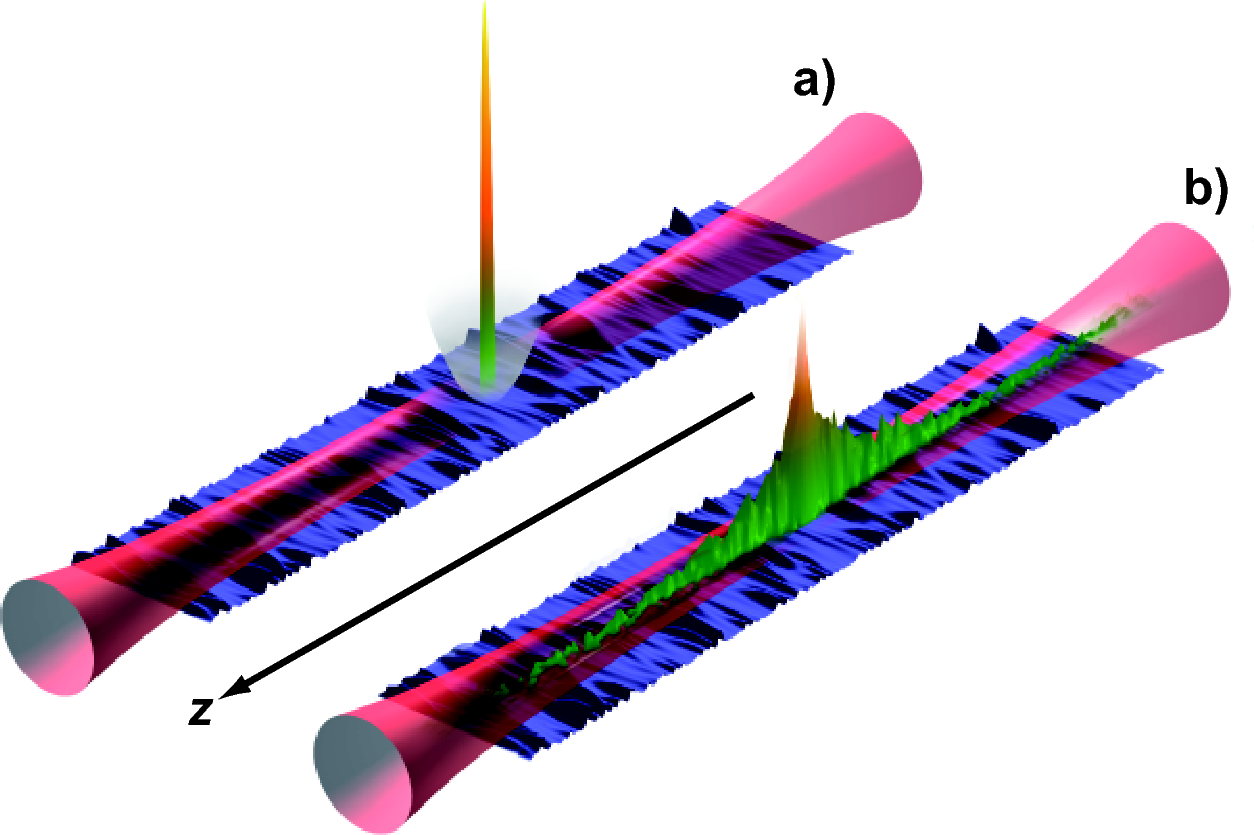}
\end{center}
\caption{\small%
Direct experimental observation of one-dimensional 
Anderson localization of an atomic matter wave in a disorder potential.
The disorder potential (represented in blue in the lower part of the
figure) is created by a speckle pattern.  (a) An initially 
localized wave packet (prepared in a harmonic trap at the center) evolves freely, diffuses
and eventually freezes at long times in a characteristic exponential
shape (b). The pink tube represents the
transverse-confinement laser 
beam that ensures an effectively one-dimensional dynamics. Reprinted from
\cite{Billy2008} (courtesy of Ph.\ Bouyer).}
\label{fig:orsay_exp}
\end{figure}

To start with a specific, state-of-the-experimental-art example, 
imagine a one-dimensional non relativistic particle
evolving in a potential $V(z)$ as depicted in
Fig.~\ref{fig:orsay_exp}. 
The evolution of the wavefunction $\psi(z,t)$ is given
by Schr\"odinger's equation: 
\belab{schroedinger.eq}
i\hbar \partial_t\psi(z,t) =  H \psi(z,t)   
\ee
with the single-particle Hamiltonian 
\belab{Hamilton1.eq}
H = \frac{p^2}{2m} + V(z).  
\ee 
Let us assume that the particle is initially prepared in a Gaussian
wave-packet. In the absence of any potential, the Gaussian wave-packet will show 
ballistic motion, where the center of mass moves at constant velocity
while the width increases linearly with time at long times. 
In the presence of a certain realization
$V(z)$ of the disorder, the wave function will take a certain form
$\psi(z,t)$. For different realizations, different wave functions
will be obtained. But we are not interested in the fine details of
each wave function. Rather, we wish to understand the generic, if not
universal, properties of the final stationary density distribution
$|\psi(z)|^2$ obtained at long times. We will see that not only
averages, but also their fluctuations contain important information.  
 
Let us forget for a moment interference effects and try to guess what
happens to a classical particle. 
If its kinetic energy is much larger than the typical 
strength of the disorder $V_0$, the particle will fly above the potential landscape, and the motion is likely
to be ballistic on the average. If on the other hand $V_0$ is
larger than the kinetic energy, 
the particle will be trapped inside
a potential well and transport over long distance is suppressed, i.e.~localization takes place.

Quantum mechanics modifies this simple picture fundamentally: 
waves can both tunnel through potential hills higher than the kinetic energy
and be reflected even by small potential fluctuations. 
So the initial wavepacket will split on each potential
fluctuation into a transmitted part and a reflected part, no matter
how large the kinetic energy with respect to the potential
strength may in detail be. 
After many scattering instances, this looks like a random walk and one
naively expects that, on average, the motion at long times
will be diffusive, with a diffusion constant depending on some 
microscopic properties of particle and potential. 

This simple model system has been recently realized
experimentally~\cite{Billy2008} using a quasi-one-dimensional atomic
matter wave, interacting with an effective optical potential created
by a speckle pattern,  see Fig.~\ref{fig:orsay_exp}. 
The experimental result is the following: at short times, the wavepacket spreads as expected, but at long times,
its average dynamics freeze, and the wavepacket takes a characteristic
exponential shape:
\begin{equation}
|\psi(z)|^2 \propto \exp \left( - \frac{|z|}{\xiloc}\right)
\end{equation}
where $\xiloc$ is called the localization length.%
\footnote{In the literature, the localization length 
is defined as the characteristic length for the decay of either
$|\psi|^2$ or of $|\psi|.$ 
The two quantities of course differ by a factor 2.
Usage of one or the other definition depends on the community, but may
also fluctuate from paper to paper. One has
to live with this source of disorder.} 
Moreover, if a different realization of the disorder is used (i.e.\ a
microscopically different, but statistically equivalent speckle
pattern), 
an almost identical shape is obtained, meaning that the phenomenon
is robust versus a change of the microscopic details.

This surprising phenomenon is known as Anderson localization, sometimes also called strong localization.
Although it was predicted on theoretical grounds in the late
50's---most famously by P.W. Anderson himself~\cite{Anderson1958}---it
has only been  observed directly rather recently. Cold atoms,
where an {\it in situ} direct observation of the wavefunction is possible, are from that point of view
highly valuable.

In these lecture notes, we present an introduction to transport properties
in disordered systems, with a strong emphasis on Anderson localization.
As an appetizer, we show in section~\ref{1d.sec} how the one-dimensional case can be exactly solved,
providing us with useful physical pictures. We then introduce in section~\ref{scalingtheory.sec} the scaling theory of localization, a typical illustration of the appearance of new concepts and parameters relevant at the long
distance and long time scale. After reviewing some of the most important experimental and numerical results
in section~\ref{experiments.sec}, we develop a microscopic description of quantum transport in 
section~\ref{microscopic.sec}, several applications of which are discussed
in sections~\ref{CBS.sec},~\ref{wl.sec} and~\ref{kicked.sec}. 

Many other, interesting questions will not be touched upon, foremost
the impact of \emph{interaction} between several identical particles. However, technically speaking averages over disorder
introduce an effective interaction. The relevant diagrammatic
approach, originally introduced in the context of quantum
electrodynamics, is quite versatile and used equally well to describe, e.g.,
interacting electrons in 
solid state samples 
or interacting atoms in  Bose-Einstein condensates. In these lecture
notes, we restrict the discusion to non-interacting particles
in a disordered medium, but the general framework and the technical
tools introduced should provide our readers with solid foundations to
follow also more advanced developments. Understanding the combined effects of
interaction and disorder has been, still is, and
doubtlessly will remain the subject of fascinating research for a long
time to come.

\section{Transfer-matrix description of transport and Anderson localization in 1d systems}
\label{1d.sec}

In order to develop some intuition on transport in disordered systems,
it is useful to study solvable models, where one can identify relevant
phenomena and mechanisms. It turns out that a one-dimensional system, 
with the specific choice of $\delta$ point scatterers put at random
positions, provides such a solvable model, that is moreover sufficiently rich 
to teach us useful lessons for more realistic disorder and higher dimensions.

\begin{figure}
\begin{center}
\includegraphics[width=0.6\linewidth]{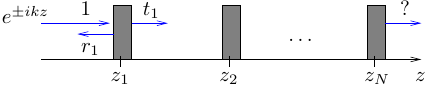}
\end{center}
\caption{\small
One-dimensional waveguide with randomly placed scatterers. We
know the reflection and transmission coefficients $r_j$ and $t_j$ of each
scatterer. What is the total transmission $T_N$ across the
whole ensemble?}
\label{d1setup.fig}
\end{figure}

Consider therefore a spinless particle confined to a 1d wave-guide 
geometry by tight transverse trapping to its ground state. 
Free propagation along the $z$-direction 
with wave vector $k$ 
is described by amplitudes $\psi_{\pm k}(z)=\exp\{\pm ik z\} $. In the following, we discuss the transmission of a
single, fixed $k$-component through a series of obstacles $j=1,2,
\dots,N$, well separated and  
placed at randomly chosen distances $\Delta z_j =z_j - z_{j-1}$ as
pictured in Fig.\ \ref{d1setup.fig}. 
 
\subsection{Scattering matrix}

Each obstacle shall be described by a potential $V_j(z)$. To fix
ideas, we may assume that it is sufficiently short-ranged
to be well approximated by a $\delta$-type impurity,
$V_j(z) = V(z-z_j) = \sigma_0 V_0 \delta(z-z_j)$, with an internal
length 
scale $\sigma_0$ that is not resolved by the propagating wave,  
$k\sigma_0\ll 1$.  
Furthermore, we suppose that 
the obstacles are well separated, i.e., their density $n=N/L$ is small
compared to the wavelength, $n\ll k$.  

Consider first scattering by a single impurity at $z=0$. We can decompose the wave
functions to the left (L, $z<0$) and right (R, $z>0$) into left- and
right-moving components: 
\begin{align}
\psi_\text{L}(z) & = \psiLin \rme^{+ikz} +  \psiLout \rme^{-ikz}\\
\psi_\text{R}(z) & = \psiRout \rme^{+ikz} +  \psiRin \rme^{-ikz}.  
\end{align}
The outgoing amplitudes are linked to the incident amplitudes by 
the reflection and transmission coefficients $r$ and $t$ 
from the left, and $r'$, $t'$ from the right: 
\be
	\begin{tikzpicture}[baseline=-5ex]
\node (inL) 			{};
\node (outR)   [right=of inL]	{};
\node (outL)   [below=of inL]	{};
\node (inR)   [below=of outR]	{};
\node (psiLin) [left=of inL] 	{$\psiLin$};
\node (psiRout) [right=of outR] {$\psiRout= t\,\psiLin + r'\, \psiRin$};
\node (psiRin) [right=of inR] 	{$\psiRin$};
\node (psiLout) [left=of outL] {$ r\,\psiLin + t'\, \psiRin = \psiLout$};
\draw[->,>=latex] (psiLin) -- (inL);
\draw[->,>=latex] (outR) -- (psiRout);
\draw[->,>=latex] (psiRin) -- (inR);
\draw[->,>=latex] (outL) -- (psiLout);
\draw[->,>=latex,blue,semithick] (inL) -- node(t)[above]{$t$} (outR);
\draw[->,>=latex,blue,semithick] (inR) -- node(t')[above]{$t'$} (outL);
\draw[->,>=latex,red,semithick] (inL) to [out=-45,in=45] node[left]{$r$} (outL);
\draw[->,>=latex,red,semithick] (inR) to [out=135,in=-135] node[right]{$r'$} (outR);
\begin{pgfonlayer}{background}
\node[draw,rectangle,black!50,inner sep=0pt,fill=black!20,fit= (outL) (t) (outR)]{};
\end{pgfonlayer}
\end{tikzpicture}

\ee
Writing these relations in matrix form introduces the scattering or S-Matrix: 
\begin{equation} 
\begin{pmatrix} 
\psi_\text{L}^\text{out}\\
\psi_\text{R}^\text{out}\\
\end{pmatrix}
= \sfS 
\begin{pmatrix} 
	\psi_\text{L}^\text{in}\\
	\psi_\text{R}^\text{in}\\
\end{pmatrix}
\qquad \text{with}
\qquad 
\sfS= \begin{pmatrix} 
		r & t' \\ 
		t & r' 
\end{pmatrix}  
.
\end{equation} 
For the present setting of a single-mode wave guide, the reflection
and transmission coefficients are complex numbers, and the
probabilities for reflection and transmission from the left 
are $R=|r|^2$ and $T=|t|^2$, respectively, and similarly from the right.  In a more
general setting of multi-mode scattering with $m$ modes or
``channels'' on the left and $m'$ modes on the right, $r$ and $t$ are matrices with $m\times m$
and $m'\times m$ entries, respectively. And for example, the total transmission
probability ``all channels in to all channels out'' then reads $T =
\sum_{m,m'} t_{m'm}t_{m'm}^* = \sum_{m'}(tt^\dagger)_{m'm'}=: 
\tr'\{t t^\dagger\}$. 
Within this section, we have only use for the
single-channel notation and refer to the literature
for the general case \cite{MelloKumar2004,Imry2002,Datta2002,Beenakker1997}

Probability flux conservation requires that $\sfS$ be unitary, $\sfS^\dagger= \sfS^{-1}$. 
From $\sfS^\dagger \sfS=\mathbbm{1}$, it follows directly that reflection and transmission  
probabilities add up to unity: $R+T=1$ and $R'+T'=1$. 
One also finds $r^*t' + t^* r' =0$ and its complex
conjugate $rt'^* + t r'^* =0$. 
From this, it follows for the single-channel case that 
$R=R'$, $T=T'$: the reflection and
transmission probabilities are the same from both sides.  

Time reversal exchanges the roles of ``in'' and ``out'' states. For a
time-reversal invariant potential $V(z)$, this implies $\sfS^* =
\sfS^{-1}$. With unitarity, this is equivalent to $\trsp{\sfS}=\sfS$ or $t=t'$, 
a symmetry called
\emph{reciprocity}. This setting defines the so-called ``orthogonal''
symmetry class of random matrix theory.  
Reciprocity is typically violated in presence of an external
magnetic field or magnetic impurities (see Secs.~\ref{dephasing.sec}
and \ref{wl.sec} below). 

\bigskip 
\begin{exercise}
Consider an elementary impurity with $V(z) =
\sigma_0V_0\delta(z)$. Solve the Schr\"odinger eigenvalue equation $-\psi''+
(2m/\hbar^2)V\psi = k^2 \psi $ at 
fixed $k$ (use the continuity of the free wave function and compute its derivative
discontinuity at $z=0$) and show that the S-Matrix in terms of $f= m
\sigma_0V_0/\hbar^2 k$ is given by \cite{Mello1994} 
\begin{equation} 
\sfS= \frac{1}{1+if}\begin{pmatrix} 
		-if & 1 \\ 
		1 & -if 
\end{pmatrix}.  
\end{equation} 
\end{exercise}

\subsection{Transfer matrix}
\label{transfer_matrix.sec}

If we now have several impurities in series, in principle the total
transmission can be calculated from the S-matrix of the whole
system. But the total S-matrix is not simply linked to the individual
S-matrices. Since the transmission depends on the incident amplitudes
from both left and right, adding a scatterer requires to recompute the
entire sequence. 
So instead of distinguishing in/out amplitudes, one prefers to decompose
the wave function into right-/left moving amplitudes, respectively:
$\psi(z) = \psi^+ \rme^{+ikz} +  \psi^- \rme^{-ikz}$, and this on both
sides R/L of the obstacles.  
The transfer matrix $\sfM$ then maps the
amplitudes from the left side of the 
obstacle to the right:  
 \begin{equation} \label{Mdef.eq}
\begin{tikzpicture}[baseline=-5ex]
\node (inL) 			{};
\node (outR)   [right=of inL]	{};
\node (outL)   [below=of inL]	{};
\node (inR)   [below=of outR]	{};
\node (psiLin) [left=of inL] 	{$\psi_\text{L}^+$};
\node (psiRout) [right=of outR] {$\psi_\text{R}^+$};
\node (psiRin) [right=of inR] 	{$\psi_\text{R}^-$};
\node (psiLout) [left=of outL] {$\psi_\text{L}^-$};
\draw[->,>=latex] (psiLin) -- (inL);
\draw[->,>=latex] (outR) -- (psiRout);
\draw[->,>=latex] (psiRin) -- (inR);
\draw[->,>=latex] (outL) -- (psiLout);
\path (inL) -- (outL) node[pos=0.5] (mid1){}; 
\path (outR) -- (inR) node[pos=0.5](mid2) {} ; 
\draw[->,>=latex,blue,thick] (mid1) -- node [below]{$\sfM$} (mid2);
\begin{pgfonlayer}{background}
\node[draw,rectangle,black!50,inner sep=0pt,fill=black!20,fit= (outL) (outR)]{};
\end{pgfonlayer}
\end{tikzpicture}
 
\qquad \text{or}
\qquad 
\begin{pmatrix} 
\psi_\text{R}^+\\
\psi_\text{R}^-\\
\end{pmatrix}
= \sfM
\begin{pmatrix} 
	\psi_\text{L}^+\\
	\psi_\text{L}^-\\
\end{pmatrix} .
\end{equation} 
One can easily determine its matrix elements in terms of
$t,t',r,r'$. For instance, $ \psiLout=  r\psiLin + t' \psiRin$  rewrites as 
$\psi_\text{L}^- = r\psi_\text{L}^+ + t' \psi_\text{R}^-$, which we
can immediately solve for $\psi_\text{R}^- =
\frac{1}{t'}\psi_\text{L}^-  - \frac{r}{t'} \psi_\text{L}^+$, and
similarly for $\psi_\text{R}^+$.  
Eliminating $r',t'$ with unitarity relations in favor of $r,t$ and their complex conjugates 
yields a simple form: 
 \begin{equation} \label{M.eq}
\sfM= 
\begin{pmatrix} 
	1/t^* & -r^*/t^*\\
	-r/t & 1/t \\
\end{pmatrix}. 
\end{equation} 

\begin{exercise}
Check the following interesting properties of the transfer matrix: 
\begin{itemize}
\item[$(i)$] $\det \sfM =1$. 
\item[$(ii)$] Current conservation (unitarity of $\sfS$) implies now that $\sfM\sigma_z\sfM^\dagger=\sigma_z$, with 
$\sigma_z= \begin{smallpmatrix} 1&0\\0&-1\\ \end{smallpmatrix}
$ the third Pauli matrix. 
\item[$(iii)$] Equivalently, $\sfM^{-1}= \sigma_z\sfM^\dagger \sigma_z$. 
\item[$(iv)$] $(\sfM^\dagger \sfM)^{-1} = \sigma_z\sfM^\dagger \sfM
\sigma_z$. Thus, the hermitian matrices $(\sfM^\dagger \sfM)^{-1}$ and
$\sfM^\dagger \sfM$ have the same (real) eigenvalues. Verify this
property by computing the eigenvalues directly using
\eqref{M.eq}. Since these eigenvalues must also be each other's
inverses, they can only be of the form $\lambda_+ = 1/\lambda_- =
\rme^{2x}$.  
\item[$(v)$] As a matrix, $2 + (\sfM^\dagger \sfM)^{-1} + \sfM^\dagger
\sfM = 4/T$. Thus, the total transmission probability is 
$T = 1/(\cosh x)^2$. 
\end{itemize}
\end{exercise}

\subsubsection{Chaining transfer matrices}

By construction, the transfer matrix maps the amplitudes from left to
right across each scatterer. Therefore, the total transfer matrix
across $N$ scatterers is obtained by multiplying them: 
\belab{sfMdef.eq}
\sfM_{12\dots N} = \sfM_N \dots \sfM_2\sfM_1. 
\ee 
Consider the simplest case of two obstacles $j=1,2$ in series, for
which  $\sfM_{12} = \sfM_2 \sfM_1$.
After matrix multiplication, one finds the transmission coefficient
\be
t_{12} = \frac{t_1 t_2}{1-r'_1r_2}
\ee
This transmission amplitude contains the entire series of repeated internal
reflection between the two scatterers: $t_{12} = t_2t_1 + t_2 r'_1r_2 t_1+ t_2
(r'_1r_2)^2 t_1 +\dots$.  
The transmission probability
reads  
\belab{T12theta.eq} 
T_{12} = \frac{T_1T_2}{|1- \sqrt{R_1R_2} \rme^{i\theta}|^2}
\ee
where $\theta$ is the total phase accumulated during one complete
internal reflection. 
Since the scatterers are placed with a random distance $k \Delta z\gg 2\pi$, the phase $\theta$
will also be randomly distributed in $[0,2\pi]$, independently of the
details of the random distribution of distance between consecutive
scatterers or their reflection phases. 
One can calculate expectation values of any
function of $\theta$ by  
\belab{avdef.eq}
\mv{f(\theta)} = \int_0^{2\pi} \frac{\rmd
\theta}{2\pi}\, f(\theta).
\ee
But as we will see in the following, a very important question is: ``what
quantity $f(\theta)$ should be averaged?''

\subsubsection{Incoherent transmission: Ohm's law}
\label{Ohm.sec}

The most natural thing seems to average directly the transmission
probability \eqref{T12theta.eq}. One finds 
\belab{avT12.eq}
\mv{T_{12}} = \frac{T_1T_2}{1 - R_1 R_2}. 
\ee 
The same transmission probability is obtained for a purely classical model where only
reflection and transmission \emph{probabilities} are combined: 
\begin{exercise}\label{RTclss.exo}
Show that the rule \eqref{avT12.eq} is also obtained if one 
uses an S-matrix propagating probabilities instead of amplitudes,    
$\mathring\sfS= \begin{pmatrix}  
		R & T \\ 
		T & R 
\end{pmatrix}, $ by determining the corresponding transfer matrix $\mathring\sfM$ and chaining it.  
\end{exercise}
In this case,
the distance $\Delta z$ of ballistic propagation between the two 
scatterers is completely irrelevant, and $T_{12}$ is given
by \eqref{avT12.eq}.  
This classical description applies to systems that are subject to 
strong decoherence, where the phase of the particle is completely scrambled by coupling to an external 
degree of freedom, while traveling between scatterers 1 and 2.

The so-called element resistance
of the obstacles, $(1-T)/T$, calculated with the classical transmission, is additive:  
\belab{addT12.eq}
\frac{1-T_{12}}{T_{12}} = \frac{1-T_1}{T_1}+\frac{1-T_2}{T_2}. 
\ee
Therefore, the classical resistance across $N$ identical
impurities distributed with linear density $n=N/L$ along a wire of
length $L$ grows like    
\belab{ohm.eq} 
\frac{R}{T}(L)  = N \frac{R_1}{T_1} =: \frac{L}{l_1} 
\ee 
where 
$l_1=T_1/(nR_1)$ is a length characterizing the backscattering
strength of a single impurity.

Within the context of electronic conduction, the result \eqref{ohm.eq}
is known as  Ohm's law, stating that the total classical resistance of a wire 
grows linearly with its length $L$. Obviously, averaging the
transmission itself at each step has wiped out completely the phase-coherence and
left us with a purely classical transport process. 
This process can be
formulated equivalently as a persistent random walk on a lattice where the
particle has uniform probability $T_1$
to continue in the same direction at each time step and probability
$R_1$ to make a U-turn. 
For long times, this random walk leads to diffusive motion with
diffusion constant 
\cite{Godoy1997}
\begin{equation}
D = v l_0 \frac{T_1}{2R_1}. 
\end{equation} 
Here $v$ is the velocity of the particle and $l_0=1/n$ the distance between
consecutive scatterers. 
Since the diffusion constant is related to the transport mean free path $l$
by the general relation $D=vl/d,$ with $d=1$ the dimension of the
system, one can identify $l_1$
as twice the transport mean free path, $l_1=2l$. 


\subsubsection{Phase-coherent transmission: strong localization}
\label{SL1.sec}

The relation \eqref{avT12.eq} cannot be easily generalized to more than two scatterers. 
Indeed, already for three scatterers, the average of the complicated product of transmission
matrices even
over independent, random phases $\theta_{12}$ 
and $\theta_{23}$ becomes very complicated.
In order to predict the behavior of transmission across long
samples, it is advantageous to find a quantity that is additive as new
scatterers are added to the wire. There is such a quantity that
becomes additive under
ensemble-averaging, namely the so-called extinction coefficient $\kappa =
-\ln T =|\ln T|$. When averaging the logarithm of \eqref{T12theta.eq}, the
denominator drops out since 
\be
\int_0^{2\pi} \frac{\rmd \theta}{2\pi}\,\ln \left|1 - \sqrt{R_1R_2} \rme^{i\theta}\right| = 0 
\ee 
due to the analyticity of the complex logarithm for all $0 \leq R_1R_2
< 1$. Thus, one immediately 
finds that the average extinction across two consecutive scatterers is strictly additive:   
\begin{equation}
\mv{\ln T_{12}} = \ln(T_1) + \ln(T_2).  
\end{equation}   
The generalization to many scatterers is now easy because $\mv{\ln T}$
is additive: 
the total extinction of a channel of length $L$ 
grows on average like $ |\mv{\ln T}| = nL |\ln T_1|$. 
With this scaling behavior, one obtains that the log-averaged transmission 
\belab{Ttyp.eq}
\exp\{\mv{\ln T}\} = \rme^{- L/\xiloc}
\ee
drops exponentially fast with increasing sample length $L$. 
In the absence of absorption, this is a hallmark of strong
localization by disorder, and we have found the localization
length $\xiloc= 1/(n|\ln T_1|)$. 
In a weak scattering situation where
$nl_1=T_1/R_1\gg 1$, we approximate $|\ln T_1|\approx 1/nl_1$ and thus
find the localization length as $\xiloc=l_1 =2l$.

What is the meaning of the log-averaged transmission \eqref{Ttyp.eq}?
It is important to realize that the 
transmission $T$ as a function of the  
microscopic realization of disorder is a
\emph{random variable}. We will show in the following sections that $T$ itself is not
a self-averaging quantity, meaning that its average $\mv{T}$ has no
resemblance to the most likely found value of $T$, called the
\emph{typical} transmission $T_\text{typ}$. We will shortly see that for long samples,
the probability distribution of $\ln T$ is very close
to a normal distribution (see \eqref{Winfty.eq} below), which is
centered at $ \mv{\ln T}=\ln T_\text{typ} $. And since the logarithm is
monotonic, the \emph{most probable value} of the transmission is indeed $T_\text{typ} = \exp\{\mv{\ln T}\} = \rme^{-
L/\xiloc}$. 
 
How does one go about to identify a properly self-averaging quantity
in the first place? --- By observing that
the transmission matrices of several obstacles  
multiply, see eq.\ \eqref{sfMdef.eq}.
\vthree{The logarithm of this random product therefore realizes a noncommuting variant of the central limit theorem, known as Furstenberg's theorem  \vthree{\cite{Furstenberg1960}}, which
implies that} the extinction or log-averaged
transmission is indeed a good candidate for a self-averaging
quantity.

\subsection{Scaling equations}

In order to substantiate the previous arguments, we should 
find the full distribution function $P(T,L)$ that permits to derive
expectation values for arbitrary functions of $T$ at length $L$. This
distribution function can be found exactly by solving recursion
relations  
that describe how the transmission is changed when a small bit $\Delta
L$ is added to a sample of length $L$: 
\begin{center}
\begin{tikzpicture}
\draw[fill=black!20,thick]
		(-4,0) node[below]{$0$} 	-- (0,0) node [below] (L) {$L$}
		-- (0,1) node {} 
		-- (-2,1) node [above] {$T_L$}
		-- (-4,1); 	 
\draw[thick,dotted,fill=black!10] (0,0)  	-- (1,0) node
[right=1cm of L.west] (dL) {$L+\Delta L$}
		-- (1,1) node {} 
		-- (0.5,1) node [above] {$T_{\Delta L}$}
		-- (0,1);
\end{tikzpicture}

\end{center}
We know already that the transfer matrices multiply, $\sfM_{L+\Delta
L} = \sfM_{\Delta L}\sfM_{L}$.  The composition law \eqref{T12theta.eq} then gives 
\belab{TDeltaL.eq} 
T_{L+\Delta L} = \frac{T_{L}T_{\Delta L}}{|1-\rme^{i\theta}\sqrt{R_L
R_{\Delta L}}|^2}. 
\ee
The idea is now to study the change in the expectation values of $T$ as the ``time'' $t = L/l_1$ grows. 
For this, we assume that the added part $\Delta L$ is long enough such
that an independent disorder average $\mv{\dots}_{\Delta L}$ in this section is meaningful. We
also assume that the scatterers are weak such that the backscattering
probability remains small:  
\be 
R_{\Delta L} = \Delta L/l_1 =:\Delta t \ll 1  . 
\ee
These assumptions are verified if $\Delta L$ is of the order of $n^{-1} = l_1 R_1$ with a
single weak scatterer on average.  
We may now expand \eqref{TDeltaL.eq} using $T_{\Delta L} = 1 - \Delta
t$ to leading order in $\Delta t$,  finding for the change
\belab{DeltaT.eq}
\Delta T = T_{L+\Delta L} - T_{L} = T_L \left[  2 \sqrt{R_L \Delta t} \cos\theta + (4 R_L \cos^2\theta -1 -R_L)\Delta t   \right].  
\ee
Thus, to order $\Delta t$, we find by averaging $\mv{\dots}_{\Delta
L}=\mv{\dots}_{\theta}$ 
that 
\belab{momentsDeltaT.eq}
\frac{\mv{\Delta T}_{\theta}}{\Delta t}  = -T_L^2, \qquad  
\frac{\mv{(\Delta T)^2}_{\theta}}{\Delta t}  = 2T_L^2 (1-T_L), \qquad 
\frac{\mv{(\Delta T)^n}_{\theta}}{\Delta t}  = 0 \quad (n\ge 3)
\ee 
So only the first two moments of fluctuations contribute. From the first
relation we can read 
off that the transmission decreases with increasing
length, an expected result. But the second relation shows that 
the relative fluctuations, while small in the beginning where
$1-T=R\ll 1$, will grow for long samples where $T\ll1$.

\subsection{Fokker-Planck equation and log-normal distribution} 

The above equations \eqref {momentsDeltaT.eq} describe a random quantity $T(t)$ whose first two
moments obey the equations $\partial\mv{T}/\partial t = \mv{A(T)}$ and
$\partial\mv{T^2}/\partial t = \mv{2TA(T)+B(T)}$ with $A(T) = -T^2$ and
$B(T)=2T^2(1-T)$. Here, we use the continuous
notations $\Delta t \to \rmd t$ and $\Delta T \to \rmd T$ while bearing
in mind the coarse-grained character of the averaged quantities. 
The theory of Brownian motion \cite{vanKampen} then teaches us that the
probability distribution $P(T,t)$ of $T$ at time $t$ obeys the
Fokker-Planck equation  
\belab{FPP.eq}  
\partial_t P = -\partial_T[AP] + \frac{1}{2}\partial^2_T[BP]. 
\ee
This equation of motion can be seen as a continuity equation,
$\partial_tP + \partial_TJ=0$, for the locally conserved probability
density with current $J = AP - \frac{1}{2}\partial_T(BP)$. Here, $A$
describes the drift, whereas $B/2$ plays the role of a diffusion
constant. 

By standard terminology, the Fokker-Planck
equation \eqref{FPP.eq} is called ``non-linear'', because $B(T)$ and $A(T)$ depend non-linearly on $T$.  
By changing the variable, one can try to simplify these coefficients.
A first option consists in using 
$T=1/(\cosh x)^2$ (remember exercise 2(\textit{v})). In the remainder
of this section, let us explore the consequences of this choice. 
Knowing that the change
of variables for a
probability density requires  
\be
P(T,t)\rmd T = P((\cosh x)^{-2},t)\left|\frac{\rmd T}{\rmd
x}\right|\rmd x = \tilde P(x,t)\rmd x,
\ee
we find the corresponding Fokker-Planck equation: 
\be
\partial_t\tilde P(x,t) =  -\frac{1}{2}
\partial_x\left[\coth(2x)\tilde P\right] + \frac{1}{4} \partial_x^2 \tilde P 
\ee
with initial condition $\tilde P(x,0)=\delta(x)$.  
With this choice of variable, the second derivative describing the
fluctuations has become most simple. Although this equation is still (too)
difficult to solve exactly, we can extract the limiting distribution
for long samples in the limit $t\gg 1$ as follows. First we rewrite
the equation as 
\be
\left[\partial_t + \frac{1}{2} \coth(2x) \partial_x\right] 
\tilde P(x,t) =  \frac{1}{(\sinh2x)^2} \tilde P  + \frac{1}{4} \partial_x^2 \tilde P 
\ee
We may interpret the derivatives on the left hand side as a Lagrangian
derivative $\partial_t + \dot x_0 \partial_x = D_t$
in a co-moving frame defined by $\dot x_0(t) = \frac{1}{2}
\coth(2x_0)$, which is solved by $x_0(t)=\frac{1}{2}\arcosh(e^{t})
\approx \frac{1}{2}t$, for large t. Developing all terms for small
deviations $\Delta=x-x_0(t)$ from this point of reference, we find a
very simple equation for $\tilde F(\Delta,t)=\tilde P(x_0(t)+\Delta,t)$: 
\be
\partial_t \tilde F(\Delta,t) =  \frac{1}{4} \partial_\Delta^2 \tilde F(\Delta,t) . 
\ee
This is the elementary diffusion equation, and the solution, perhaps
best known as the heat kernel, is readily
obtained by Fourier transformation: 
\be
\tilde F(\Delta,t) = \frac{1}{2\sqrt{\pi t}} \exp\left\{-\frac{\Delta^2}{4t}\right\}
\ee
Going back to the transmission using $\ln T=-2x$ valid for large $x$, 
we thus find the limiting distribution 
\belab{Winfty.eq}
F_\text{log-norm}(\ln T,t) = \frac{1}{2\sqrt{\pi t}} \exp\left\{-\frac{(\ln
T+t)^2}{4t}\right\}
\ee
for small deviations around the most probable value $\ln T_0(t) =
-2x_0(t)=-t$. 
We have successfully demonstrated that indeed the logarithm of the
transmission is a normally distributed random
quantity. It is characteristic for disordered channels that the two
defining moments
\belab{momentslogT.eq}
|\mv{\ln T}| = t,\qquad \text{var}(\ln T) = \mv{(\ln T)^2} - \mv{\ln
T}^2 = 2t  
\ee
are determined by a single parameter, namely the length $t=L/2l=L/\xiloc$ of the
one-dimensional wire in units of the localization length. Clearly, the relative fluctuations 
$\text{var}(\ln T)/\mv{\ln T}^2=2/t$ 
decay with system size. Thus we are assured that the 
transmission logarithm 
is a self-averaging quantity, with moreover a normal probability
distribution whose most probable value is equal to the mean.  

But careful! Even in the limit $t\gg1$, this does unfortunately \emph{not} imply that one may use
the log-normal distribution \eqref{Winfty.eq} indiscriminately to calculate
moments of the transmission. A striking counterexample is 
\belab{mvTwrong.eq}
\mv{T}_\text{log-norm} = \mv{e^{\ln T}}_\text{log-norm} = \int \rmd y  F_\text{log-norm}(y,t)e^y
= 1 \qquad \text{(wrong)},   
\ee
and quite obviously so. What goes wrong here? We will have a second
look at the end of the next section 
once we know the exact solution.

\subsection{Full distribution function}
\label{full_distrib.sec}

\begin{figure}
\begin{center}
\includegraphics{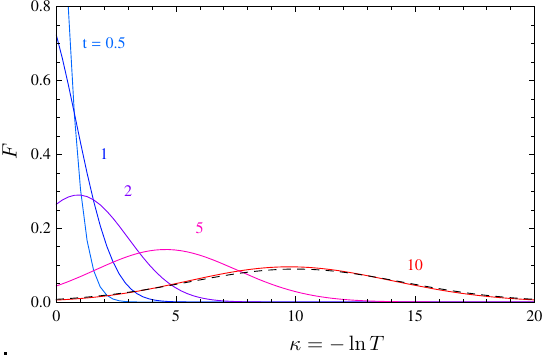}
\end{center}
\caption{\small%
Probability distribution function $F(\kappa,t) = W(e^\kappa,t)e^\kappa$ for the extinction 
$\kappa=\ln\rho=-\ln T$ across a one-dimensional channel of length $t=L/\xiloc=0.5,1,2,5,10$. Dashed
line at $t=10$: log-normal distribution
\eqref{Winfty.eq}.}
\label{Wofz.fig}
\end{figure}

Another choice of variable is $\rho=T^{-1}$, 
the dimensionless total
resistance of the channel. The Fokker-Planck equation \eqref{FPP.eq}  for its
probability distribution $ W(\rho,t) = P(\rho^{-1},t)\rho^{-2}$ reads 
\belab{FPW.eq}
\partial_t W = \partial_\rho\left[\rho(\rho-1)\partial_\rho W\right] 
\ee
with initial condition $W(\rho,0) = \delta(\rho-1)$, 
i.e., a wire of zero length has perfect transmission. 
The solution
can be calculated in closed form
\cite{Abrikosov1981}:  
\belab{w_exact.eq}
W(\rho,t) = \frac{\exp\{-t/4\}}{\sqrt{\pi} t^{3/2}} \int_{\arcosh\sqrt{\rho}}^\infty 
\frac{\exp\{-y^2/t\}\rmd(y^2)}
{\sqrt{(\cosh y)^2 - \rho}}. 
\ee
Figure \ref{Wofz.fig} shows how the distribution function 
$F(\kappa,t)= W(e^\kappa,t)e^\kappa$ for the extinction $\kappa=\ln\rho=-\ln T$ 
moves from a
$\delta$-distribution  with growing system size $t$ to the log-normal distribution
\eqref{Winfty.eq}, drawn as a dashed line at $t=10$.  

The full distribution function permits to calculate all moments
$\mv{\rho^{-n}}=\mv{T^n}$ of the transmission \cite{Abrikosov1981}. 
For $t\gg1$, one finds the asymptotic expression 
\belab{Tmoments.eq}
\mv{T^n} = \frac{\pi^{3/2}\Gamma(n-\frac{1}{2})^2}{2\Gamma(n)^2}\,
t^{-3/2}e^{-t/4}, 
\ee
showing that \emph{all} moments decay with the same dependence
$t^{-3/2}e^{-t/4}$. The first two moments are then 
\belab{T12moments.eq}
\mv{T} = \frac{\pi^{5/2}}{2}\,
t^{-3/2}e^{-t/4}, \qquad 
\mv{T^2} = \frac{1}{4} \mv{T} \ll 1.  
\ee
Although the average transmission and its fluctuations decay
exponentially, as expected for a strongly localizing system, 
the \emph{relative} fluctuations of the
transmission itself grow very quickly:
$\text{var}(T)/\mv{T}^2\propto t^{3/2}e^{t/4}$.

And why does the blind application \eqref{mvTwrong.eq} of the limiting
log-normal distribution \eqref{Winfty.eq} predict $\mv{T}_\text{log-norm}=1$ instead 
of the correct decrease \eqref{T12moments.eq}? 
Well, in \eqref{mvTwrong.eq}, we have used the normal distribution on
the entire real axis for $y=\ln T$, without paying attention to the
constraint that the physically admissible transmission is
$T\leq1$. To take this into account, a popular recipe consists in using a \emph{truncated} log-normal
distribution on the half-line $\kappa\geq0$ 
\cite{Beenakker1997,Evers2008}
\belab{Wtln.eq}
F_\text{log-norm}^+(\kappa,t) = \frac{\Theta(\kappa)}{C(t)\sqrt{\pi t}}
\exp\left\{-\frac{(\kappa-t)^2}{4t}\right\} 
\ee
with a normalization $C(t)=1+\mathrm{erf}(\sqrt{t}/2) \approx 2$. 
The moments of this distribution depend entirely on the value at
truncation, $\lim_{\kappa\to 0^+}F_\text{log-norm}^+(\kappa,t)= (2\sqrt{\pi})^{-1} t^{-1/2}e^{-t/4}$,
that is the probability for perfect transmission $T=1$. 
The transmission moments for $t\gg1$ are  predicted to be 
\be
\mv{T^n}_\text{log-norm}^+ \approx
\frac{1}{(2n-1)\sqrt{\pi}}t^{-1/2}e^{-t/4}.  
\ee
Comparing them with the exact moments \eqref{Tmoments.eq}, we see that the
truncated log-normal distribution can describe the
\emph{leading exponential decay}, but fails to capture the 
algebraic dependence correctly. Mathematically, this is due to the fact that the
log-normal distribution overestimates the probability of perfect
transmission
 $T=1$, which is really only $W(1,t)= (\pi^{3/2}/2)
t^{-3/2}e^{-t/4}$ for $t\gg1$.

\begin{figure}[b]
\begin{center}
\includegraphics{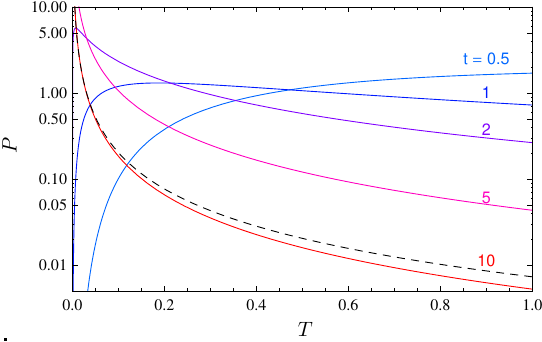}
\end{center}
\caption{\small%
Probability distribution function $P(T,t)$ for the transmission $T$ across
a one-dimensional channel of length $t=L/\xiloc$. Dashed
line at $t=10$: log-normal distribution
\eqref{Winfty.eq}.}
\label{PofT.fig}
\end{figure}

On physical grounds, this limited applicability of the log-normal
distribution emphasizes the difficulties one faces when dealing with
broad distributions. Figure \ref{PofT.fig} shows the transmission
probability distributions at different lengths $t=L/l$. It is
instructive to look at the last curve for the altogether moderate
system size $t=10$. The \emph{most probable} or ``typical'' value for the
transmission is quite small, $T_\text{typ}= \exp\{\mv{\ln T}\}=e^{-10} \approx 4.5\cdot10^{-5}$. The exact \emph{average
value} is much bigger, $\mv{T}\approx 1.06 \cdot10^{-2}$, indicating
that this average is to a large extent determined by very rare events
with anomalously large transmission. 
The truncated log-normal distribution, shown in
dashed, overestimates the frequency of large transmissions and predicts
$\mv{T}\approx 1.28 \cdot10^{-2}$. 

Let us close this section by emphasizing once more that the
typical transmission --- averaged over all possible relative
phases accumulated between consecutive scatterers --- displays
exponential decay at large distance as stated in eq.\ \eqref{Ttyp.eq}. This is 
in sharp contrast to the classical transmission \eqref{ohm.eq}, where
classical probabilities, not amplitudes, are combined to 
an algebraic decay. Clearly, averaging over disorder implies averaging
over quantum mechanical phases globally, but is not equivalent to
removing phase-coherence and interference effects locally from the
very start. 
This is a striking example of a \emph{mesoscopic} effect, a rather
counter-intuitive phenomenon, where microscopic 
phase coherence has macroscopic physical consequences 
that survive averaging over quenched disorder.

\section{Scaling theory of localization} 
\label{scalingtheory.sec} 

\subsection{What is a scaling theory?}

A scaling theory describes the relevant properties of physical systems by
considering their behavior under changes of size $L\mapsto b L$. 
Quantitative scaling arguments were invented in quantum field theory
in the context of renormalization. Scaling arguments became widely
popular in statistical
physics by the mid-60's for describing phase transitions and 
critical phenomena \cite{Kadanoff1967}. The immense success
of renormalization-group techniques developed in the 70's
\cite{Wilson1982} rapidly radiated to the field of disorder-induced phase
transitions that Anderson's celebrated paper had founded 
\cite{Anderson1958}. After pioneering work by Wegner
\cite{Wegner1976}, 
a scaling theory of localization was formulated by Abrahams,
Anderson, Licciardello, and Ramakrishnan \cite{Abrahams1979}, 
a quartet that became known as the ``gang of four''.

A scaling theory can hope to capture those features that are important on macroscopic
scales, but will be insensitive to microscopic details. This  means
that its predictions are only semi-quantitative, in the sense that it
cannot furnish the precise location of a
critical point in parameter space nor provide any system-specific
data. In return, if one feeds it with the microscopic data (such as
the transport mean-free path), it can give general, and surprisingly accurate, 
predictions of universal character.

Let us have a look at the different lengths characterizing 
quantum transport in a disordered material:
\begin{center}
\begin{tikzpicture}
\draw[->,>=latex] (0,0) node(left) {}	
		-- (1,0) node (lambda){$|$}
		-- (2,0) node (lcor){$|$}
		-- (4,0) node (L1){} 
		-- (5,0) node (l){$|$}
		-- (6,0) node (L2){}
		-- (7,0) node (xiloc) {$|$}
		-- (8,0) node (L3){} 
		-- (9,0) node[right](right){size}; 	 
\node [anchor=base](lambdalab) at ($(lambda)+(0,0.35)$) 	{$\lambda$};
\node [anchor=base](lcorlab) at ($(lcor)+(0,0.35)$)	{$\lcor$};
\node [anchor=base](l) at ($(l)+(0,0.35)$) 	{$l$};
\node [anchor=base](xiloclab) at ($(xiloc)+(0,0.35)$)	{$\xiloc$};
\draw[blue!50,dashed] ($(L1)-(0,0.35)$) node[below] {$L$} -- ($(L1)+(0,0.6)$); 
\draw[blue,dashed] ($(L2)-(0,0.35)$) node[below] {$L$} -- ($(L2)+(0,0.6)$); 
\draw[blue!50!black,dashed] ($(L3)-(0,0.35)$) node[below] {$L$} -- ($(L3)+(0,0.6)$); 
\node[draw,dotted,green,below,fit=(lambda) (lcor) (lambdalab) (lcorlab)](smallbox) {};
\node[draw,dotted,red,below,fit=(L1) (xiloc) (xiloclab) (L3)](bigbox) {};
\end{tikzpicture}
\end{center}
As short-scale lengths (in the left dotted box) one has the wavelength $\lambda=2\pi/k$ of the
propagating object and the correlation length $\lcor$ of the
disorder. If $\lcor\ll \lambda$, the details of the disorder are
unimportant, and models of $\delta$-correlated scatterers are
appropriate. If  $\lcor\gg \lambda$, the disorder correlation can be
resolved by the wave; this is typically the case for our example of optical speckle
potentials probed with ultracold atoms \cite{Billy2008}. 

As larger scales (in the right dotted box)  one has the transport mean free path $l$ and the
localization length $\xiloc$.  We have already seen in
section \ref{SL1.sec} that  in $d=1$ these two
lengths are practically identical, $\xiloc=2l$. In $d=2$ the localization length is much
larger than the transport length, as will be discussed in
sections~\ref{scaling_2d.sec} and~\ref{sc.sec} below, and in  $d=3$ it may well be infinite.  Depending on the
system size $L$, one can distinguish three basic transport regimes: \emph{ballistic}
transport through small samples with $L<
l$, \emph{diffusive} transport for $l<L<\xiloc$ with, possibly, weak-localization
corrections, and finally \emph{strong localization}
for large samples with $\xiloc<L$. In $d=1$, there is no 
room for diffusion between $l$ and $\xiloc$, and strong localization
is basically a single-scattering effect.%
\footnote{See \cite{Lugan2009,Gurevich2009} for
a situation where backscattering of an atom by a smooth speckle
potential is zero at lowest order, but localization
still prevails due to higher orders in perturbation theory.} 
The scaling theory of
localization has the purpose of describing the transition between
these regimes as function of system size $L$~\cite{Anderson1980}.  

\subsection{Dimensionless conductance} 

Traditionally, the scaling theory of localization is formulated in
terms of a channel's \emph{proper conductance}, a dimensionless parameter
defined as $g=T/R$ by 
transmission and reflection probabilities. Equivalently, one may
consider the channel's \emph{proper resistance} $g^{-1}$. A perfectly
transmitting channel $T=1$ has a proper conductance of $g=\infty$, and a perfectly resisting
channel with $T=0$ has $g=0$, which seems a rather sensible
definition. Moreover, we have seen in section \ref{Ohm.sec} that
this resistance is additive when classical subsystems are chained in series. 
Alternatively, one could define the total resistance as $\rho=1/T=1+g^{-1}$,
where the additional 1 represents the ``contact resistance'' due to the leads connecting
the sample to the external world.

In the previous section, we also learned that the transmission of a
disordered channel is a random variable with a broad
distribution around a most probable, typical value
$T_\text{typ}=\exp\mv{\ln T}$. Therefore also $g=T/(1-T)$ is a broadly
distributed random variable, fluctuating around the \emph{typical conductance} 
$g_\text{typ}=T_\text{typ}/(1-T_\text{typ})$. 
In all of the following, we discuss the behavior of $g_\text{typ}$, but in order not to
overburden the notation, we will simply write  $g_\text{typ}=g$.

\begin{figure}
\begin{center}
\includegraphics{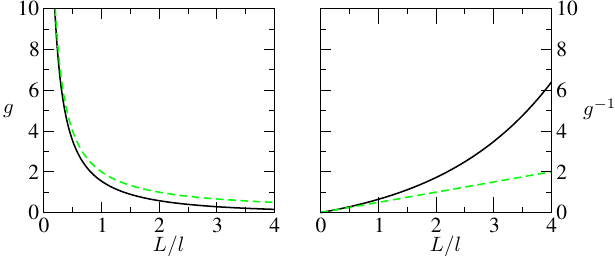}
\end{center}
\caption{\small%
Typical conductance (left panel) and resistance (right panel) of a 1d channel as function of
sample length $L/l$. Short channels have a resistance that grows
linearly as expected by Ohm's law (green dashed, eqn \eqref{g1ballist.eq}), whereas long channels show
exponentially large resistance (full line, eqn \eqref{g1loc.eq}).}
\label{g1.fig}
\end{figure}

In order to get used to this vocabulary, let us reformulate the
results of section \ref{1d.sec} for the typical
conductance. The exact exponential behavior \eqref{Ttyp.eq} of the typical
transmission translates into 
\begin{numcases}
{g(L)= \frac{1}{\exp\{L/2l\}-1} = }	
		2l/L,		& $L\ll l$, \label{g1ballist.eq}\\
		\exp\{-L/2l\} 	& $L \gg l$. \label{g1loc.eq}
\end{numcases} 
This conductance together with the resistance $g^{-1}$ is plotted in Fig.~\ref{g1.fig}. 
Only the conductance of short, ballistic channels is given by the
classical expression \eqref{g1ballist.eq}, that we have already
encountered as Ohm's law in section \ref{Ohm.sec}. 

The results of scaling for the conductance can be reformulated for
other quantities if those seem more convenient. One of the most popular,
and useful, quantities is the \emph{diffusion constant} 
$D = v l/d$, the product of velocity and transport mean free path 
divided by the number of dimensions $d$, a convention whose rationale will
become clearer below. 
For matter waves with wave vector $k$,
 the velocity is $v=\hbar k/m$, and the diffusion constant can also be
written $D=\frac{\hbar}{dm} kl$, i.e., the product of an elementary 
diffusion constant $(\hbar/dm)$ by the dimensionless quantity $k
l =2\pi l/\lambda$. This ratio 
describes the effective disorderedness of the medium: $kl\gg1$ means
that the wave can travel over many periods before suffering
scattering. We will see in Sec.\ \ref{wl.sec} below that $kl$ is a crucial
parameter for 
transport and localization properties. 

In a metallic sample with usual electrical conductivity, the Drude
formula $\bar\sigma=ne^2\tau/m$ 
establishes the direct proportionality between the conductivity $\bar\sigma$ and
the classical diffusion constant $D=v^2\tau /d$ of
charge carriers with mean free path $l = v\tau$.
We are thus led to define a dimensionless classical conductivity%
\footnote{This definition, as well as \eqref{conductance.eq}, uses
$\hbar/m$ available 
for quantum matter waves. It should be adapted to any other specific
transport problem under study, along the same lines. The somewhat
arbitrary factor of 2 is included for future convenience.}
\belab{conductivity.eq}
\class{\sigma} = \frac{2mD}{\hbar}. 
\ee 
The conductance $\class{g}$ of a sample of linear size $L$ in $d$ dimensions is
the ratio of $\class{\sigma}$ to $L^{2-d}$. 
To see this, picture a metallic block where the
voltage $U$ is applied along one dimension to give $E=U/L$, whereas the current
density $j=\bar\sigma E$ over the transverse area $L^{d-1}$ yields the total current
$I=L^{d-1}j$, which results in the dimension-full conductance $G=I/U=L^{d-2}\bar\sigma$.  
In order to define a 
dimensionless conductance $\class{g}$, 
one has to compensate the factors of $L$ with another length scale. The simplest choice is the
inverse of the wave number $k$. One can thus define a classical dimensionless conductance as
\belab{conductance.eq}
\class{g}(L) = (kL)^{d-2} \class{\sigma}
\ee
In 1d, this definition gives
$\class{g}(L)= 2l/L,$ i.e., is fully compatible with the known
exact result at short distance, eq.~\eqref{g1ballist.eq} 
(this is the reason for the factor 2 introduced in eq.~\eqref{conductivity.eq}).
It is of course no accident that the two definitions of dimensionless
conductance --- through the diffusion constant or through the
transmission across a sample --- coincide. The Landauer
formula~\cite{Landauer}  makes the connection explicit.

In any dimension, the classical dimensionless conductance can be rewritten as:
\belab{conductance_clas.eq}
\class{g}(L) = \frac{2kl}{d} (kL)^{d-2}
\ee
In particular, in dimension 2 we have $\class{g} =  kl$ itself, independently of the system size.

\subsection{Scaling in 1D systems} 

\label{scaling_1d.sec}
Since we wish to follow how the dimensionless conductance $g$ evolves with system size, 
we make use of the $\beta$-\emph{function},
\belab{betaofg.eq}
\beta = L \frac{\rmd \ln g}{\rmd L} = \frac{\rmd \ln g}{\rmd \ln(L/L_0)}. 
\ee
$\beta =0$ means that $g$ does no change with $L$. Actually, 
$\beta=cst$ implies purely algebraic dependence 
$g(L) \propto L^\beta$. The celebrated function $\beta(g)$ has been
introduced originally by Callan and
Symanzik  to describe the change of a coupling constant
under a change of scale within quantum field theory
\cite{PeskinSchroeder}. Let us familiarize ourselves with the
$\beta$-function, arguably the most important single object of
scaling theory, in the case $d=1$, for which we know already
everything exactly. It is a matter of elementary calculus to find 
\belab{betag1ofL.eq}
\beta(L) = - \frac{L}{2l} \frac{1}{1-\exp\{-L/2l\}}. 
\ee
Since $g(L)$ is a monotonous function of $L/2l$, one can easily invert
this dependence and express $\beta$ as function of the conductance
alone: 
\belab{betag1.eq}
\beta(g) = - (1+g) \ln\left[1+g^{-1}\right]. 
\ee
This result can also be derived directly as follows: the linear
scaling of the typical-transmission logarithm implies $T_\text{typ}(bL) =
[T_\text{typ}(L)]^b=\exp\{b \ln T_\text{typ}(L)\}$. Writing 
$T_\text{typ}^{-1}=1+g^{-1}$, differentiating with respect to $b$, and setting $b=1$ at the end
leads to \eqref{betag1.eq}. 
The fact that $\beta(g)$
can be expressed as function of $g$ instead of the original
length scale $L$ does not seem very profound in $d=1$
\cite{Abrikosov1981}. However, in field theory this property is vital for renormalizability
\cite{PeskinSchroeder}, and in statistical
physics it guarantees that
$\beta(g)$ can describe universal behavior close to a
phase transition.

\begin{figure}
\begin{center}
\includegraphics{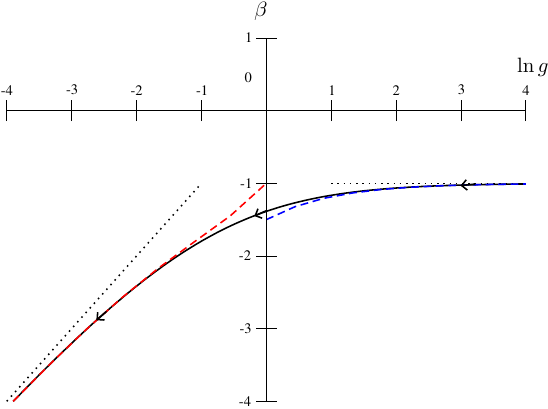}
\end{center}
\caption{\small
Conductance scaling $\beta$-function in $d=1$,
eq.~\eqref{betag1.eq}. Arrows show the flow from the 
Ohmian behavior \eqref{betag1asymp1.eq} for
$g\gg1$ in short samples to the exponential localization \eqref{betag1asymp2.eq} for
$g\ll1$ in long samples.}
\label{betag1.fig}
\end{figure}

Figure \ref{betag1.fig} shows the $\beta$-function for $d=1$, plotted
as function of $\ln(g)$ together with its asymptotics. For short
samples $L\ll l $, the conductance $g\propto L^{-1}$ is large, and
$\lim_{g\to\infty}\beta(g)=-1$. 
More precisely, one has the following asymptotic behavior:
\belab{betag1asymp1.eq}
\beta(g) = -1-\frac{1}{2g} + O(g^{-2})
\ee
which shows a \emph{weak-localization} correction (see section \ref{wl.sec}
below).
In the opposite limit of a large sample, $g\ll 1$ is exponentially
small, and 
\belab{betag1asymp2.eq}
\beta(g) = \ln g - g (|\ln g|+1) + O(g^2) . 
\ee
The transition between the two asymptotic regimes occurs around
$g=1$. 
The function $\beta(g)$ entirely describes how the dimensionless conductance evolves
with system size $L.$ Indeed, if $g$ is known for some small size $L$, it can be deduced for 
any other size by solving the differential equation
\eqref{betaofg.eq}, moving along the arrows shown on the curve of figure 
\ref{betag1.fig}. This is the so-called ``renormalization flow'' followed
by the system when its size is increased towards macroscopic scales.   
For $d=1,$ $\beta$ is always negative, implying that $g$ always
decreases with increasing system size, and thus 
the renormalization flow is unidirectional from right to left.
Characteristically, when $g$ decreases, $\beta$ becomes more negative, which
makes $g$ decrease even faster, until it finishes 
by dropping exponentially fast.

A crucial asset of scaling theory is that its predictions are valid for an
arbitrary 1d system, although the  
specific form of $\beta(g)$ was deduced using a specific model. Suppose you
have a large-size complex disordered system and that 
you want to study its conductance. You may start with a small system
for which you can calculate 
the conductance microscopically using a method of your choice. By following the renormalization flow,
you are then provided, almost magically, with the conductance at any
scale. Moreover, you find the
localization length $\xiloc$ as the 
system size for which $g(\xiloc)=O(1).$
Of course, there is no real magic here: your initial 
calculation
yields the mean free path $l$ and, as
this is the only macroscopic  
length scale relevant for transport in a
1d system, you finally have everything.  

\subsection{Quasi-1D systems} 

The previous results may also be applied to quasi-one-dimensional
systems that consist of several parallel channels $i=1,\dots
,N_\perp$. 
One may either have in mind channels that are literally
built parallel to each other \cite{Lahini2009} or a multi-mode
waveguide with spatially overlapping transverse modes. If there is no
coupling between channels, then the purely 1d description of section
\ref{1d.sec} applies. For
weakly coupled channels, which arises naturally by the
disorder present, an equation of motion for the full distribution
function of transmission eigenvalues very similar to \eqref{FPW.eq}
has been derived by Dorokhov and independently by Mello, Pereyra, and Kumar, 
known as the DMPK equation 
\cite{Beenakker1997}.
 Also the scaling picture remains
essentially the same. Keeping the number of transverse modes $N_\perp $
fixed, the short-scale conductance is 
$g = N_\perp 2l/L$, as expected for
parallel resistors. Thus, the
initial condition for the scaling flow on the curve $\beta(g)$ is changed, but the transition
to the localized regime is the same. Since the crossover again occurs
at $L=\xiloc$ with $g(\xiloc)=O(1)$, we simply find that the localization
length is increased toward $\xiloc= 2N_\perp l$. 
 
\subsection{Scaling in any dimension} 
\label{scalinganyd.sec}

In arbitrary dimension $d$, one changes the system size $L\mapsto bL$ in
\emph{all} directions, but still looks at the transmission along one
chosen direction. In the ballistic regime $L\ll l$, we start again from the
classical behavior, eq.~\eqref{conductance_clas.eq}, 
where  $g(L)
\propto L^{d-2}$. One
therefore expects to find $\lim_{g\to\infty}\beta(g)=d-2$, and
\belab{betagdasymp1.eq}
\beta(g) = d-2-\frac{c_d}{g} + O(g^{-2}) 
\ee
where a microscopic calculation is required to find the coefficient
$c_d$ that describes weak localization corrections. 

In the strongly localized regime $L\gg\xiloc$,
exponential localization prevails. And since adding parts to the system
beyond the localization length in the perpendicular direction cannot change its longitudinal
transport, we still expect the power law $T_\text{typ}(bL) =
[T_\text{typ}(L)]^b$ to hold in each channel. 
Thence follows 
the asymptotic behavior  
\belab{betagdasymp2.eq}
\beta(g) = \ln(g/g_d) 
\ee
in the strongly localized regime in any dimension, with a constant $g_d$ of order unity. 

Taking into account that the number of transverse
channels scales as $b^{d-1}$, we would 
obtain the simple scaling relation $T_\text{typ}(bL) \approx b^{d-1}
[T_\text{typ}(L)]^b$, if there were strictly no coupling between the
channels. Then, the same calculation than for 1d would give 
\belab{betagdapprox.eq}
\beta(g) = (d-1) - (1+g) \ln\left[1+g^{-1}\right],  
\ee
that is a simple vertical shift of \eqref{betag1.eq} by $d-1$. 
In particular, this would imply that the weak localization correction
$-c_d/g$ is the same in all dimensions, a result known to be wrong,
see Sec.~\ref{wl.sec}. It nevertheless remains true that the shape of the true $\beta(g)$
curves,  interpolating smoothly between the known asymptotics, 
is qualitatively given by \eqref{betagdapprox.eq}, see
also Fig.~\ref{betad.fig}. 
Although the scaling description encompasses arbitrary dimensions, its
consequences are radically different in $d=2$ and $d=3$, meriting a
separate discussion.

\subsection{\texorpdfstring{$d=2$}{d = 2}} 
\label{scaling_2d.sec}
\begin{figure}
\begin{center}
\includegraphics{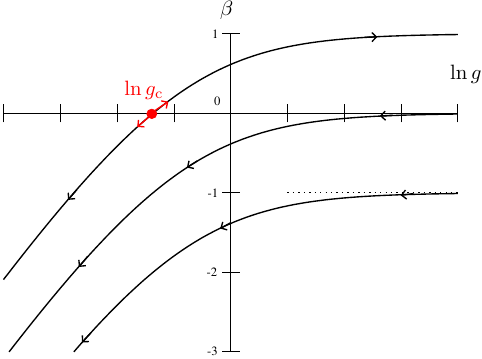}
\end{center}
\caption{\small%
Schematic plot of the $\beta$-function in $d=1,2,3$, showing a smooth
interpolation from the metallic regime  \eqref{betagdasymp1.eq} for
$g\gg1$ in short samples to the localized regime \eqref{betagdasymp2.eq} for
$g\ll1$ in long samples. Note the existence of an unstable fix-point at
critical $g_c$ in $d=3$.}
\label{betad.fig}
\end{figure}

In the ballistic limit of short samples with typical conductance $g\gg
1$, $\beta(g)\approx0$ describes scale-independent 
conductance of $N_\perp\propto L$ transverse channels, each with element conductance
$g\propto L^{-1}$. But then, $\beta(g)$ is not exactly zero. Starting
the flow at the finite conductance $g_0$ of a sample of length $L_0$, a slightly negative
$\beta(g)=-c_2/g$ makes $g$ decrease with size (it will be shown in section \ref{wl.sec} below that
$c_2=2/\pi$). We can integrate the
flow equation 
\be
\beta(g) = -\frac{c_2}{g} = \frac{1}{g} \frac{\rmd g}{\rmd \ln(L/L_0)}
\ee
by elementary means to find 
\belab{gofLd2.eq}
g(L) = g_0 - c_2\ln(L/L_0). 
\ee
To fix ideas, we can chose $L_0=l$, a scale on which transport is
classical, such that, from eq.~\eqref{conductance_clas.eq}, $g(L_0)=\class{g}=kl\gg
1$. 
The transition to the strong localization regime occurs at
$g(\xiloc)=O(1)$. Together with \eqref{gofLd2.eq}, this 
predicts an exponentially large localization length 
\belab{xiloc_2d.eq}
\xiloc \sim L_0 \exp\{g_0/c_2)\} = l \exp\{kl/c_2\}. 
\ee
The prediction of scaling theory for noninteracting particles in $d=2$
is therefore that all states are localized. This transcends also from
the scaling flow depicted in figure \ref{betad.fig}. However, the localization
length can be extremely large if the system is only weakly disordered
with $kl\gg1$. Let us take some figures from the Orsay experiment
\cite{Billy2008}. With $l=100\,\mu$m and $k=2.5\,\mu$m$^{-1}$, one
finds the rather large localization length $\xiloc\approx le^{400}$,
which would surely overstretch the possibilities of even the most capable
experimentalist. The take-home message here is: In order to observe 2d localization, $kl$ should be
chosen as close to unity as possible.   
In turn, this implies that the classical diffusion constant $D=\hbar kl/2m$ must be of the order 
of $\hbar/m$, which  
for a typical cold atomic gas is  a rather small quantity of the order of $10^{-9}\,
\mathrm{m}^2/\mathrm{s}$. 
In order to observe 2d Anderson localization with atomic matter waves,
the experimentalist must be capable to observe the dynamics 
for a long time while keeping phase coherence, a challenging task indeed.

\subsection{\texorpdfstring{$d=3$}{d = 3}} 

\label{scaling_3d.sec}

In $d=3$, we encounter a qualitatively new situation: the
$\beta$-function is positive for large $g$. So if we start with some $g\gg1$, the
conductance flow will take us to even larger values of $g$. In
renormalization-group terms, the behavior of the system in the
thermodynamic limit $L\to\infty$ is described by the ``infra-red stable
 fix-point'' $g=\infty$ (as in
statistical physics, but in contrast to quantum field theory, 
we are interested in the large-distance behavior,
i.e., the infra-red asymptotics with respect to momentum). Since large conductance characterizes a good
conductor, this is also known as the ``metallic'' fix-point. 

By contrast, if we start with some $g\ll1$, the negative
$\beta$-function will drive the system towards the stable insulating
fix-point $g=0$ with exponentially small conductance at finite
length. Between these two extrema, the $\beta$-function, 
assumed to be continuous, 
must have a zero at some $g_c$. A zero of
$\beta(g) \propto \rmd g/\rmd L$ is also a fix-point, but in this case an unstable one. This
unstable fix-point $\beta(g_c)=0$ marks the critical point and shows
the possibility of a \emph{metal-insulator phase transition} at some critical strength
of disorder. Although a scaling theory, with its roughly
interpolating $\beta$-function, cannot predict 
the precise position of the critical point, it can give a semi-quantitative estimate.
Indeed, a microscopic calculation of the transport mean free path $l$ provides us with
the dimensionless conductance $g(l).$ At such a scale, interference effect are unimportant,
and thus, eq.~\eqref{conductance_clas.eq} can be used, giving $g(l)\approx 2(kl)^2/3.$
As the critical point is such that $\gc\sim 1,$ we obtain that the
threshold for Anderson localization is given by:
\belab{ioffe-regel.eq}
kl\sim 1
\ee
an equation known as the Ioffe-Regel criterion for localization.
The precise value
of the critical $kl$ depends on microscopic details and is thus not universal.

Let us assume that the microscopic physics involves disorder whose
strength is measured by some parameter $W$, typically the width of the
disorder probability distribution (cf.\ Sec.\ \ref{anderson_model_1d.sec}).
Even though scaling theory does not predict
the precise position of the critical point, the  
behavior of the $\beta$-function around the critical point yields
precious information about the large-scale physics: it permits
to calculate \emph{critical exponents} that are the
hallmark of universality. In their 1979 paper \cite{Abrahams1979}, Abrahams et al.\ showed
that the localization length diverges close to the transition for 
$W>\Wc$ as 
\belab{xilocnu.eq}
\xiloc \sim (W-\Wc)^{-\nu}, 
\ee
where the critical exponent $\nu=1/s$ is determined by the slope of the
$\beta$-function at the transition, $s=\left[\rmd\beta/\rmd \ln
g\right]_{\gc}$. 

The calculation leading to this prediction is
elementary, but quite instructive in order to appreciate the power of a
scaling description. Let us start at some length $L_0$ with some value
$g_0<\gc$ on
the localized side of the fix-point. The $\beta$-function always
allows us to
calculate any other $g(L)$ implicitly by integration: 
\be
\ln\left( \frac{L}{L_0}\right) = \int_{\ln g_0}^{\ln g}
\frac{\rmd \ln g'}{\beta(g')}.
\ee
Using the linearized form $\beta(g)= s \ln(g/\gc)$ around the
fix-point leads to
\be
\left( \frac{L}{L_0}\right)^s = \frac{\ln(\gc/g)}{\ln(\gc/g_0)}. 
\ee
Now we are free to choose $L_0=\xiloc$ for which $g_0=O(1)$ such that 
$\xiloc \sim L \left[\ln(\gc/g)\right]^{-1/s}$. Because the
microscopic physics on small scales ignores the critical behavior on
large scales and can involve only smooth dependencies, 
one can always write $\ln(g/\gc) \approx (g-\gc)/\gc \propto
(\Wc-W)$ close enough to the critical conductance
$\gc$, and we finally end up with \eqref{xilocnu.eq}.

For the simplest possible interpolation \eqref{betagdapprox.eq},
one finds $\nu \approx 1.68$. This value is not disastrously far from the
true value $\nu=1.58\pm 0.01$ that is known today from extensive
numerical simulations~\cite{Slevin,Lemarie2009b}, cf.\ Sec.\ \ref{NumAnderson_d3.sec}. 

The ``metallic'' side of the transition can also be studied using a similar approach,
but following the metallic branch $\beta >0$ of the renormalization
flow. For smaller-than-critical disorder strength $W<W_c,$ 
the microscopically computed $g$ at some size $L_0$ will be slightly larger than $\gc.$
It is left as an exercise for the reader to show that this results at
large scale in a diffusive (i.e., metallic) behavior with a diffusion constant
\be
D \propto (W_c-W)^\nu. 
\ee 
The continuous (algebraic) vanishing of diffusion constant and 
conductance on the metallic side of the 
Anderson transition is characteristic of a continuous second order phase transition.

\subsection{\texorpdfstring{$d>3$}{d > 3}} 

The Anderson transition is expected to take place in any dimension
$d\ge 3$.
According to the simple scaling theory sketched above, the transition
point will shift to lower and lower $\gc$, requiring a more strongly
scattering medium to observe localization, and thus a 
Ioffe-Regel criterion, eq.~\eqref{ioffe-regel.eq}, with a smaller constant.

Contrary to conventional phase transitions, the Anderson
metal-insulator transition does not have a finite upper critical
dimension above which fluctuations would be unimportant and critical
exponents simply given by their mean-field values 
\cite{Evers2008}. This is compatible with the observation that as the
dimension $d$ increases, the zero of the
$\beta$-function must shift more and more to the asymptotic $\ln(g)$-wing
where the slope tends towards $s=1$. Thus, from the scaling description
it is tempting to surmise that the critical exponent tends towards
$\nu=1$ only continuously as $d\to\infty$. We will see in section
\ref{kicked.sec} below that this 
observation is not only a theoretician's spleen but may be put to
experimental testing.

\section{Key numerical and experimental results}

\label{experiments.sec}
Over the past 50 years, a wealth of numerical and experimental results has been accumulated on localization phenomena,
especially on Anderson localization in dimension 1, 2, 3 and beyond. 
In the following, we present a selection, necessarily subjective and limited, of the most remarkable results. 
 
\subsection{\texorpdfstring{$d=1$}{d=1}} 
\label{exp1d.sec}

Anderson localization is a generic feature in phase-coherent 1d and quasi-1d systems, as explained in Section~\ref{SL1.sec} above.  
Any amount of disorder, even very small, will eventually localize a wavepacket, independently
of how large its energy is. Of course, the localization length
can be huge if the energy is large compared to the 
disorder; see Section~\ref{loc_length_1d.sec} for a quantitative estimate.

\subsubsection{Localization of cold atoms}
\begin{figure}
\begin{center}
\includegraphics[width=0.45\linewidth]{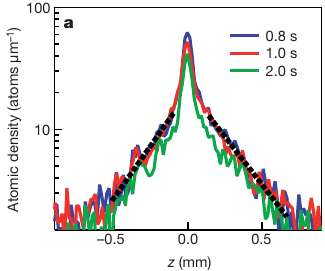}
\includegraphics[width=0.45\linewidth]{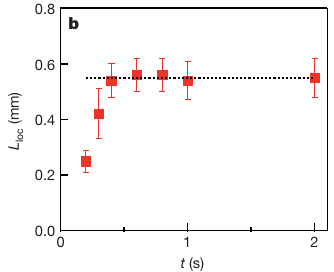}
\end{center}
\caption{\small%
a) The atomic density of a BEC expanding in a quasi-1d optical speckle
potential, shown in logarithmic scale at various times, displays clear
exponential localization in the wings, from which the localization
length is extracted by a fit (dashed line).  
b) The localization length first increases linearly with time, then saturates in the stationary regime. 
Reprinted from \cite{Billy2008} (courtesy of Ph. Bouyer).} 
\label{time_dynamics_orsay.fig}
\end{figure}

Concerning the experiment described in the Introduction, there is thus no surprise that a quasi-1d atomic wavepacket
displays localization in an optical speckle potential. 
Figure~\ref{time_dynamics_orsay.fig}a) shows the experimentally measured
spatial shape of the wavepacket at various times. One clearly
distinguishes an exponential decrease in the wings, from which a
localization length is extracted by a fit to
$\exp\{-2|z|/L_\text{loc}\}$. As shown in b), this localization length
first increases with time, then settles for a stationary value after
about $500\,$ms.  
In the stationary regime, the wavepacket displays spatial fluctuations
which are different for each single realization of the disorder. In
addition, there remains a large fraction 
of the atoms still trapped near the original location of the wavepacket. 

How can we understand these experimental results? 
The initial wavepacket is not monochromatic at all: it contains plane waves with a large 
dispersion in the wave-vector $k$ and consequently in the kinetic
energy $\hbar^2k^2/2m$ (the added optical potential also contributes
to the total energy, but is a small correction here). 
The initial, free expansion of an interacting Bose-Einstein condensate released from a harmonic trap leads to  
a population of the various $k$ classes that is given by an inverted parabola \cite{LSP07}:
\begin{equation}
\Pi_0(k) = \frac{3(k_\text{max}^2-k^2)}{4 k_\text{max}^3} \Theta(k_\text{max} - |k|),  
\end{equation}
where $k_\text{max}$ is the maximum $k$ value, related to the initial chemical potential by
$\mu = \hbar^2k_\text{max}^2/2m$. 

Because the disordered is ``quenched'' or stationary, energy is 
conserved, and each $k$-component of the wavepacket evolves independently.
When averaged over time, the interference terms between different energy components will be smoothed out, leaving the
averaged wavepacket as the \emph{incoherent} superposition of all energy components. From Section~\ref{SL1.sec}, we know that each $k$-component localizes with a localization length $\xiloc(k)$
equal to twice the transport mean free path. For the fastest atoms, this localization length is much larger than the initial spatial extension of the wave packet. 
Thus, we predict the stationary spatial distribution once localization sets in  to be roughly given by 
\begin{equation}
\label{wave_packet_final}
\langle |\psi(z)|^2\rangle  = \int_{-k_\text{max}}^{k_\text{max}}{\frac{\Pi_0(k)}{2\vthree{\xiloc(k)}} \exp\left(-\frac{|z|}{\xiloc(k)}\right)\ \rmd k}.
\end{equation}
Since $\xiloc(k)$ is an increasing function of $|k|,$  see section~\ref{loc_length_1d.sec},
the asymptotic decrease at large distance is dominated by the largest $|k|$ values,
such that $\langle |\psi(z)|^2\rangle \propto \exp[-|z|/\xiloc(k_\text{max})].$
The low-$k$ components have short localization lengths and thus produce the large bump near the origin in the
final density. Using only the wings of the experimentally measured density, it is possible
to estimate the localization length, for which we derive in
Sec.~\ref{loc_length_1d.sec} a theoretical prediction.

\subsubsection{Localization of light: a ten-Euro experiment}
The reasoning in section~\ref{SL1.sec} is entirely based on the construction of a $2\times2$ transfer matrix which can be chained;
any randomness in the transfer matrix then leads to localization. The fact that our starting point was a quantum matter wave 
and the underlying wave-particle duality of quantum mechanics
are not central to this argument. Indeed, the transfer-matrix description applies to all physical situations governed 
by a 1d (or quasi-1d) linear wave equation. Consequently, Anderson localization has been observed for many different types of non-quantum waves:
microwaves, elastic waves in solids, optical waves, to cite a few.

\begin{figure}
\begin{center}
(a)\includegraphics[width=0.45\linewidth]{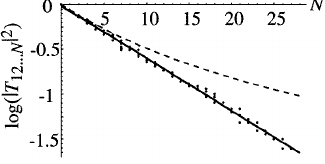}
(b)\includegraphics[width=0.45\linewidth]{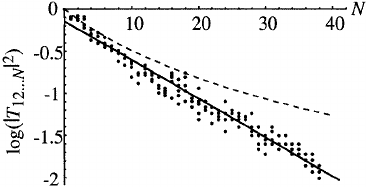}
\end{center}
\caption{\small%
Logarithmic
transmitted intensity across stacks of $N$ plastic films with mean thickness (a) 0.25\,mm;  (b) 0.1\,mm.
Dots: experimental data. Full curve: best fit to the data. Dashed curve:
prediction of incoherent transmission (Ohm's law). 
Reprinted from \cite{Berry1997} (courtesy of M.V.\ Berry).} 
\label{transparencies.fig} 
\end{figure} 

Even a poor man's experiment using viewgraph transparencies, 
i.e., plastic films made of polyester carbonate, 
allows to observe Anderson localization.  
A stack of several transparencies parallel to each other, separated by
air layers of randomly varying thickness, realizes the simple model 
shown in fig.~\ref{d1setup.fig}. The transmission and reflection
coefficients for an individual film can be computed from its index of
refraction and its thickness. If the randomness in the film spacing is
larger than an optical wavelength, we have a truly disordered system. 
Transport and localization can be observed by illuminating the stack
of transparencies with a plane light wave (or rather a good
approximation of a plane wave, 
namely, the light from a simple commercial He-Ne laser) and recording the transmission. 
Fig.~\ref{transparencies.fig} shows the transmission vs.\ the number $N$ of films or thickness of the sample.
It displays a clear exponential decay, one of the signatures of Anderson localization, and markedly differs from the 
linear decay (Ohm's law) predicted for the incoherent transport of intensities.

It is important to realize that absorption in the transparencies would also result in an exponential decay of the transmitted intensity. 
In all Anderson localization experiments, it is crucial to ensure that absorption is negligible, especially when working with electromagnetic waves. 
In this specific case, the bulk absorption coefficient of polyester
carbonate is known to be negligible here. In principle, one could also double-check that no photon is lost by measuring the reflection coefficient
of the sample and verifying that $R+T=1$. In condensed-matter experiments with electrons or cold atoms, 
number conservation of massive particles makes absorption less relevant and simpler to monitor.  
 
Another crucial requirement in the experiment is that it remains a 1d system, i.e., that there is a single transverse
electromagnetic mode involved. Light polarization is not an issue (for perpendicular incidence, scattering is independent of polarization),
but surface roughness or lack of parallelism can cause scattering into other transverse modes. This coupling into loss channels 
eventually destroys localization. Some indication of this loss 
is visible in Fig.~\ref{transparencies.fig}(b): the decay is not really exponential, but bent upwards, towards the prediction for incoherent transmission.

The experimental data show fluctuations of the transmission for various realizations of the experiment. 
This is not surprising, on the contrary, according to Section \ref{full_distrib.sec}, the fluctuations are even expected to be large. However, it turns out that
the observed fluctuations are smaller than predicted: in the plot, the transmission logarithm should appear as a cloud of points whose variance, eqn.\ \eqref{momentslogT.eq}, increases like $N$, which is not clearly the case.
Most probably, this is due to the experimental imperfections mentioned above that couple several transverse modes 
and consequently attenuate the fluctuations.

\subsubsection{Fluctuations}
As already emphasized several times, the existence of large fluctuations of the transmission in a characteristic
feature of Anderson localization. A key advantage of measuring relative fluctuations is that they are not much affected by absorption,
which merely induces a global decay of the whole transmission distribution. Consequently, in the last few years,
much progress has been made in calculating and measuring fluctuations in diffusive and localized systems. Fluctuations provide us with an unambiguous way of characterizing Anderson localization, even in the presence of absorption. 

\begin{figure}
\begin{center}
\includegraphics[width=5.6cm]{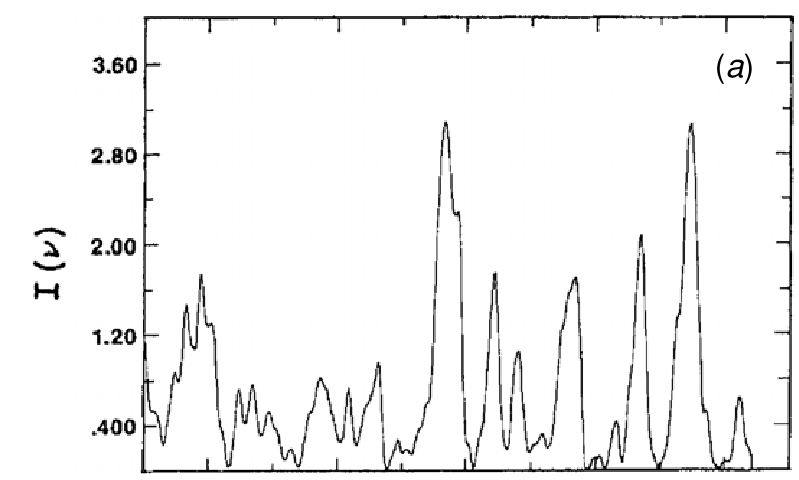}
\raisebox{-1.75em}{\includegraphics[width=6cm]{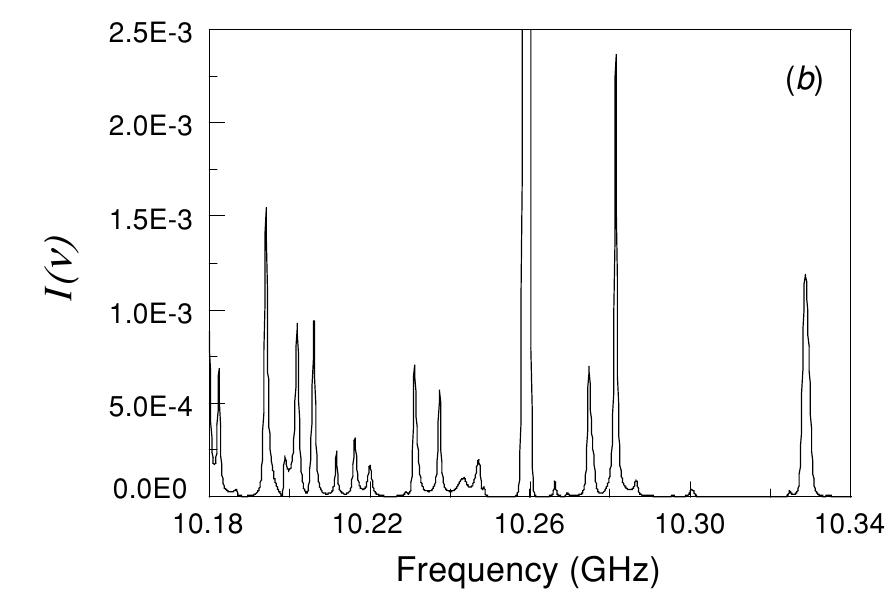}}
\end{center}
\caption{\small%
Microwave intensity transmitted across a copper tube filled with scattering aluminium spheres
vs.\ the microwave frequency.
In the diffusive regime (a) (frequency around 17 GHz), fluctuations
are comparable to the average value. 
In the localized regime (b),
huge fluctuations are visible, a hallmark of Anderson
localization. Reprinted from~\cite{Genack2005} (courtesy of
A.\ Z.\ Genack).} 
\label{genack_fluctuations.fig} 
\end{figure}

In order to illustrate this claim, we show in Fig.~\ref{genack_fluctuations.fig} the transmission of microwaves across a quasi-1d sample composed of aluminium spheres randomly disposed in a long copper tube
(cooled with liquid nitrogen so that absorption is negligible), as a function of the microwave frequency \cite{Genack2005}. 
The transport mean free path depends on the resonant scattering cross section of the aluminium spheres, and thus varies strongly with frequency, implying large changes in the relative sample length $t=L/l$.   
Plot (a) is obtained in the diffusive regime, where the localization length is longer than the sample size: there, the transmitted
intensity fluctuates in an apparently random way, but the fluctuations are relatively small, the rms deviation
being comparable to the mean. This is expected in the diffusive regime for relatively short samples, where the transmission amplitude itself is expected
to behave like a complex random number, whose real and imaginary parts are independent, normally distributed variables.
In contrast, in the localized regime shown in plot (b), the fluctuations are much larger, the transmission being most of the time
very small with some rare events of exceptionally high transmission, as predicted in Section~\ref{full_distrib.sec}. 

Visual inspection reveals immediately that
plots (a) and (b) are obtained in different regimes. While plot (a) has relatively
small fluctuations, characteristic of a diffusive regime, 
where the fluctuations are comparable
to the mean, plot (b) suggests some kind of huge (log-normal) fluctuations, typically associated
with the localized or critical regime.
The take-home message here is: don't rely solely on exponential decay to prove the existence of localization, look also
at the fluctuations, they are better indicators. Since the
relative fluctuations are insensitive to moderate absorption,
they may even provide a quantitative criterion whether the
strong localization threshold has been reached or not, and this under
circumstances when the exponentially decreasing transmission alone
could not be a reliable signature 
\cite{Chabanov2000,Chabanov2001a}.

\subsubsection{The Anderson model: a free (numerical) experiment} 
\label{anderson_model_1d.sec}

Although the Anderson model was originally introduced as a tight binding model for electrons
in a disordered crystal, it is of broader interest and has become a paradigm for one-body localization effects.

Up to now, we have considered continuous models where a wave propagates along a continuous
1d axis, encountering a set of discrete objects that 
scatter the wave backward and forward. The transfer-matrix game is just to combine
the discrete scatterers with the proper phases. One can even go one step further, disregard the ballistic propagation, and  
build a completely
discrete model, where the wave lives on a 1d lattice. 
One of the simplest  discrete models
is certainly the Anderson model whose Hamiltonian is
\belab{anderson_h}
H=\sum_{n=-\infty}^{+\infty}\left(w_n\ket{n} \bra{n} + t |n\rangle \langle n+1| + t |n+1\rangle \langle n| \right)
\ee
where the state $|n\rangle$ is the occupation amplitude of site $n$ with 
$w_n$ its on-site energy.  $t$ is the so-called ``tunneling'' matrix element coupling neighboring sites, traditionally 
taken with numerical value $t=1$. In a concrete physical realization, typically also the $t$'s are random variables and thus define ``off-diagonal disorder''. However, it is enough to take diagonal disorder to observe localization.    
The precise value of $t$ is irrelevant, as long as it is not zero, and may be used to 
define the energy scale of the problem.
If $w_n=0,$ it is easy to check that the eigenstates are discrete plane Bloch waves $\psi_n=\exp{ikn},$
for $k \in [-\pi,\pi[,$ and the energy is given by the dispersion relation $E(k)=2\cos k$ of this single-band model. 

Disorder is introduced by allowing the on-site energies $w_n$ to be random variables. 
The standard choice is to take $w_n$ to be uncorrelated random variables uniformly
distributed in the interval $[-W/2,W/2]$, with $W^2 =12\mv{w_n^2} $  measuring the disorder strength.
A noteworthy property---simplifying analytic calculations---is
that the spatial correlation length
of the disorder is zero (see Sec.~\ref{Speckle.sec} for a discussion of spatial
correlations arising in optical speckle potentials). 
Actually, the 1d tight-binding Anderson model can be solved for
almost any distribution~\cite{Luck}, with simple closed expressions
for the Cauchy-Lorentz on-site distribution. 
The techniques developed in Sec.~\ref{microscopic.sec} permit to show that the localization
length is 
at lowest order in $W$ given by 
\belab{xiloc_anderson_model.eq}
\xiloc = \frac{4 \sin^2 k}{\mv{w_n^2}} = \frac{12(4-E^2)}{W^2}.
\ee

\begin{exercise}
Few properties of the Anderson model.
\begin{itemize}
\item[$(i)$] Show that the equations obeyed by an eigenstate of the Anderson model at energy $E$
can be put in the following matrix form:
\begin{equation}
\begin{pmatrix} 
\psi_{n+1} \\
\psi_n \\
\end{pmatrix}
= T_n
\begin{pmatrix} 
\psi_{n} \\
\psi_{n-1} \\
\end{pmatrix}
\end{equation}
with a transfer matrix:
\begin{equation}
T_n = 
\begin{pmatrix} 
	E-w_n & -1\\
	1 & 0 \\
\end{pmatrix}. 
\end{equation}
Show that the amplitudes of the left and right propagating plane waves
in a disorder-free region can
be expressed as simple linear combinations of $\psi_n$ and $\psi_{n+1}$.
Show that, in consequence, it is possible to construct a transfer matrix $\sfM$ as in 
eq.~\eqref{Mdef.eq}.
This shows that the general results of section~\ref{transfer_matrix.sec} can be used and that exponential localization
is expected. 
\item[$(ii)$] 
Consider a continuous model of a 1d particle in a disordered potential $V(z).$
By discretizing the Schr\"odinger equation on a lattice with sufficiently small spacing
(much shorter than the de Broglie wavelength and the correlation length of the potential),
show than one recovers the Anderson model, but with spatially correlated $w_n.$
\end{itemize}
\end{exercise}

Numerical simulations of the Anderson model are extremely easy, at least in dimension 1.
Indeed, the previous exercise shows that the time-independent Schr\"odinger equation reduces, for an eigenstate $|\psi\rangle =
\sum_n{\psi_n |n\rangle}$ with energy $E$, to the three-term recurrence relation
\begin{equation}\label{Anderson_discrete_1d.eq}
\psi_{n+1} + (w_n-E) \psi_n + \psi_{n-1} = 0
\end{equation}
which can be solved recursively. In Fig.~\ref{pythonlist}, we give an 
example of a simple script, written in the \vthree{Python} language, 
that solves this equation at some arbitrary energy
across a random sample of arbitrary length.
\newgeometry{top=0.5cm,left=2.5cm,bottom=1cm}
\thispagestyle{empty}
\begin{figure}
\lstset{basicstyle=\ttfamily\footnotesize,language=Python,commentstyle=\color{red},breaklines=false,showstringspaces=false} 
\begin{lstlisting}
#!/usr/bin/python
from __future__ import print_function
import math
import random
import sys
# compute_AL1D.py
# Authors: Dominique Delande and Cord A. Mueller
# Release date: May 25, 2016
# License: GPL3 
# -----------------------------------------------------------------------------------------
# This script models localization in the 1D Anderson model with box disorder, 
# i.e. uncorrelated on-site energies w_n uniformly distributed in [-W/2,W/2].  
# The script computes the wave function and resulting transmission as function of energy,  
# system size L, and for nr number of realizations.  
# Without disorder, the dispersion relation is energy=2*cos(k), with support on [-2,2]. 
# The equations to be solved are:  psi_{n+1} + psi_{n-1} + (w_n-E) psi_n = 0
# They are solved in the backward direction, starting from an outgoing wave of unit modulus,
# with normalization to unit incident flux at the end of the computation. 
# Optionally, either the values of -log(T) are printed in the file logT.dat  
# or the intensities |psi_n|**2 are printed in the file logpsi2.dat.
# The localization length is 12*(4-energy**2)/W**2 to lowest order in W.
# -----------------------------------------------------------------------------------------
if len(sys.argv) != 6:
  print('Usage (5 parameters):\n compute_AL1D.py L W energy nr savepsi')
  sys.exit()
# switch savepsi: '1' for saving log|psi_n|^2 (with a small number of realizations) 
#                 '0' for saving -logT (with a large number of realizations)
L = int(sys.argv[1])
W = float(sys.argv[2])
energy = float(sys.argv[3])
nr = int(sys.argv[4])  
savepsi=int(sys.argv[5])  
if savepsi == 1:
  filename='logpsi2.dat'
elif savepsi == 0:
  filename='logT.dat'
else:
  sys.exit("Please choose between '1' for saving log|psi|^2 or '0' for saving -logT")

k = math.acos(0.5*energy)
exp_i_k = complex(math.cos(k),math.sin(k))
psi=[0.0]*(L+1)

def computepsi():
  psi[L] = exp_i_k
  psi[L-1] = 1.0
  for n in range(L-1,0,-1):
    w_n = W*(random.random()-0.5)
    psi[n-1] = (energy-w_n)*psi[n] - psi[n+1]
  return psi

def minuslogt(psi):
  psi_minus_1 = energy*psi[0]-psi[1]
  incident = (0.5*abs(psi_minus_1-exp_i_k*psi[0])/math.sin(k))**2
  return math.log(incident)
  
f=open(filename,'w')
for j in range(nr):
  psi = computepsi()
  if savepsi == 1:
    logpsi2 = [2*math.log(abs(x)) for x in psi] 
    for n in range(L):
      f.write('%6d %12.8g\n' % (n,logpsi2[n]-logpsi2[0])) 
    f.write('\n \n') 
  else:
    f.write('%6d %12.8g\n' % (j,minuslogt(psi)))
f.close()
print("Done, result saved to",filename,"!")
\end{lstlisting}
\caption{\small%
\vthree{Python script for solving the 1D Anderson model, Eq.~\eqref{Anderson_discrete_1d.eq}, also 
available at 
\url{http://www.lkb.upmc.fr/Anderson-localization-in-a-one}.
The script is run in a shell by typing  
``\texttt{python compute\_AL1D.py L W E NR svpsi}'' where the parameters \texttt{L,W,E,NR,svpsi} are respectively the system size, the disorder strength,
the energy, the number of disorder realizations, and the switch between an output of $\log|\psi_n|^2$ or $-\log T$, respectively. Typical results are exemplified in Fig.~\ref{numerics_1d.fig}. 
}}
\label{pythonlist}
\end{figure}
\restoregeometry

Note that the boundary condition used, $\psi_N=1$ and $\psi_{N+1}
= e^{ik}$, describes a purely outgoing wave with wavevector $k$ and
amplitude $1$ on the right end of the sample. The boundary condition
on the left is actually more complicated, because there the
incident wave interferes with the reflected wave, whose amplitude
depends on the microscopic realization of disorder of the entire
sample. The Schr\"odinger recursion equation is thus better solved
backwards from the far end of the sample,  yielding on average an
exponentially \emph{increasing} solution toward the left. This is in
agreement with the fact, shown in exercise 2$(iv)$ above, that the two
eigenvalues of the transfer matrix are of the form $\lambda_\pm=e^{\pm
2x}$. So starting with this boundary condition and an arbitrary value
of $k$ (and thus $E$), one has, with probability one, a finite overlap
with the eigenvectors of the larger eigenvalue and therefore
numerically picks up an exponentially growing solution. This solution
is then at the same time physically acceptable for the transmission
experiment, viz., decreasing on average exponentially from left to
right. Since eq.~\eqref{Anderson_discrete_1d.eq} is linear, one can
always normalize the solution to unit 
incoming 
flux and thus finally find the transmission probability as the
outgoing flux on the right side, calculated as a linear combination of
$\psi_N$ and $\psi_{N+1}$ (cf.~the \vthree{Python} script in Fig.~\ref{pythonlist} and exercise 4($i$)). 

\begin{figure}
\begin{center}
\includegraphics[width=6cm]{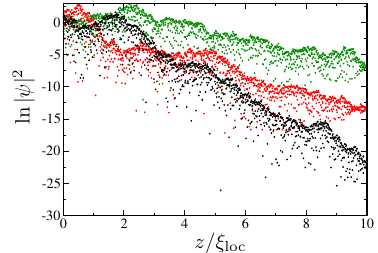}
\includegraphics[width=6cm]{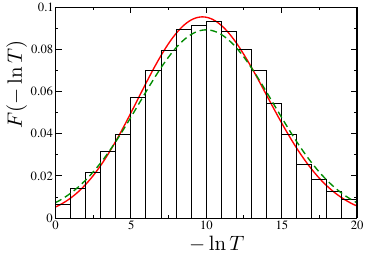}
\end{center}
\caption{\small%
Results of numerical experiments on the 1d Anderson model, with uncorrelated
uniform distribution of disorder. The left plot shows the intensity $|\psi(z)|^2$ inside the disordered medium---on a
logarithmic scale---for three different realizations of the disorder. 
Note the overall exponential decrease, decorated by huge fluctuations: the transmission
across a sample of size $L=10\xiloc$ fluctuates by more than 3 orders of magnitude.
This illustrates why the typical transmission differs from the average one.
The right plot shows the full probability distribution $F(-\ln(T),t)$ (histogram
over 10000 realizations) for $t=z/\xiloc=10,$ together with the
prediction eq.~\eqref{w_exact.eq} (full red line), which is close to a
truncated Gaussian (dashed). 
}\label{numerics_1d.fig} 
\end{figure}

Figure~\ref{numerics_1d.fig} (left plot) shows the intensity $\ln|\psi_n|^2$ 
for three different realizations of the disorder at energy
$E=0.5$ and disorder strength $W=0.6$ for a moderately large 
sample of 1250 sites, corresponding to a length $L=10\xiloc$. 
Although the decay is on the average exponential, huge
fluctuations 
from one realization to another are visible with the naked eye.  
Moreover, even for this pure
transmission experiment from left to right, the intensity is not monotonously decreasing at all, 
sometimes increasing by factors larger than 10. 
The histogram of the
extinction (or transmission logarithm) over 10000 realizations is shown in the right plot.
Its width clearly visualizes the huge fluctuations
characteristic for the localized regime. 
The agreement with the predicted distribution function, eq.~\eqref{w_exact.eq},
shown with a red full line, is excellent. One also observes the
convergence toward the truncated normal distribution (dashed), implying the
truncated log-normal distribution \eqref{Wtln.eq} for $T$ itself.

The reader is strongly encouraged to play with this script
to experiment personally with Anderson localization. We recommend the following numerical experiments: 

\begin{exercise} 
Numerical study of 1d Anderson localization
\begin{itemize}

\item[$(i)$] Use a single realization of the disorder and look at the wavefunction (or rather $|\psi|^2$)
inside the medium. Try different sample lengths and different energies, but avoid the band center $E=0$. Indeed,
the Anderson model is singular at this value. There is still Anderson localization, but the localization length slightly
differs from eq.~\eqref{xiloc_anderson_model.eq}, a so-called Kappus-Wegner singularity~\cite{Luck}.
\item[$(ii)$] Using a few hundred or thousand realizations, compute the statistical
distribution of the transmission. Compare with the exact prediction
eq.~\eqref{w_exact.eq} as well as with a truncated normal
distribution. 
\item[$(iii)$] Modify the script, for example for a Gaussian or Cauchy distribution
of disorder and run additional numerical experiments. You may also introduce correlated disorder to simulate
e.g. cold atoms in a speckle potential (see~\cite{Lugan2009} for generation of a realization of the disorder with proper
correlation functions).
\end{itemize}
\end{exercise}

A slightly different approach consist in diagonalizing the Hamiltonian for a large system numerically. Choosing strict boundary conditions on both ends of the sample yields normalizable eigenstates, centered at random positions within the sample and decreasing from there in both directions (similar to Fig.\ \ref{time_dynamics_orsay.fig}(a)),  
together with the discrete set of corresponding eigenvalues. In the thermodynamic limit, these eigenenergies form a dense, but still pure-point spectrum.

An important message to keep in mind is that there is very little difference between a continuous and a discrete system,
as far as localization on large spatial scales is concerned. For example, the exercise shows how a particle in a continuous 1d random potential
(e.g., an optical speckle) can be mapped to a variant of the Anderson model.
This universality of the Anderson model cannot really surprise because localization is an asymptotic property taking place at large
distance; whether the underlying configuration space is discrete or continuous plays only a minor role.

\subsection{\texorpdfstring{$d=2$}{d=2}} 

Scaling theory predicts $d=2$ to be the lower critical dimension for Anderson localization.
In dimension $d=2+\epsilon$ (which can be numerically studied by constructing an Anderson model
on a fractal set), a critical point should exist where $\beta(\gc)=0$, separating a diffusive phase
from an insulating one.  
Strictly at $d=2,$ scaling theory predicts localization provided there is a weak localization correction
with $c_2>0$, see section~\ref{scaling_2d.sec}. We will see in section~\ref{wl.sec} that this is indeed
what a microscopic approach predicts in spinless time-reversal invariant systems.
Scaling theory does not pretend to be an exact theory, there is thus a real interest in 
knowing whether there is localization in 2 dimensions for specific systems. 

Experiments with cold atoms are expected to be much more difficult than in 1d. Indeed, the localization length,
eq.~\eqref{xiloc_2d.eq},
is predicted to increase \textit{exponentially} with the parameter $kl$, instead of linearly in 1d.
Detailed theoretical studies~\cite{Kuhn2007} have shown that experimental observation requires at the same time
a speckle potential with a very short correlation length (comparable to what has been done in 1d, 
but in 2 directions) and a long atomic de Broglie wavelength, that is very cold atoms.
Altogether, satisfying all conditions is far from easy, making 2d Anderson localization of ultra cold atoms an interesting challenge.  

A metal-insulator transition has been observed for electrons in clean semiconductor samples~\cite{Kravchenko1994}.
It is generally acknowledged that the Coulomb electron-electron interaction---much stronger than the atom-atom interaction
in a dilute cold atomic gas---plays a major role in this transition, which is thus qualitatively different
from the pure Anderson transition and sometimes referred to as the Mott-Anderson transition \cite{Belitz1994}. 

\begin{figure}
\begin{center}
\includegraphics[width=8cm]{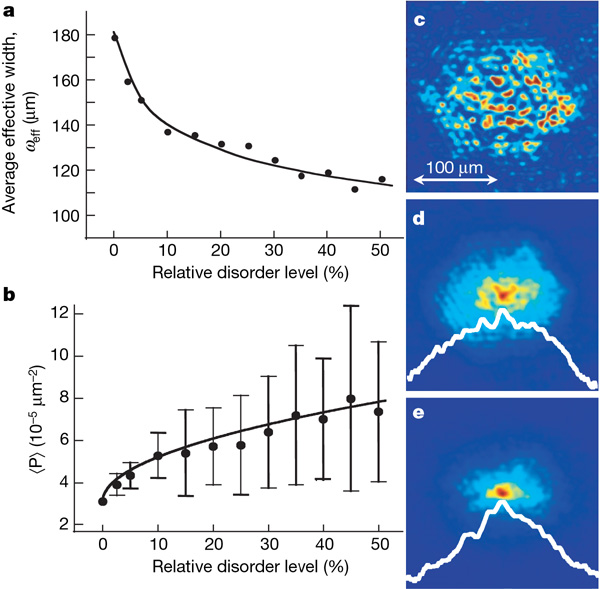}
\end{center}
\caption{\small%
Experimental results on the propagation of light across 
a transversally disordered 2d lattice of photonic wave guides, mimicking the 2d Anderson model. 
As the strength of the disorder is increased, the dynamics evolves from 
ballistic (a and c)  to diffusive (b and d), characterized by a Gaussian shape
of the wavepacket,
and eventually to Anderson localization, with a wavepacket of characteristic exponential shape (e),
when the disorder is sufficiently strong to make the localization length comparable
to the extension of the wavepacket. Reproduced from~\cite{Segev2007} (courtesy of S. Fishman).}
\label{segev.fig} 
\end{figure} 

Other types of waves have been successfully used in 2d systems. For example, using conveniently engineered optical fibers,
one can create a 2d ``photonic lattice" composed of parallel optical guides along which 
the light can freely propagate. Thanks to the photorefractive material used,  
its index of refraction can be adjusted by an external light source.  
Also the transverse coupling between the optical guides can be adjusted
at will, as well as the disorder due to small variations of the index of refraction in each guide.
As the light propagates at roughly constant velocity along the guides, the spatial propagation 
mimics the temporal evolution of the Anderson model,
each guiding mode playing the role of a site. Using such a device, the evolution from ballistic motion (on a scale shorter 
than the mean free path) to diffusive motion and eventually to strong localization has been
experimentally observed~\cite{Segev2007}, see Fig.~\ref{segev.fig}.

The Anderson model itself, described in section~\ref{anderson_model_1d.sec}, can be trivially
extended to any dimension by adding hopping terms to nearest neighbors in a (hyper)cubic lattice.
The numerical study is slightly more difficult than in 1d. The basic idea is to study first the quasi-1d propagation on a strip with a fixed number $M$ of transverse sites, imposing for example periodic boundary conditions along this direction. 
One can write a $2M\times2M$ transfer matrix for this quasi-1d system and calculate its asymptotic properties as the length  $N$ goes to infinity, extracting the quasi-1d localization length $\xiloc(M)$. 
Next, one studies the behavior of $\xiloc(M)$ as $M$ is sent to infinity. 
If $\xiloc(M)$ diverges without bounds, one concludes that the system is not localized. 
If on the other hand $\xiloc(M)$ tends to a finite limiting value, one concludes that the system localizes with 
$\xiloc= \lim_{M\to\infty}\xiloc(M)$. 

\begin{figure}
\begin{center}
\includegraphics[width=5cm]{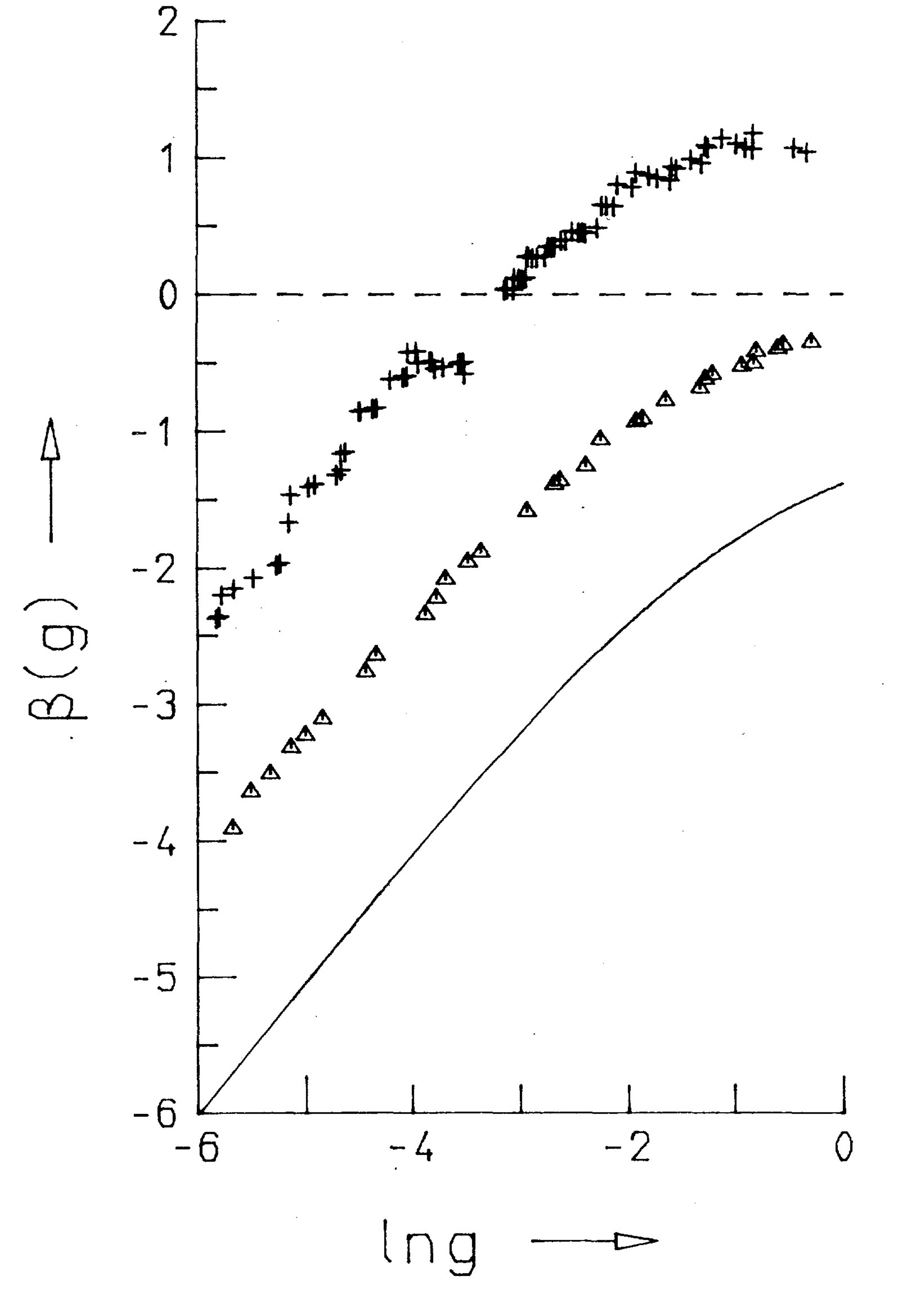}
\end{center}
\caption{ \small%
Scaling function $\beta(g)$ reconstructed from numerical simulations
of the Anderson model. The solid line is for dimension 1, the triangles for dimension 2 and the crosses for dimension 3.
Different sets of microscopic parameters   
produce data lying on the same curve, which can be considered an 
``experimental proof'' that the scaling approach is valid.
In 2d, $\beta(g)$ is always negative, proving that the system is in the localized regime. In 3d, depending
on the disorder strength, the system may be localized ($\beta(g)<0$,
strong disorder) or diffusive ($\beta(g)>0$, weak disorder). Reprinted
from~\cite{McKinnon1981} (courtesy of A. McKinnon and B. Kramer).}
\label{betag_numerics.fig} 
\end{figure}

Powerful numerical techniques, such as finite-size scaling \cite{finite-size-scaling}, make it possible to extrapolate
properties of the infinite system from numerical experiments on limited systems.
Especially, the scaling function $\beta(g)$ can be reconstructed, see Fig.~\ref{betag_numerics.fig}.
The fact that various data, computed for various values of the system parameters (energy, disorder
strength, system size), lead to the very same $\beta(g)$ strongly indicates that the scaling approach is valid, 
and thus corroborates the existence of universal properties independent of the
microscopic details.
In 2d, the numerically computed $\beta(g)$ is always negative, as expected and its shape is in good agreement
with the naive prediction, eq.~\eqref{betagdapprox.eq}.

\subsection{\texorpdfstring{$d=3$}{d=3}} 
\label{NumAnderson_d3.sec}
Dimension 3 is arguably the most interesting, because scaling theory there predicts a transition between
diffusive behavior for small disorder and Anderson localized behavior at large disorder.
Consequently, much experimental and numerical effort has been spent to observe this Anderson transition.
Numerical simulations of the 3d Anderson model are a very valuable tool, especially to locate the
critical point where $\beta(\gc)=0$ and to characterize its vicinity.
The results in Fig.~\ref{betag_numerics.fig} very clearly show the existence of the two regimes
and the fact that $\beta(g)$ behaves smoothly across the
transition. This constitutes a clear-cut proof that the Anderson
transition is a continuous phase transition of second order.  
 Note the absence of data for $g$ just below $\gc;$ this corresponds to localized systems with a localization length too large to be reliably measured in the
numerical simulations. As shown in Sec.~\ref{scaling_3d.sec}, the
slope $\rmd \beta/\rmd \ln g|_{\gc}$ at the critical point  is the inverse of the critical exponent $\nu$ of the Anderson transition.
Although the slope  at the critical point
cannot be accurately measured on these data, it is without any doubt
smaller than unity---the value of the asymptotic slope in the deep localized regime $\ln g \to -\infty.$ This implies that the critical
exponent $\nu$ is larger than unity. Recent numerical studies on much larger
systems fully confirm this point, the current best estimate being $\nu=1.58\pm 0.01$~\cite{Slevin,Lemarie2009b}.

Direct experimental observation of Anderson localization in 3d is even more difficult than in 2d,
because it requires an even more strongly scattering system ($kl$ smaller than 1 from the Ioffe-Regel
criterion, eq.~\eqref{ioffe-regel.eq}, instead of $kl$ of the order of few units). Moreover, creation
of a sufficiently disordered potential can be technically much more difficult in 3d: for a speckle potential, this would require to send plane waves with random phases from a large solid angle. Thus, Anderson localization of atomic matter
waves in a disordered potential has not yet been observed. However, using the equivalence of
a quasi-periodically kicked rotor with a 3d Anderson model, the Anderson transition with atomic matter waves
has been observed, and its critical exponent experimentally measured, as discussed in section~\ref{kicked.sec}.

\begin{figure}
\begin{center}
\includegraphics[width=0.5\linewidth]{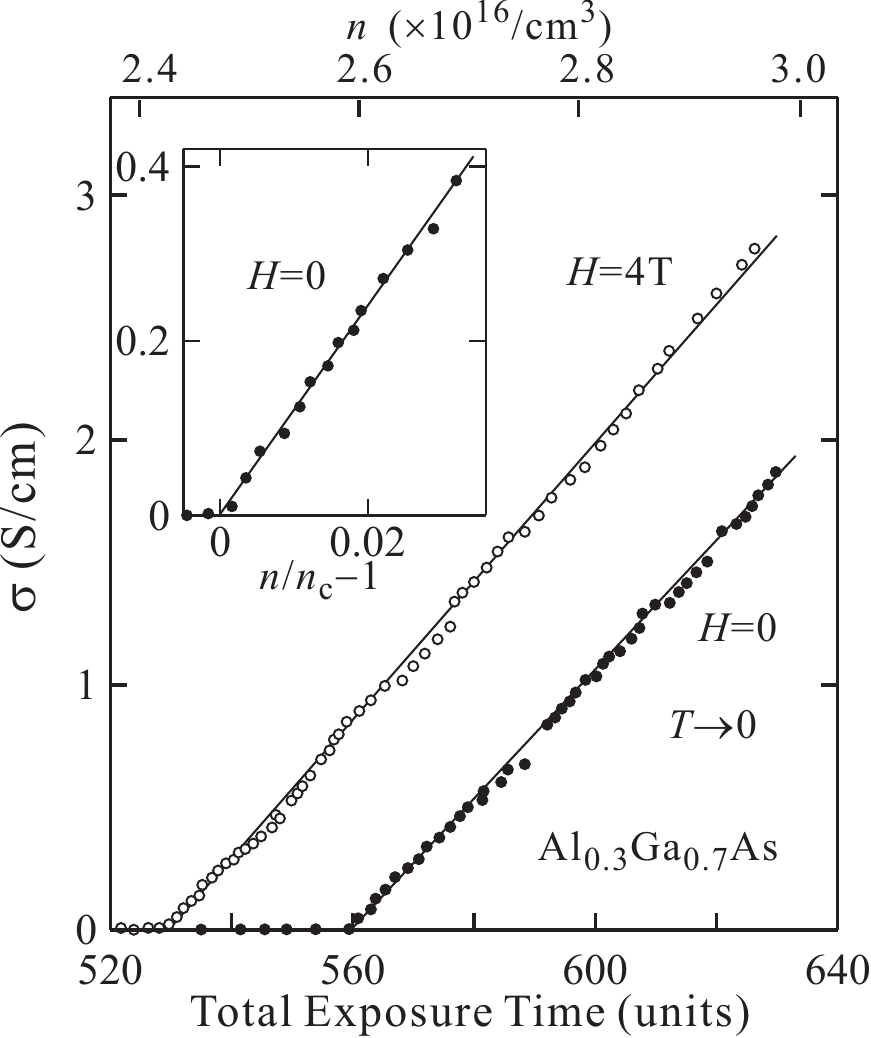}
\end{center}
\caption{\small%
Experimentally measured conductivity of a Si-doped AlGaAs 3d crystal vs.\ electron 
concentration (upper horizontal scale), showing a clear metal-insulator transition. The critical exponent
is very close to unity, a value not compatible with the universal value $\nu=1.58$ of the pure 3d Anderson
transition. Electron-electron interaction is probably responsible for
the difference. Adapted from~\cite{Katsumoto1987} (courtesy of
S. Katsumoto).}
\label{katsumoto.fig}
\end{figure} 

Electronic transport in solids, the field where localization theory was originally developed, provides 
also interesting experimental results. Metal-insulator transitions can be observed in solid state samples,
but it is never easy to identify the microscopic mechanism. This is because electron-electron interactions play an essential role. Whether the observed transition is a one-body effect like the Anderson transition or a many-body one like the Mott transition~\cite{Greiner2002} is not easily proved. We are not aware of any unambiguous observation of the pure Anderson transition. 
Figure~\ref{katsumoto.fig} shows the experimentally measured conductivity of a Si-doped AlGaAs crystal vs.\ a parameter
essentially representing the electronic Fermi energy. A clear insulator-to-metal transition is observed.
It seems that the curve is almost linear in the metallic regime, which---because conductivity is essentially a measure
of the diffusion constant ---means that the measured critical exponent is $\nu \approx 1.$ This markedly differs
from the exponent of the pure Anderson transition, indicating that interaction effects are probably important.

\begin{figure}
\begin{center}
\includegraphics[width=6cm]{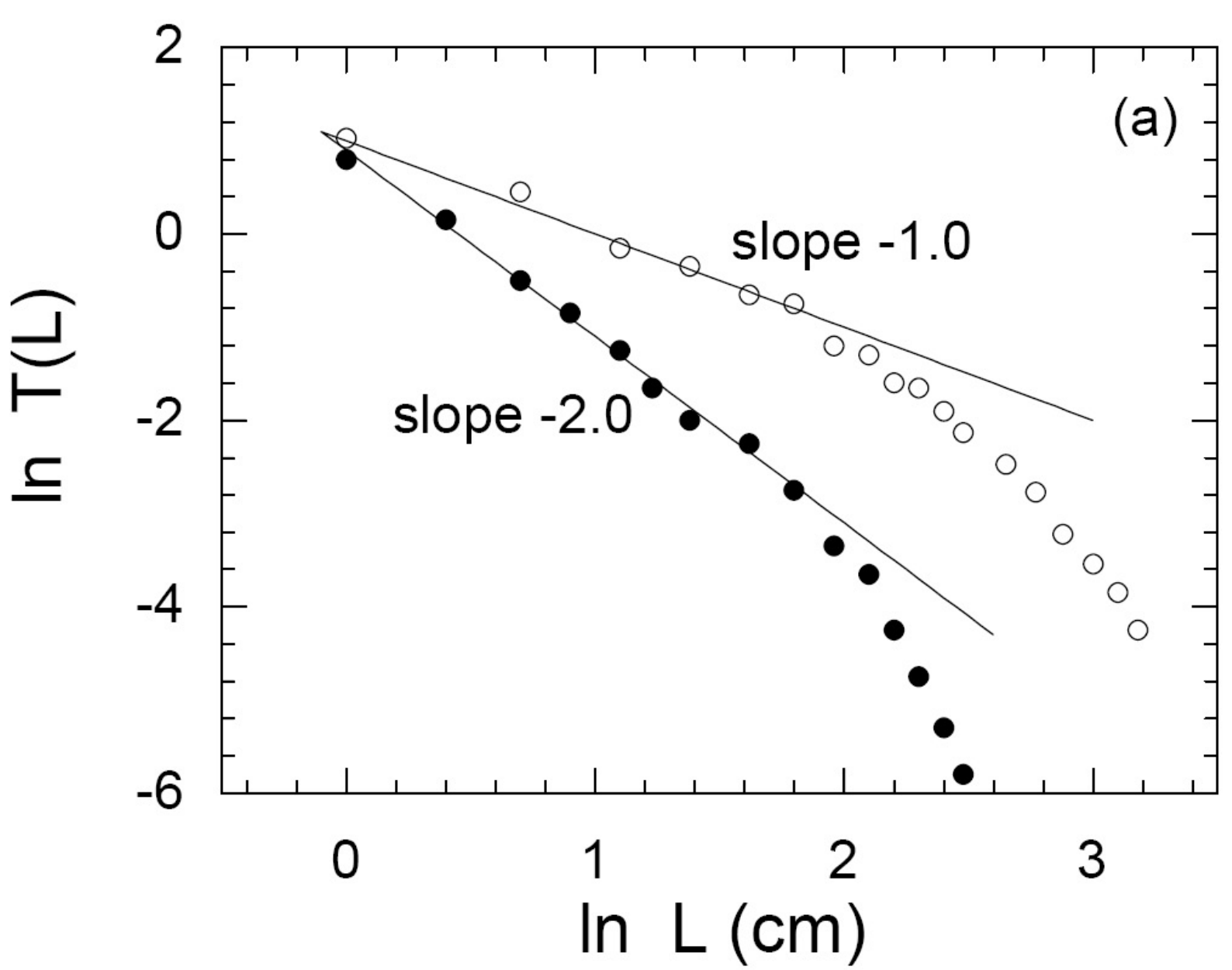}
\end{center}
\caption{\small%
Experimentally measured transmission of microwaves through a 3d 
strongly scattering medium vs. the size of the medium (doubly logarithmic scale). In the usual diffusive
regime (open symbols), a $1/L$ decrease is observed, in agreement with classical transport theory
(Ohm's law). At the critical point of the Anderson transition (filled symbols) a characteristic
$1/L^2$ behavior is observed, in agreement with the scaling theory of localization. Note that,
because of residual absorption, the signal drops at large size. Reprinted from~\cite{Genack2001} (courtesy of A.Z. Genack).
}
\label{genack_critical.fig} 
\end{figure}

One may also turn to other type of waves, for example ultrasonic waves~\cite{Page2008}  or electromagnetic waves. Direct measurement of the
electromagnetic field inside the disordered medium is not straightforward, and transmission experiments
are easier. As mentioned earlier, absorption induces an exponential decay of the intensity,
which must be carefully discriminated from the same effect being produced by Anderson localization. 
Thus, experimentalists have turned to measuring tell-tale properties right at the critical point.
There, according to scaling theory, the dimensionless conductance has the constant value $\gc,$ independently
of the system size, whereas the classical dimensionless conductance, eq.~\eqref{conductance_clas.eq}, 
increases linearly with the system size $L$. This additional power of $L$ makes the total transmission
across the sample evolve from a $1/L$ behavior (Ohm's law) in the diffusive regime to a $1/L^2$ 
scaling law at the critical point, and eventually to the exponential decay in the localized regime.
Any spurious absorption is likely to transform the critical $1/L^2$ behavior into an exponential decrease.
Thus, the existence of an $1/L^2$ may be considered a sensitive test of observing the Anderson transition.
Fig.~\ref{genack_critical.fig} shows the experimental result obtained on the propagation of microwaves in a disordered medium, in the diffusive and critical regimes. The existence of a range with $1/L^2$ power law---before
absorption wins at even larger size---is a convincing proof. 

Similar results
have been obtained in the optical regime~\cite{Wiersma1997}, where strong scattering is provided by oxide powders, but the
role of absorption has been discussed controversially~\cite{Maret1999}. Recently, time-resolved transmission experiments,
where absorption has less impact, have shown a slowing down of classical transport~\cite{Stoerzer2006}, which
gives strong evidence for Anderson localization.  

In the last few years, several numerical and laboratory experiments have characterized the fluctuations
appearing in the vicinity of the Anderson transition. 
 In particular, numerical experiments on the 3d Anderson model
have shown that the critical eigenstates have a multi-fractal structure, implying the coexistence
of regions where the wavefunction is exceptionally large together with regions where
it is exceptionally small. 
This is presently a very active field of research~\cite{Page2009}, whose description is beyond the
scope of these lectures. The reader may refer to the recent review paper of Evers and Mirlin~\cite{Evers2008}.
In the near future, it is very likely that experiments on localization of atomic matter waves will concentrate
on the existence and properties of fluctuations.

\section{Microscopic description of quantum transport}
\label{microscopic.sec}

\subsection{Diagrammatic perturbation theory} 

In section \ref{1d.sec}, we have seen that localization in (quasi-)one-dimensional
systems can be very efficiently described by a transfer-matrix
approach. In higher dimensions, we have resorted to the scaling
arguments presented in section \ref{scalingtheory.sec}. We now wish to
give an introduction to a microscopic description of quantum transport
in disordered systems. 
 The main advantage of diagrammatic perturbation theory 
lies in its versatility. It applies in arbitrary 
dimensions $d$ and to any
model with a Hamiltonian of the form 
\belab{modelH0plusV}
H = H_0 + V, 
\ee
in which $H_0$ describes regular propagation in an ordered
substrate, and $V$ is the disorder potential that breaks translational
invariance. 
On a microscopic level, ``disorder'' refers to  degrees
of freedom whose detailed dynamics are not of interest and whose properties
are only known statistically. In the following, we will consider 
\emph{static} or \emph{quenched} disorder that remains frozen
on the timescale of wave propagation under study (as sole 
exception of this rule, we mention in section
\ref{dephasing.sec} the dephasing effect
of moving impurities). 

A first model of type \eqref{modelH0plusV} describes a single quantum
particle in an external potential,
\belab{Hsingleparticle}
H = \frac{p^2}{2m}+ V(r), 
\ee
 with direct bearing on  experiments with non-interacting matter waves
like \cite{Billy2008}, but equally applicable to other massive
particles like electrons, neutrons, etc. Note that the plain Hamiltonian \eqref{Hsingleparticle} 
operating in Hilbert space 
describes the same single-particle 
physics as the more fanciful many-body version 
\belab{Hmanybody}
H = \int \rmd^d r \Psi^\dagger(r)\left[ -\frac{\hbar^2}{2m}\nabla^2+
V(r)\right]\Psi(r),  
\ee
defined in terms of particle creation and annihilation operators
$\Psi^{(\dagger)}$ in Fock space.

Since we assume that $H_0$ is translation-invariant (if only by
discrete translation on a lattice), the following equivalent
formulation in Fourier space is also useful:
\belab{HmanybodyFourier}
H = \sum_k\eps_k^0 a^\dagger_k a_k + \sum_{k,q}V_q a^\dagger_{k+q}a_k .
\ee
Here, wave vectors $k$ are used as good quantum numbers labeling the
eigenstates of $H_0$. If there is an
underlying lattice, one has to include also a Bloch band index. 
$\eps_k^0$ is the free dispersion relation; for matter waves, 
 $\eps_k^0=\hbar^2k^2/2m$. Particle annihilators $a_k =
L^{-d/2}\int \rmd^dre^{ik\cdot r}\Psi(r)$ and creators $a^\dagger_k$
fulfill the canonical commutation relations
$[a_k,a^\dagger_{k'}]_\pm=\delta_{kk'}$. 

The disorder potential breaks translation invariance by scattering
particles $k\to k'$ with an amplitude given by
its Fourier component $V_q = \bra{k+q}V\ket{k} = 
L^{-d} \int \rmd^dre^{-iq\cdot r}V(r) $, conveniently represented by 
\belab{Vq.eq}
V_q= \Vq . 
\ee

This model Hamiltonian \eqref{HmanybodyFourier} is not limited to matter waves. By appropriate
changes in $\eps_k^0$, different physical systems can be described. For
instance, photons and other mass-less excitations have a linear
dispersion $\eps=\hbar c k$ with characteristic speed $c$. 
Yet another realization is provided by the elementary excitations of Bose-Einstein
condensates, featuring the Bogoliubov dispersion $\eps_k=
\sqrt{\eps_k^0(\eps_k^0+2\mu)}$,  
that interpolates between a linear sound-wave
dispersion at low energy and a quadratic
particle-like dispersion $\eps_k^0=\hbar^2k^2/2m$ at high energy. 
The general formalism to be introduced below applies to all these
cases, provided the scattering potential $V_q$ is known. 

The basic model can be made richer, depending on the
circumstances and effects one wishes to describe. For example, spin
often  plays an important role, for instance via spin-orbit effects, 
due to coupling of spin and direction of propagation. Also spin-flip
processes can be of interest, as in electronic spin-flips induced by
magnetic impurities or photon polarization flips induced by
Zeeman-degenerate atomic dipole transitions.  
A typical spin-flip process $(m,\sigma)\mapsto(m',\sigma')$
changes the spin of the propagating object from $\sigma$ to $\sigma'$,
while the impurity spin undergoes $m\mapsto m'$. Processes
of this type can store information about the path traveled,
and generally act as a source of strong decoherence (as discussed in 
Sec.~\ref{dephasing.sec} below). 

The main drawback of the diagrammatic Green function approach is its
perturbative character. Most results are obtained from an expansion in
powers of $V$ and are valid only for small enough potential
strength. If one is interested in truly strong-disorder effects, it may be worthwhile to start from the
opposite situation where the propagation described by $H_0$ is small;
Anderson's original method of a ``locator expansion''
\cite{Anderson1958} 
is an example for such a weak-coupling perturbative approach.  
In any case, it takes considerable effort to derive non-perturbative
results using controlled approximations. Yet, the basic
diagrammatic technique is a prerequisite for more powerful,
field-theoretic methods involving, e.g., replica methods,
renormalization-group analysis and supersymmetry~\cite{Efetov}.  

\subsubsection{Quantum propagator}

Let us then start by calculating the Green function for the
single-particle Hamiltonian \eqref{Hsingleparticle} that determines
the time evolution of a state $\ket{\psi}$ in Hilbert space according to the Schr\"odinger equation
$i\hbar\partial_t\ket{\psi}=H\ket{\psi}$. For $t>0$, the forward-time
evolution operator $\GR(t) =
-\frac{i}{\hbar}\theta(t)\exp\{-iHt/\hbar\}$  solves the
differential equation 
\be
\left[ i\hbar \partial_t - H\right]\GR(t)=\delta (t). 
\ee
Obviously, $\GR(t)$ is the retarded Green operator for the
Schr\"odinger equation. It encodes the same
information than its many-body version
$\GR_{kk'}(t)=
-\frac{i}{\hbar}\theta(t)\mv{\left[a_k(t),a^\dagger_{k'}\right]}$
that one would use starting from  
\eqref{HmanybodyFourier}; in the following, we stick to the simpler
form, referring the reader to the literature for the more
advanced presentation
\cite{BruusFlensberg,NegeleOrland,Mahan,AltlandSimons}. 
Going from time to energy by Fourier transformation, one defines the \emph{resolvent}
\belab{GRdef.eq}
\GR(E)= \lim_{\eta\to 0^+} \int \rmd t e^{i(E+i\eta)t/\hbar}\GR(T) =
\lim_{\eta\to 0^+}\left[E-H+i\eta\right]^{-1} =: \left[E-H+i0\right]^{-1}.
\ee
The limiting procedure $\eta\to 0^+$ guarantees that indeed the retarded
Green operator is obtained, different from zero only for $t>0$.  The
advanced Green operator $\GA(t)$ is obtained by taking $\eta\to
0^-$. 
In the basis where $H$ is diagonal, $H\ket{n}=\eps_n\ket{n}$, the
resolvent is also diagonal and thus admits the spectral decomposition 
$G(z)=\sum_n \ket{n}[z-\eps_n]^{-1} \bra{n}$ for any argument
$z\in\mathbbm{C}$ outside the spectrum of $H$. The resolvent's matrix elements are
called ``propagators''. For example, in the position
representation  $\langle r\ket{n}=\psi_n(r)$, 
\be
\GR(r,r';E) = \bra{r'}\GR(E)\ket{r} = \sum_n
\frac{\psi_n(r')\psi^*_n(r)}{E-\eps_n+i0} = \Grrprime . 
\ee
This propagator contains precious information: As function of $E$, it
has singularities on the real axis that are precisely the
spectrum of $H$ and thus encode all possible evolution
frequencies. Furthermore, the residues at these poles provide information about the eigenfunctions. 

The total Hamiltonian \eqref{Hsingleparticle} contains the disorder
potential so that we cannot write down its 
eigenfunctions and eigenvalues analytically  (numerically, one may of course
calculate eigenfunctions and eigenvalues for each realization of
disorder). We start therefore with the free
Hamiltonian $H_0$. Its  resolvent
$G_0(z)=[z-H_0]^{-1}$ is diagonal in momentum representation,
$\bra{k'}G_0(z)\ket{k}=\delta_{kk'}G_0(k,z)$ with 
\be
\GR_0(k;E) = \frac{1}{E-\eps_k^0+i0} = \arprop{k} . 
\ee
Now we are ready to describe the perturbation due to the potential
$V$: Using $G(E) = [E-H_0-V]^{-1} = [(E-H_0)\{1- (E-H_0)^{-1}V\}]^{-1}$,
we express 
\begin{align}
G(E) 	& =  [ 1-G_0V]^{-1} G_0 \\
	& =  G_0 + G_0VG_0 + G_0VG_0VG_0  +\dots \label{Born.eq}
\end{align}
 as the Born series in powers of $V$. For notational brevity, we have
already suppressed the energy argument on the right-hand side. Still, if one
tries to write out a matrix element $\bra{k'}G(E)\ket{k}$, the operator
products convert into cumbersome expressions that tend to obscure the series'
simple structure: 
\be
\bra{k'}G(E)\ket{k} = \delta_{kk'} G_0(k) + G_0(k') V_{k'-k} G_0(k) 
+ \sum_{k''} G_0(k') V_{k'-k''} G_0(k'') V_{k''-k} G_0(k)  +\dots 
\ee
At this point, we are well advised to use the graphical representation
known as ``Feynman diagrams'', for which we have already all
ingredients at hand: 
\belab{Borndiag}
\bra{k'}G(E)\ket{k} = \delta_{kk'} \arprop{k} + \GoVGo +
\GoVGoVGo+\dots 
\ee
Already, we achieve a much more compact notation, aided by the fact
that we do not need to label the dangling impurity
lines, defined in \eqref{Vq.eq}, since their momentum is automatically determined by the incident and
scattered momenta. Also, we henceforth use the prescription that all
internal momenta have to be summed over, here $k''$ in the last
contribution.

\subsubsection{Ensemble average}

In principle, the Born series \eqref{Borndiag} permits to calculate
the full propagator perturbatively. However,
the result will be different for each realization of disorder. We are
really only interested in suitable expectation values and thus have to
understand how to perform the ensemble average over the disorder
distribution.  

The potential $V(r)$ as a function fluctuating in space is a \emph{random
process}. As such, it can be completely characterized by its
moments or correlation functions $\mv{V_1}$, $\mv{V_1V_2}$, 
$\mv{V_1V_2V_3}$, etc., with the short-hand notation $V_i=V(r_i)$. 
We will assume that the process is \emph{stationary} or, preferring the
spatial dictionary, \emph{statistically homogeneous}, which means that correlation
functions can only depend on coordinate differences $r_{ij} = r_i-r_j$. We can therefore
define the following correlation functions and corresponding diagrams:
\begin{alignat}{2}
\mv{V_1} 	&= \mv{V} 				\\ 
\label{defP.eq} \mv{V_1V_2} 	&= P(r_{12})   	&&= \VPr{1}{2}	\\
\mv{V_1V_2V_3} 	&= T(r_{12},r_{23})	&&= \VTr{1}{2}{3}	 	
\end{alignat}  
and so on for arbitrary $n$-point correlation functions. Without loss
of generality, one may always take $\mv{V}=0$ by defining a centered
potential $V\mapsto V-\mv{V}$ while redefining the zero of
energy $E-\mv{V}\mapsto E$. 
In Fourier representation, these correlation functions are 
\belab{Pofq.def}
 P(q)   	= \VPk{q},\qquad 
 T(q,q')	= \VTk{q}{q'}, \qquad \text{etc.}	 	
\ee 
Depending on the specific type of disorder, these general 
correlation functions can take different forms, and it may be
instructive to discuss two of them in detail. 

\subsubsection{Gaussian disorder}
\label{GaussianDisorder.sec}

As a first example, let us consider Gaussian-distributed disorder 
that is completely defined by its first two moments $\mv{V}=0$ and $ P(r) =
V_0^2 C(r)$. Here, one conveniently factorizes the one-point variance
$V_0^2=\mv{V_1^2}$ from the spatial correlation function  $C(r)$ that obeys $C(0)=1$
by construction. The characteristic property of a Gaussian process is that all
higher-order correlation functions completely factorize into pair
correlations. Indeed, a simple property of the Gaussian integral
implies that the moments of a normally distributed, centered scalar
random variable $X$ are $\mv{X^{2n}}=\mathcal{C}_n\mv{X^2}^n$ where
$\mathcal{C}_n=(2n)!/(2^nn!)$ is the number of pairs that can be
formed out of $2n$ individuals. Similarly, the Gaussian moment theorem
applies to a Gaussian-distributed random potential: 
\be
\mv{V_1 \cdots V_{2n}} = \frac{1}{2^nn!}\sum_\pi \mv{V_{\pi(1)}V_{\pi(2)}} \cdots \mv{V_{\pi(2n-1)}V_{\pi(2n)}}
\ee
where $\pi$ denotes the $(2n)!$ permutations. Pictorially, this
implies also a complete factorization of $2n$-point potential
correlation into products of pair correlations. The first interesting example is
$n=2$ with 
\belab{Gaussiandecomp.eq}
\VQ = \Vgaussone + \Vgausstwo + \Vgaussthree
\ee
and so on for higher orders. 

Such a Gaussian potential can be constructed with arbitrary
spatial correlation $C(r)$. A popular choice here is often
to model it as a  
Gaussian as well, $C(r) = \exp\{-r^2/2\sigma^2\}$, such as in \cite{Hartung2008}, 
because this is easy to implement
numerically (it suffices to draw uncorrelated random variables $V_i$ on
a discrete grid and convolute by a Gaussian correlation function afterwards). 
Moreover, this choice leads to simple analytical calculations because also
the $k$-space pair correlator is Gaussian, $P(q) =V_0^2\sigma^d
(2\pi)^{d/2}\exp\{-q^2\sigma^2/2\}$. 
In the limit of low momenta $q\sigma\ll 1$, the potential details
cannot be resolved and it appears 
$\delta$-correlated. Then, everything can be expressed in terms of 
$P(0)=(2\pi)^{d/2}\sigma^dV_0^2$. 

\subsubsection{Speckle}
\label{Speckle.sec}

A slightly more interesting example is provided by the optical speckle
potential used recently for matter-wave
Anderson localization \cite{Billy2008,Clement2006}. The atoms are subject to an
optical dipole potential $V(r)= K |E(r)|^2$ created by the local
field intensity of far-detuned laser light. $K$ contains the
frequency-dependent atomic
polarizability besides some constants~\cite{AllenEberly}. 
With a laser beam that is blue-detuned from the optical resonance, one
has $K>0$ and thus expells atoms from high-intensity regions. This potential
landscape features repulsive peaks with $\mv{V}>0$. Conversely,
a red-detuned laser leads to $K<0$, and one
finds a potential landscape with attractive wells and  $\mv{V}<0$.   
To create a disorder potential, the laser beam is focused
through a diffuse glass plate, whose randomly positioned individual grains act as
elementary sources for the emitted field. 
The electric field $E(r)$ at some far point
then is the sum of a large number of complex amplitudes. By virtue of
the central limit theorem, it is a \emph{complex Gaussian random variable}
with normalized pair correlator
\belab{spcklcorr}
\gamma_{ij}=\gamma(r_i-r_j) = \frac{\mv{E^*(r_i)E(r_j)}}{\mv{|E|^2}} =
\spcklcorr{i}{j} 
\ee
with the obvious properties $\gamma_{ij}^*=\gamma_{ji}$ and
$\gamma_{ii}=1$. In a 1d-geometry, the pair correlator takes its
simplest form in Fourier components: 
\belab{gammaq_1d.def}
\gamma(q) = \pi\lcor\Theta(1-|q|/k_\lcor)
\ee
where $k_\lcor=k \alpha$ is the maximum wave-vector that can be built
from a monochromatic laser source with wave vector $k$ 
seen under an optical aperture $\alpha$. This
eqn~\eqref{gammaq_1d.def} simply says
that the random field contains all wave vectors inside the allowed
interval with equal weight. In real space, this Fourier
transforms to $\gamma(r) = \sin(r/\lcor)/(r/\lcor)$ with the
correlation length $\lcor= 1/k_\lcor=1/(\alpha k)$. 

Other pair correlations such as $E_iE_j$ and $E^*_iE^*_j$ have
uncompensated random phases and average to
zero. The Gaussian moment decomposition now applies to arbitrary moments of
the speckle disorder potential $V_i=KE_i^*E_i$. An $n$-point potential correlation is
really a $(2n)$-field correlation, which 
decomposes into all possible pair correlations
\eqref{spcklcorr}. As in a conventional ballroom dancing situation
involving  $n$ couples, all possible heterosexual pairings between the
$E^*_i$s and $E_j$s are allowed. This gives for the 2-point potential
correlator 
\belab{V12def3.eq}
\mv{V_1V_2} 	= K^2 \mv{E^*_1E_1E^*_2E_2} 
		= \mv{V}^2\left[\gamma_{11}\gamma_{22} +
\gamma_{12}\gamma_{21}\right] .
\ee
Setting $r_1=r_2$ shows that $\mv{V^2}=2\mv{V}^2$, which means that
the potential variance is equal to its mean square, 
$\mathrm{var}(V)=\mv{V^2}-\mv{V}^2 = \mv{V}^2$. 

The shift $V\mapsto V-\mv{V}$ to the centered potential removes the
first term in the bracket in \eqref{V12def3.eq}. The same applies to all diagrams with field
self-contractions: \be
\gammaself=0.
\ee 
So henceforth, we can neglect those diagrams by considering a centered
potential  $\mv{V}=0$. 
Altogether, we have as a first building block the speckle potential pair correlator 
\belab{Pspeckle.eq}
\mv{V_1V_2} = P(r_{12}) = V_0^2 C(r_{12}) 
= V_0^2 \Vspckltwo
\ee
Here, the potential strength $V_0^2=\mathrm{var}(V)$ is factorized from 
the dimensionless correlation function $C(r)=|\gamma(r)|^2$ that is normalized to
$C(0)$=1.  In $d=1$, from
\eqref{gammaq_1d.def}, we have the real-space intensity correlator 
$C(r) = [\sin(r/\lcor)/(r/\lcor)]^2$. In higher dimensions and in an
isotropic setting, the Fourier
transformation of the simple $k$-space field correlator yields $C(r)=
[2J_1(r/\lcor)/(r/\lcor)]^2$ in $d=2$ and $C(r)=
[\sin(r/\lcor)/(r/\lcor)]^2$ again in $d=3$ \cite{Kuhn2007}.  

An interesting effect occurs for potential correlations of odd order
$(2n+1)$. They are really field correlation of twice the order,
which is even and thus different from zero. The first example of this
kind is (since the fields $\ast$ and $\circ$ will always appear
together, we note $\circledast=\bullet$ from now on)
\be
\mv{V_1V_2V_3} = V_0^3 2 \Re \{\gamma_{12}\gamma_{23}\gamma_{31}\} =
V_0^3 \Vspcklthree 
\ee
Diagrams of this type can only contain closed loops of field
correlations (because field self-contractions no longer appear). Since
the loops can be closed both clockwise and
counterclockwise, there are two contributions complex conjugate of
each other. 

\subsubsection{Average propagator: self-energy}

Now we are in position to take the ensemble average of the single-particle propagator \eqref{Born.eq}: 
\be
\mv{G} = G_0 + G_0 \mv{VG_0V} G_0  + G_0 \mv{VG_0VG_0V} G_0 + \dots  
\ee
or 
\be
\mv{G} = \prop + \GoPGo+ \GoTGo + \dots  
\ee
The precise form of potential correlations depends on the model of
disorder. As shown by the example of the Gaussian model
\eqref{Gaussiandecomp.eq}, 
starting from the fourth-order
term there appear completely factorized contributions. 
Before writing all possible combinations down, we had better introduce  
one of the cornerstones of diagrammatic expansions: 
the \emph{self-energy} $\Sigma(E)$  defined by the \emph{Dyson equation} 
\belab{Dyson.eq}
\mv{G} = G_0 + G_0 \Sigma \mv{G} . 
\ee
Introducing the self-energy invariably prompts the following 
frequently asked questions: 
\begin{enumerate}
\item Why is the self-energy convenient for perturbation theory? 
\item How do I calculate $\Sigma$?
\item What is the physical meaning of $\Sigma$?
\item Is there a simple example?
\end{enumerate}
Let us answer them in turn. 

1.~By iterating the Dyson equation \eqref{Dyson.eq}, one finds that the average propagator expands as 
\newcommand{\sgm}{\tikz{\node[circle,draw,tight]{$\Sigma$};}}
\be
\mv{G} =  \prop +  \prop \sgm \prop + \prop \sgm \prop  \sgm \prop + \prop \sgm \prop  \sgm \prop \sgm \prop +\dots 
\ee
By construction, there are no disorder correlations between the different self-energies appearing here. 
In return, this implies that the self-energy contains \emph{exactly
all} correlations that cannot be completely factorized by removing a
free propagator $G_0$ in between. These non-factorizable terms are
called ``one-particle irreducible'' (1PI). Moreover, the self-energy
contains only the correlations and internal propagators, but is
stripped off the external propagator lines (``amputated''). 
This makes the self-energy the simplest object describing all relevant disorder correlations. 

2.~
Due to statistical homogeneity, the self-energy is diagonal in
momentum and thus only depends on $k$ and $E$. 
The self-energy matrix element $\Sigma(k,E)$ is calculated by applying so-called Feynman
rules to evaluate the diagrams. As a specific example, let us give the
Feynman rules for the self-energy of the retarded single-particle
propagator $\mv{\GR(k,E)}$ in momentum representation for the case of
the speckle potential:  
\begin{enumerate}
\item[(i)] Draw all amputated 1PI diagrams with incident momentum $k$:

\be
\Sigma(k,E) =  \Sigmatwo + \Sigmathree + \dots 
\ee

\item[(ii)] Convert straight black lines to free propagators 
$$\arprop{k}= \GR_0(k,E) = [E-\eps_k^0+i0]^{-1}.$$ 

\item[(iii)] Convert disorder correlation lines to $\fieldcorr{q} =
\gamma(q)$. The precise functional dependence $\gamma(q)$ depends on
dimension and geometry. 

\item[(iv)] For each scattering vertex, multiply by one power of the potential strength and 
the conservation of momentum: 
\be
\vrtx = V_0 \delta_{k+q, k'+q'}
\ee
\item[(v)] Sum over all free momenta after respecting momentum conservation.  
\end{enumerate}
For other types of disorder, these rules have to be adapted slightly
to the precise shape of diagrams, correlation functions and vertex factors. But in all cases, the
general idea of writing all possible combinations, respecting momentum
conservation and integrating out the free momenta is the same.   

3.~One can rewrite
the Dyson equation \eqref{Dyson.eq} as  
$[1-G_0 \Sigma ]\mv{G} = G_0$ 
and solve formally for the average propagator: 
$\mv{G} = [1-G_0 \Sigma ]^{-1}G_0 =  [G_0^{-1}-\Sigma]^{-1}.$ Thus, its matrix elements are  
\be
\mv{\GR(k,E)} = \frac{1}{E-\eps_k^0-\Sigma(k,E)}.
\ee 
We recognize that the self-energy modifies the free dispersion
relation. Generally, the self-energy is a complex quantity with a real
as well as an imaginary part. The \emph{modified dispersion relation} 
\be
E_k = \eps_k^0 + \Re \Sigma(k,E_k)
\ee 
is an implicit equation for the new eigen-energy $E_k$ of the mode
$k$. So one effect of the disorder is to shift the energy levels.   
\footnote{Alternatively, one can also solve for $k_E$ as the modified
$k$-vector of an excitation with given energy $E$.}  
But plane waves with fixed $k$ are no longer proper eigenstates of the
disordered system. This is encoded in the imaginary
part. Writing $\Gamma_k= -2\Im\Sigma(k,E_k)$ and using the fact that
the self-energy varies smoothly with $k$ and $E$, one finds a 
\emph{spectral
density}  
\be 
A(k,E) = -2 \Im \mv{\GR(k,E)} = \frac{\Gamma_k}{(E-E_k)^2+\Gamma_k^2/4}. 
\ee 
This spectral function is the probability density that an excitation
$k$ has energy $E$. Its wave-number integral is the average density
of states per unit volume,  
\belab{dos.def}
N(E) = \frac{1}{2\pi}\int\frac{\rmd^dk}{(2\pi)^d}A(k,E). 
\ee
For the free Hamiltonian, $A_0(k,E)=
2\pi\delta(E-\eps_k^0)$.
The disorder introduces a finite spectral width
$\Gamma_k$, which translates into a finite lifetime
$\vthree{\tau_k=\ }\hbar\Gamma_k^{-1}$.   
Equivalently, this finite lifetime translates into a finite scattering
mean-free path $\ls$ for the spatial matrix elements 
of the average propagator,  
\belab{avGreal}
\mv{G(r-r',E)} = \int \frac{\rmd^dk}{(2\pi)^d} e^{ik\cdot(r'-r)}
\mv{G(k,E)} = G_0(r-r',E) e^{-|r'-r|/2\ls}, 
\ee
showing an exponential decay \vthree{over $\ls= v_k \tau_k$ 
where group velocity $v_k=\hbar^{-1}\partial_k\eps_k^0$ and lifetime $\tau_k$ are 
evaluated at the wave vector 
$k$ such that $E_k=E$}.

4.~The simplest possible example is the calculation of the lifetime
from the lowest-order, so-called \emph{Born approximation} 
\belab{SigmaBorn}
\Sigma(k,E) =  \Sigmaonewithk 
\ee
for some potential with correlation function $P(q)$.  
To lowest order in $V_0$, we can use $E_k=\eps_k^0$ and thus find 
\vthree{the spectral width}  
\belab{GammakBorn}
\vthree{
\Gamma_k 
= \int \frac{\rmd^d k'}{(2\pi)^d}
P(k-k')
 2\pi\delta(\eps_k^0 - \eps_{k'}^0).
}
\ee
This is precisely the result that one gets
from a straightforward application of Fermi's Golden Rule for the
average probability of scattering out of the mode $k$ by the external
potential $V_q$. The interest of the full-fledged diagrammatic
expansion is of course that one is in principle able to calculate
corrections to the lowest-order estimate, and to tackle more
complicated potentials. There exist literally hundreds of other
applications in the most diverse physical systems. Let us mention two
examples from our own experience.  

For \vthree{matter waves with quadratic dispersion $\eps_k^0=\hbar^2k^2/2m$ in} two-dimensional Gaussian correlated potentials such as the one
introduced  in section \ref{GaussianDisorder.sec}, the scattering rate evaluates to 
\belab{GammaGauss}
 \frac{1}{k\ls} =
\frac{\Gamma_k}{2\eps_k^0} = \frac{2\pi V_0^2}{k^2\sigma^2 E_\sigma^2}
e^{-k^2\sigma^2}I_0(k^2\sigma^2)
\quad (\text{Gauss}, d=2)
\ee
where $E_\sigma=\hbar^2/m\sigma^2$ is a characteristic correlation
energy and $I_0$ a modified Bessel function. 
\vthree{In} a one-dimensional speckle potential, we can use
\eqref{SigmaBorn} and \eqref{GammakBorn} with the speckle potential 
correlation function \eqref{Pspeckle.eq}. In $d=1$, the only
contributions can come from forward scattering $k'=k$ and backward
scattering $k'=-k$, such that
\belab{GammaSpeckle}
\frac{1}{k\ls} = \frac{\Gamma_k}{2\eps_k^0}  = \frac{V_0^2k}{{\vthree{4}\eps_k^0}^2} 
\left[P(0)+ P(2k)\right] \quad (\text{speckle}, d=1)
\ee
in terms of the $k$-space pair correlator $P(2k) = \pi \zeta
(1-|k\lcor|)\, \Theta(1-|k\lcor|)$.  

The 
estimates \eqref{GammaGauss} and \eqref{GammaSpeckle} 
can only be trusted if $\Gamma_k/\eps_k^0\ll 1$ or equivalently $k\ls\gg1$, otherwise
the assumption of a small correction to the free dispersion is no
longer valid. Since the scattering rates diverge at low
$k$, we find that the perturbative approach breaks down at low
energy. A closer analysis shows that a sufficient criterion for weak
disorder is $E_k\gg V_0^2/E_\sigma$ \cite{Kuhn2007}.  

Sometimes, also the real part of the self-energy is of importance. For
example, one can calculate the speed of sound in
interacting Bose-Einstein condensates, and especially the shift due to
correlated disorder by the same Green function formalism 
\cite{Gaul2009b}. Incidentally, for sound waves the scattering
mean-free path grows as $k\to 0$, and the perturbative approach stays
valid even at very low energy.

\subsection{Intensity transport} 

We would like to calculate the ensemble-averaged density
$n(r,t)=\mv{\bra{r}\rho(t)\ket{r}}$ (or, its many-body form
$\mv{\mv{\Psi^\dagger(r)\Psi(r)}}$) in the limit of long time. In the
Schr\"odinger picture, the state evolves as
$\rho(t)=U^\dagger(t)\rho_0U(t)$. After transforming the time
evolution operators to Green functions as in \eqref{GRdef.eq}, we need a theory for the
ensemble-averaged product $\mv{\GA(E)\GR(E')}$. 
This is known as the average intensity propagator. In most
experimental situations---be it with electromagnetic or matter
waves---one measures 
intensities (see for example
the average transmission through a 1d disordered system of length $L$ studied in section~\ref{1d.sec}); 
the average
intensity propagator is thus the fundamental quantity of interest. 
Before going into
details, we propose to have a look at what we should expect to be the
result. 

\subsubsection{Density response}

The generic behavior that one may expect for transport in a disordered
environment is \emph{diffusion}. Indeed, diffusion follows from two
very basic and rather innocuous hypotheses. Firstly, one generally has
a local conservation law, for instance for particle number, 
taking the form of a \emph{continuity
equation}:
\belab{continuity}
\partial_t n + \nabla \cdot j = s
\ee
where $j(r,t)$ is the current density associated with $n(r,t)$, 
and $s(r,t)$ is some source function. Secondly, one assumes a
\emph{linear response} in the form of Fourier's law 
\belab{linearresponse}
j = - D \nabla n,  
\ee
saying that a density gradient induces a current that tries to
reestablish global equilibrium. The diffusion constant $D$ appears
here as a linear response coefficient. Inserting
\eqref{linearresponse} into \eqref{continuity}, we immediately find as
a consequence the \emph{diffusion equation}
\belab{diffusion}
[\partial_t - D \nabla^2 ] n(r,t)  = s(r,t). 
\ee
This equation can be solved by Fourier transformation%
\footnote{which was invented right
for this purpose by Joseph Fourier, namesake of the
French university in Grenoble hosting the Les Houches school.}.
The solution for a unit source $s(r,t)=\delta(r)\delta(t)$ is the
Green function for this problem, viz., the density
relaxation kernel 
\belab{Phizero.eq}
\Phi_0(q,\omega) = \frac{1}{-i\omega+ D q^2}. 
\ee
Its temporal version 
\be
\Phi_0(q,t) = \int\frac{\rmd \omega}{2\pi} e^{-i\omega t}
\Phi_0(q,\omega) = \theta(t) \exp\{-Dq^2 t\}  
\ee
shows that the relaxation $\exp\{-t/\tau_q\}$ with characteristic time
$\tau_q=1/Dq^2$ becomes very slow in the large-distance limit $q\to 0$ 
because of the local conservation law. In real space and time, the
relaxation kernel reads 
\be
\Phi_0(r,t) = \int\frac{\rmd^d q}{(2\pi)^d} e^{iq\cdot r}
\Phi_0(q,\omega) = \theta(t)[4\pi D t]^{-d/2} \exp\{-r^2/4Dt\}.  
\ee
This relaxation kernel describes diffusive spreading with $\mv{r^2}
= 2 d D t$. 

This is the ``hydrodynamic'' description of dynamics on large
distances and for long times, accessed by small momentum $q$ and
frequency $\omega$.
A microscopic theory is then only required to calculate the linear response
coefficient $D$. 

\subsubsection{Quantum intensity transport}

In complete analogy to the Dyson equation \eqref{Dyson.eq} for the
average single-particle propagator, one may
write a structurally similar equation for the intensity propagator 
$\Phi = \mv{\GR\GA}$, known as the Bethe-Salpeter equation: 
\be
\Phi = \mv{\GR} \mv{\GA} +  \mv{\GR} \mv{\GA} U \Phi. 
\ee
Here, one splits off the known evolution with uncorrelated, average
amplitudes 
\belab{mvGRmvGA}
 \mv{\GR(k,E)} \mv{\GA(k',E')} =  \mvGRmvGA.  
\ee
The upper part of intensity diagrams describes the retarded propagator,
called ``particle channel'' in condensed-matter jargon,
whereas the lower part contains the advanced propagator or ``hole
channel''. 
All scattering events that couple these amplitudes are contained in 
the intensity scattering operator $U$. By construction, this
``particle-hole irreducible'' vertex 
contains exactly all diagrams that cannot be factorized by removing a 
propagator pair \eqref{mvGRmvGA}. Its detailed form
again depends on the model of disorder. In all cases, $U_{kk'}(E)$ is
essentially the differential cross-section for scattering from $k$ to
$k'$ and generally has the following structure: 
\belab{Udiag}
U(k,k';E) = \Uboxwithk = \ \UGone\ + \UGtwoA +\UGtwoB +\UGtwoC + \dots 
\ee
Linear-response theory shows that this scattering vertex permits to
calculate the transport mean-free path $l$, in close analogy to the calculation
of the scattering mean-free path $\ls$ from the self-energy. Their
ratio is expressed as 
\belab{lsldef}
\frac{\ls}{l} = 1-\mv{\cos\theta}_U
\ee 
where $\theta$ is the scattering angle between $k$ and $k'$, and the
brackets $\mv{.}_U$ indicate an average over the scattering
cross-section $U$. 
The physical interpretation of the transport mean free path $l$ is the
following: while the scattering mean free path $\ls$ measures the distance after which the memory of
the initial phase of the wave is lost, $l$ is the distance over which the 
direction of propagation is randomized.

\subsubsection{Diffusion}

The scattering processes encoded in $U$  are perhaps more
easily visualized in real space. We will draw a full line for every
amplitude $\psi$ propagated by $\GR$ (upper lines in \eqref{Udiag}) and a dashed line for every
$\psi^*$ propagated by $\GA$  (lower lines in
\eqref{Udiag}). Impurities are represented by black dots as before. Then,
the  first  contribution to $U$ describes the single-scattering process 
\belab{UB}
U_\text{B} :  \UBreal
\ee
in which both $\psi$ and $\psi^*$ are being scattered by the same
impurity at position $r_1$. This process is insensitive to phase
variations and could just as well take place for classical particles. 
So this Boltzmann contribution
$U_\text{B}$ describes classical diffusion with diffusion constant
\be
\DB = \frac{v \lB}{d}
\ee 
The Boltzmann transport mean-free path is calculated
by inserting $U_{\text{B}kk'}= V_0^2P(k-k')$ into \eqref{lsldef}: 
\belab{lslB}
\frac{\ls(k)}{\lB(k)} = 1 - \frac{\int\rmd\Omega_d \cos\theta P(2k|\sin(\theta/2)|)}{\int\rmd\Omega_dP(2k|\sin(\theta/2)|)}
\ee

Depending on the microscopic scattering process, $\lB$
can be longer than $\ls$, if forward scattering is dominant,
$\mv{\cos\theta}_{\UB}>0$. This is the case for matter waves in
spatially correlated potentials. For isotropic scattering  with
$\mv{\cos\theta}_{\UB}=0$, these two length scales coincide,
$\ls=\lB$. 

By combining eq.~\eqref{lslB} with eq.~\eqref{GammakBorn} giving the
scattering mean free path, one can easily compute, in the Born
approximation, the transport 
mean free path and consequently the classical Boltzmann diffusion constant,
using only microscopic ingredients: the dispersion relation of the free wave
and the correlation function of the scattering potential. 

\subsubsection{Localization length in 1d systems}
\label{loc_length_1d.sec}

\begin{figure}
\begin{center}
\includegraphics[width=6cm]{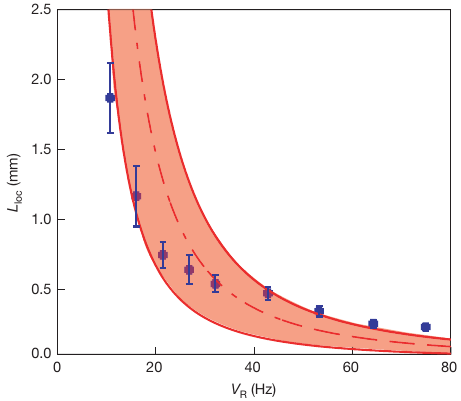}
\end{center}
\caption{\small Comparison between the experimentally measured
localization length for a quasi-1d atomic wave-packet launched
in the disordered optical potential created by a speckle pattern, and the theoretical
prediction, eq.~\eqref{xilocborn_speckle}, when the strength of the disordered
potential is varied. 
There is no adjustable parameter. Reprinted from \cite{Billy2008} (courtesy of Ph. Bouyer).} 
\label{orsay_th_exp.fig}
\end{figure}

As we have already seen repeatedly in previous sections, in 1d the
transport mean free path is (up a factor 2) equal to the localization
length, $\ell=\xiloc/2$, an identity that can also be verified
microscopically \cite{Thouless1973}, at least to lowest order $V_0^2$ in
perturbation theory.%
\footnote{Whether this holds to all orders in perturbation theory is
to our knowledge an open, and also interesting question, especially in
optical speckle potentials \cite{Lugan2009}.} So now we are in a
position to give a microscopic prediction for the 1d localization
length for arbitrarily correlated potentials, namely taking twice the
backscattering contribution  from \eqref{GammaSpeckle}, selected by the $(1-\cos\theta)$-factor
in \eqref{lslB}: 
\belab{xilocborn}
\frac{1}{k\xiloc} = \frac{V_0^2k}{4{\eps_k^0}^2} P(2k). 
\ee
Figure~\ref{orsay_th_exp.fig}, taken from \cite{Billy2008}, shows this prediction for
$L_\text{loc}=2\xiloc$ as a dashed line together with the results of a fit to the
intensity measured in the real experiment (see
Fig.~\ref{time_dynamics_orsay.fig}). Here $P(2k) =
\pi\zeta(1-k_\text{max}\lcor) \Theta(1-k_\text{max}\lcor)$ 
with $k_\text{max}$ the largest $k$-value present in
the expanding wave packet, resulting in:
\belab{xilocborn_speckle}
\xiloc = \frac{\hbar^4 k_\text{max}^2}{\pi m^2 V_0^2 \lcor (1-k_\text{max}\lcor)}.
\ee
There is no adjustable parameter, and the agreement is rather
satisfactory.  
Significant deviations are visible both for small disorder (there the
localization length becomes too large 
and experimental limitations start to show) and for large disorder,
where the lowest-order theoretical estimate, or Born approximation
\eqref{xilocborn_speckle} becomes insufficient. Moreover, for strong disorder, 
atom-atom interaction for the strongly localized cloud may no longer
be negligible and induce some delocalization.  

An interesting scenario occurs in speckle potentials when
the fastest atoms have a wave vector $k_\text{max}$ larger than the
most rapid spatial fluctuations, with wave vector $1/\lcor$. Then,
$P(2k_\text{max})=0$, and \eqref{xilocborn_speckle} predicts prima facie $\xiloc=\infty$ or absence of
localization, which would signal the
existence of a mobility edge in this 1d random potential, contradicting rigorous mathematical theorems
stating that all states are exponentially localized 
\cite{Kotani1987}. 
In fact, exponential localization still prevails, but requires more
than a single scattering event by the smooth random potential. Going
to higher orders in perturbation theory, beyond the Born
approximation, one can show in excellent quantitative agreement with
numerics, that the localization length is always finite 
\cite{Gurevich2009,Lugan2009}. However, 
for weak disorder, the localization length can become much larger than the system
size, such that numerical or experimental results show an \emph{apparent
mobility edge}. For other types of long-range correlations, one does find
mobility edges \cite{deMoura1998}. There seems to be no obvious way of
deciding, for a certain class of potentials, whether true exponential
localization exists or not, and correlated potentials are still
actively investigated in different contexts, see \cite{Izrailev2005} and
references therein.

Despite the nice agreement shown in Fig.~\ref{orsay_th_exp.fig}, some caution is indicated.  
The experimental observation involves averaging over $k,$ and also
over several different realizations of the disorder (single-shot results look similar, just more noisy).
This means that the experimental data resembles the 
average transmission $\mv{T(z)}$ as a function
of sample thickness $z$.\footnote{The situation is actually more complicated, because
this expansion experiment strictly speaking 
does not measure the transmission across a sample. Instead, one starts with
an atomic density inside the medium and see how it propagates. The boundary conditions are thus different from those of a transmission experiment with its connection to outside leads. Still, huge fluctuations must exist in the localized regime, implying that the average transmission deviates from a pure exponential, as discussed in Section~\ref{full_distrib.sec}.} 
In section~\ref{SL1.sec}, we showed that it is the \textit{typical} transmission
$T_\text{typ}(z)=\exp(\mv{\ln T(z)})$ which decays exponentially, not the average transmission. At very large $z$,
this can make a huge difference, see section~\ref{full_distrib.sec}. Fortunately, for $z$ of the order of
the localization length ($t$ or order unity in the language of \ref{full_distrib.sec}), the fluctuations have not yet built up, and the difference between the typical and the average value is still small,  
$\ln{\mv{T(z)}} \approx -z/\xiloc$,   
making the pure exponential decay an acceptable approximation. 
Further in the wings, one expects deviations of the average density from a pure exponential decay,
see eq. \eqref{T12moments.eq}. This takes place however in the region where fluctuations are huge, so that a typical
experiment may not measure the \emph{average} value of the density, but rather its \emph{typical} value.


\subsubsection{Weak-localization correction}

The first corrections to the classical, incoherent scattering process
\eqref{UB} shown in \eqref{Udiag} involve one more scatterer and
several possibilities of intermediate propagation. The most well-known
type of correction stems from the diagram with two crossed lines. In
real space, the scattering process is 
\belab{Umcreal}
\Umcreal
\ee
This is an interference correction with a phase shift $\Delta\varphi$
between $\psi$ and $\psi^*$ that depends on the impurity positions $r_1$
and $r_2$. Contributions of this type are ensemble-averaged
to zero---or rather, almost averaged to zero. Indeed, if the starting
and final point of propagation come close, $r\approx
r'$, the phase shift picked up by the two counter-propagating
amplitudes becomes smaller:  
\belab{Umcrealclose}
\Umcrealclose
\ee
At exact backscattering $r=r'$ and in the absence of any dephasing
mechanisms, the phase difference is exactly zero. Vanishing phase
difference means constructive interference and therefore enhanced
backscattering probability to stay at the original position. This holds true no
matter how many scatterers are visited on the path. One is led to
consider all maximally
crossed diagrams: 
\belab{Umc}
\UC = \UGtwoA + \UGthreeA + \dots  
\ee
These diagrams were first considered in the electronic
context \cite{Langer1966} and became known as the 
\emph{Cooperon} contribution. 
This contribution is peaked around backscattering
$k=-k'$. Therefore, one may resort to a diffusion approximation and
sum up all contributions with the help of the diffusion kernel \eqref{Phizero.eq}: 
\be
\frac{1}{l} = \frac{1}{\lB}\left[1+\frac{1}{\pi N_0} 
	\int \frac{\rmd^d q}{(2\pi)^d}\frac{1}{-i\omega+\DB q^2}\right]_{\omega\to 0}
\ee
Writing this in terms of the diffusion constant, one arrives at the 
weak-localization correction 
\belab{Dwl}
\frac{1}{D} = \frac{1}{\DB}\left[1+\frac{1}{\pi N_0 \DB} 
	\int \frac{\rmd^d q}{(2\pi)^d}\frac{1}{q^2-i0}\right]
\ee
The quantum correction of the Cooperon makes $D<\DB$, and we have thus
found the microscopic reason for the weak-localization correction that
was first mentioned in the scaling section \ref{scalinganyd.sec}. 
Before looking in more details at this correction in section~\ref{wl.sec},
we should like to understand it better by selectively probing the
Cooperon contribution. In Optics, this is indeed possible and is
developed in the next section.

\section{Coherent backscattering (CBS)} 

\label{CBS.sec}

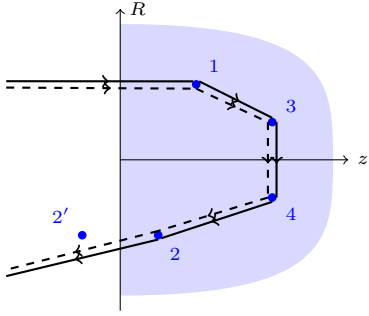
\begin{figure}
\begin{center}
\begin{tikzpicture}[]
  \path[fill=blue!15] (0,1.8) .. controls (2.8,1.8) and (2.8,1) .. (2.8,0) 
		.. controls (2.8,-1) and (2.8,-1.8) .. (0,-1.8);
  \draw[->] 	(0,0) -- (3,0) node[right] {$z$};
  \draw[->] 	(0,-2) -- (0,2) node[right] {$R$};
  \path node[tight] (in) at (-1.5,1) {}
	node[vertex,blue,label={[blue]45:1}] (p1) at (1,1) {}
	node[vertex,blue,label={[blue]15:$3$}] (p3) at (2,0.5) {}
	node[vertex,blue,label={[blue]-15:$4$}] (p4) at (2,-0.5) {}
	node[vertex,blue,label={[blue]-60:$2$}] (p2) at (0.5,-1) {}
	node[tight] (out) at (-1.5,-1.5) {};
   \draw[arprop] (in.north) -- (p1.north west); 
   \draw[arprop] (p1.north east) -- (p3.north);
   \draw[arprop] (p3.east) -- (p4.east); 
   \draw[arprop] (p4.south) -- (p2.south east);
   \draw[arprop] (p2.south) -- (out.south);
   \draw[arprop,dashed] (in.south) -- (p1.south); 
   \draw[arprop,dashed] (p1.south) -- (p3.west);
   \draw[arprop,dashed] (p3.west) -- (p4.west); 
   \draw[arprop,dashed] (p4.west) -- (p2.north);
   \draw[arprop,dashed] (p2.north) -- (out.north);
\node[vertex,blue,label={[blue]130:$2'$}] (p2') at (-0.5,-1) {};
\end{tikzpicture}
\caption{\small%
Half-space geometry of a backscattering
experiment with cylindrical coordinates $r=(R,z)$. 
$2'$ is the image point of the exit scatterer $2$ used to
construct the half-space propagator for the incoherent intensity. As
an example, the contribution of scattering from four scatterers is depicted.}
\label{CBSgeom.fig}
\end{center}
\end{figure}

One can probe the specific geometry of scattering paths like
\eqref{Umcreal} by using a source of plane waves together with a
collection of randomly positioned scatterers in a half-space geometry
(fig. \ref{CBSgeom.fig}). Hereafter, we suppose normal incidence and
detection close to the backscattering direction; generalizing to
arbitrary incident and detection angles changes nothing to the central argument. 

\subsection{Theory}
\label{CBStheo.sec}

The picture in Fig.\ \ref{CBSgeom.fig} shows scattering by 
four impurities, contributing to the incoherently transported
intensity. The corresponding intensity diagram is 
\belab{ladderfour}
\Ladderfour. 
\ee
The sum of all such diagrams with a distinct ladder topology yields
the intensity propagator or \emph{diffuson}, whose long-distance and long-time form is 
precisely the diffusion kernel \eqref{Phizero.eq}, evaluated with the
Boltzmann diffusion constant. 
\belab{PhiB}
\Phi_\text{B} = \frac{1}{-i\omega+ \DB q^2}. 
\ee
The total back-scattered diffuse intensity per unit surface is given by
summing contributions from all possible starting and end
points: 
\belab{IL1}
I_\text{L} \propto \int \rmd z_1\ e^{-z_1/\ls}\int\rmd z_2 \, e^{-z_1/\ls}
\int\rmd^2 R \, \Phi_\text{B}(r_1,r_2). 
\ee
The exponential attenuation factors describe the
propagation of intensities from the surface to the first scatterer and
back out again with average propagator $\mv{\GR(z_i)}$, \eqref{avGreal}, featuring the scattering
mean free path $\ls$.  
Moreover, translation invariance along the surface direction has been
used, leaving only the surface integral over
the lateral distance $R= R_1-R_2$. 

The propagation inside an infinite disordered medium would occur with
the bulk kernel 
\eqref{PhiB} and thus have a time-integrated diffusion probability of
$\Phi_\text{B}(r) = \int \rmd t \Phi_\text{B}(r,t) = [4\pi \DB
r]^{-1}$. 
This expression leads to a diverging integral over $R$ in 
\eqref{IL1}. But the starting and end points $r_1$ and $r_2$ lie rather close to the
surface, namely typically one scattering mean-free path $\ls$ away from
it. So for calculating the back-scattered intensity \eqref{IL1}, we 
have to worry about appropriate boundary conditions. 
The complete integral equation for intensity propagation in a half-space
geometry of a scalar wave and isotropic scatterers, known by the name Milne equation, can be solved exactly
\cite{MorseFeshbach,Nieuwenhuizen1993}, albeit with considerable
mathematical effort. For a simple solution involving the diffusive
bulk propagator valid far from the boundary, one can employ the method of images that is
often used in electrostatics. Since photons reaching the surface would
escape prematurely from the medium, one can exclude these events by 
subtracting the contribution of propagation to an image point $r_{2'} = (R_2,-z_2)$ 
mirrored to the outside of the sample: 
\belab{PhiImage}
\Phi_\text{B}(r_1,r_2) = \frac{1}{4\pi \DB} \left[ \frac{1}{r_{12}} -
\frac{1}{r_{12'}}\right]. 
\ee
This half-space propagator behaves like $R^{-3}$ at large $R$ and thus
permits to carry out the integration. The final result is some number
$I_L$ and gives the incoherent background on top of which we now study the
interference contribution. 

Each multiple-scattering diagram like \eqref{ladderfour} has an
interference-correction 
counterpart such as 
\be
\Crossedfour 
\ee
in which the conjugate amplitude travels along the same
scatterers, but in opposite direction. The contribution of such 
maximally crossed diagrams to the back-scattered intensity 
can be accounted for along the same lines. First of all, incident and
scattered amplitudes now pick up a phase between the surface and the
scattering end points.
Namely, the amplitude $\psi$ picks up $\exp\{ik\cdot r_1\}$ at
the entrance and $\exp\{ik'\cdot r_2\}$ at the exit of the medium. The
path-reversed complex conjugate amplitude picks up $\exp\{-i[k\cdot
r_2+k'\cdot r_1]\}$. So after all, there is a total phase difference
of $\Delta \varphi = (k+k')\cdot(r_1-r_2)$. Exactly
toward the backscattering direction, $k'=-k$, this phase difference
vanishes. Close to backscattering, for a small angle $\theta\ll 1 $, one has
$|k+k'|\approx k_\perp =k\sin\theta \approx k\theta$, and the phases
differ by $\Delta \varphi = k R \theta$. 

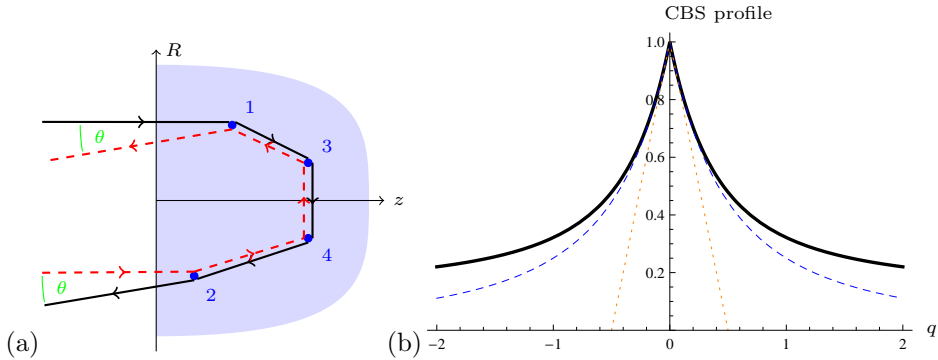
\begin{figure}
(a)
\begin{tikzpicture}[baseline=-2cm]
  \path[fill=blue!15] (0,1.8) .. controls (2.8,1.8) and (2.8,1) .. (2.8,0) 
		.. controls (2.8,-1) and (2.8,-1.8) .. (0,-1.8);
  \draw[->] 	(0,0) -- (3,0) node[right] {$z$};
  \draw[->] 	(0,-2) -- (0,2) node[right] {$R$};
  \path node[tight] (in) at (-1.5,1) {}
	node[vertex,blue,label={[blue]45:1}] (p1) at (1,1) {}
	node[vertex,blue,label={[blue]15:$3$}] (p3) at (2,0.5) {}
	node[vertex,blue,label={[blue]-15:$4$}] (p4) at (2,-0.5) {}
	node[vertex,blue,label={[blue]-60:$2$}] (p2) at (0.5,-1) {}
	node[tight] (out) at ($(p2)+(190:2)$) {};
   \draw[arprop] (in.north) -- (p1.north west); 
   \draw[arprop] (p1.north east) -- (p3.north);
   \draw[arprop] (p3.east) -- (p4.east); 
   \draw[arprop] (p4.south) -- (p2.south east);
   \draw[arprop] (p2.south) -- (out.south);
   \path node[tight] (in') at (-1.5,-1) {}
	 node[tight] (out') at ($(p1)+(190:2.5)$) {};
   \draw[arprop,dashed,red] (in'.north) -- (p2.north); 
   \draw[arprop,dashed,red]  (p2.north) -- (p4.west);
   \draw[arprop,dashed,red] (p4.west)  -- (p3.west); 
   \draw[arprop,dashed,red]  (p3.west) -- (p1.south);
   \draw[arprop,dashed,red] (p1.south) -- (out'.south);
   \draw[green] (p1)++(-2,0) arc (180:190:2);
   \draw[green] (p1)++(185:1.75) node{$\theta$}; 
   \draw[green] (p2)++(-2,0) arc (180:190:2);
   \draw[green] (p2)++(185:1.75) node{$\theta$}; 
\end{tikzpicture}
\hspace{-2em}
(b)\includegraphics{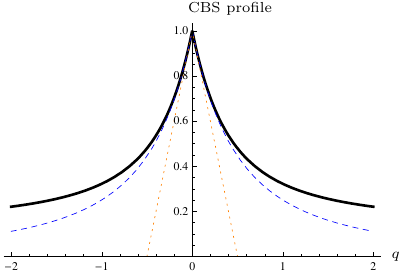}
\caption{\small%
Coherent backscattering (CBS)  (a) Schematic picture of a
four-scatterer path; the constructive interference of
path-reversed amplitudes in the backscattering direction $\theta=0$
leads to an observable intensity enhancement. Away from
backscattering, the phase differences average out and leave only the
background intensity.
(b) CBS profile as function of reduced scattering angle
$q=kl\theta$, normalized to the value at $\theta=0$. Solid black: exact
solution \eqref{ICexact.eq}. Dashed blue: Diffusive solution
\eqref{ICdiff.eq}. 
Dotted orange: Linear solution \eqref{ICsmallq.eq} with characteristic
slope discontinuity at backscattering.
} 
\label{CBS.fig}
\end{figure}

Thus, each path acts like Young double-slit interferometer with the two
end-point scatterers playing the role of the two slits. The larger
the transverse distance $R$ between the scatterers, the finer the
interference fringes. The only point where all fringes are bright
is the symmetry point $\theta=0$ toward backscattering. Sufficiently far away from this
direction, the sum of random fringe patterns averages out to zero. The
sum of all interference term is again the integral over all end points with the appropriate
weight furnished by the intensity propagator \eqref{PhiImage} (which
must be modified if 
some additional dephasing processes are at work, see
section~\ref{dephasing.sec} below). This simple
calculation predicts a relative interference enhancement over the
background  
\belab{ICdiff.eq}
\frac{I_\text{C}(\theta)}{I_\text{L}} \approx \frac{1}{(1+kl|\theta|)^2}
\ee  
The interference-induced enhancement, shown in figure \ref{CBS.fig} as
a dashed blue
line, survives in an angular range
$\Delta\theta = 1/kl = \lambda/(2\pi l)$ around backscattering. Very
characteristically, this peak features a triangular cusp at
backscattering (plotted in dotted orange), 
\belab{ICsmallq.eq}
\frac{I_\text{C}(\theta)}{I_\text{L}} = 1 - 2 |q| +O(q^2)
\ee  
where $q=kl\theta$ is the reduced momentum transfer. 

The exact solution for scalar waves and isotropic point scatterers 
can be calculated solving the Milne equation of
intensity transport. The CBS profile can then be expressed as the 
integral  \cite{Nieuwenhuizen1993}
\belab{ICexact.eq}
\frac{I_\text{C}(\theta)}{I_\text{L}} = \frac{1}{C} 
	\exp\left\{-\frac{2}{\pi} \int_0^{\pi/2}\rmd \beta 
\ln\left[ 1-\frac{\arctan\sqrt{q^2+\tan^2\beta}}{\sqrt{q^2+\tan^2\beta}}\right] \right\}
\ee
where $q=kl\theta$ and a constant $C= 
	\exp\left\{-\frac{2}{\pi} \int_0^{\pi/2}\rmd \beta 
\ln\left[1-\beta\cot\beta\right] \right\}\approx 8.455$ 
such that
at the origin $I_\text{C}(0)= I_\text{L}$. 
This profile is plotted as
a black curve in Fig.~\ref{CBS.fig}. It becomes apparent that the
diffusive solution \eqref{ICdiff.eq} gives a very good description for
small scattering angles. Notably, its slope at
$\theta=0$ is precisely equal to the exact value that can be extracted
from \eqref{ICexact.eq}. This was to be expected since diffusion
should be valid for long-distance bulk propagation, and long scattering
paths have widely separated end points that contribute to the small
transverse momenta making up the top of the CBS 
peak. Indeed, the diffusion prediction \eqref{ICdiff.eq} is precisely recovered
by replacing the exact propagation kernel under the logarithm by its
diffusion approximation: 
\be
A(Q)=\arctan(Q)/Q \approx 1-\frac{1}{3}Q^2
\ee
valid at small $Q=\sqrt{q^2+\tan^2\beta}$. 

The agreement between the diffusive profile
\eqref{ICdiff.eq} and the exact result \eqref{ICexact.eq}
deteriorates at larger angle $q=kl\theta$. This
discrepancy is due to low scattering orders that are not accurately
captured by the diffusion approximation and the imaging method used to
mimic the exact boundary conditions. Indeed, only the very tip of the
CBS peak stems from long paths reaching far into the bulk. The larger
part of the total signal is due to contributions from rather short
paths, for which scattering inside the surface skin layer is crucial.  

One can calculate the contribution of scattering orders
$n=1,2,\dots$ to the total incoherently back-scattered intensity,
measured in units of the incident flux by the so-called bistatic coefficient
$\gamma(\cos\theta',\cos\theta)$ \cite{Ishimaru} 
that depends on the angles $\theta'$
and $\theta$ of incidence and observation. For exact backscattering
and normal incidence ($\theta=\theta'=0$), one has $\gamma=\sum_{n\ge
1}\gamma_n$. The largest contribution comes from single scattering
with $\gamma_1=1/2$ followed by double-scattering with
$\gamma_2=\ln(2)/2\approx 0.35$ and so on, with an asymptotic decrease
as $\gamma_n\sim n^{-3/2}$ \cite{Nieuwenhuizen1993}.

When the CBS peak was first observed in the beginning of the 1980's 
\cite{Kuga1984,Albada1985,Wolf1985}, the diffusive theory used 
an image point placed at $z_{2'}=-(2z_0+z_2)$ such that the diffuse
propagator vanishes at a distance $z_0=2/3$ outside the sample. This
translates to the boundary condition that the total incident diffusive
flux on the surface vanishes \cite{MorseFeshbach,AkkermansMontambaux}
and leads to a diffusive CBS peak shape of 
\belab{ICAkk.eq}
\frac{I_\text{C}(q)}{I_\text{L}} = \frac{1}{(1+|q|)^2}
\frac{1}{1+2z_0}
\left[1+\frac{1-\exp\{-2z_0|q|\}}{|q|}\right] . 
\ee  
This diffusive solution predicts a 
different slope at the origin, viz., $-2[1+z_0^2/(1+2z_0)]$, which is
off by more than 20\% from the exact value, although one would expect
the diffusion solution to get this value right
\cite{Nieuwenhuizen1993}. This is all the more disturbing as fits to
the diffuse CBS peak shape are generally used to measure the transport
mean-free path. Also at larger angles this solution cannot
convince because the diffusion profile
decays as $q^{-2}$, whereas the exact solution decreases like
$|q|^{-1}$. This asymptotic behavior is known to come from the
double-scattering contribution. 

\begin{figure}
\begin{center}
\includegraphics{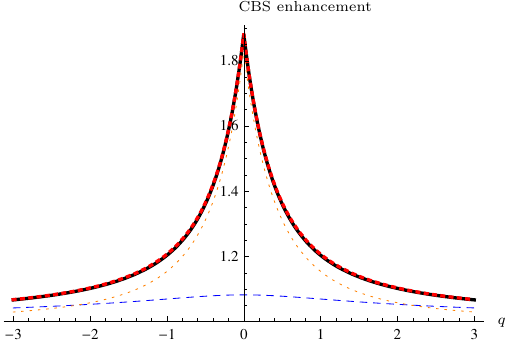}
\caption{\small%
CBS intensity enhancement in units of the background intensity 
as function of reduced scattering angle
$q=kl\theta$. Solid black: exact
solution \eqref{ICexact.eq} minus the single-scattering value
$\gamma_1=1/2$.
 Dashed blue: double scattering
contribution
\eqref{cbsdouble}. 
Dashed red: sum of double scattering and best diffusive solution,
\eqref{ICdiffit.eq}
with $\alpha^*\approx 0.86$ and
$z_0^*\approx0.81$. 
Dotted orange: Traditional diffusive CBS peak, eq.~\eqref{ICdiffit.eq}, with $\alpha=1$ and
$z_0=2/3$ shown for comparison. 
Even for this time-reversal invariant case, the backscattering
enhancement is slightly smaller than 2 because of the single-scattering
contribution to the background that is absent in the CBS signal. 
} 
\label{CBSplot2.fig}
\end{center}
\end{figure}

Bart van Tiggelen
has noticed \cite{BvTPhD} that the diffusion approximation becomes virtually exact
if single- and double scattering are included separately since 
\be
\frac{1}{1-A(q)}=1+A(\vthree{q}) + \frac{A(q)^2}{1-A(q)} \approx 1+A(q)+ \frac{3\alpha}{q^2}
\ee
both \emph{for small and large} $q$, with a numerical coefficient 
$\alpha=1$ for $q\to 0$ and $\alpha=\pi^2/12\approx 0.822$ for
$q\to \infty$. 
Therefore, the best approximation to the exact solution is obtained by
first taking the exact double-scattering profile \cite{Nieuwenhuizen1993,Mueller2001}
\belab{cbsdouble}
\gamma_2(q)= \frac{1}{\pi} \int_0^{\pi/2}\rmd\beta
A\left(\sqrt{q^2+\tan^2\beta}\right) 
=
\frac{2\cosh^{-1}\left(1/|q|\right)-\cosh^{-1}\left(1/q^2\right)}{2\sqrt{1-q^2}}
\ee
where  $\cosh^{-1}(x)$ is the inverse hyperbolic cosine function, 
then adding the diffusive solution   
\belab{ICdiffit.eq}
\gamma_\text{diff}(q)= \frac{3\alpha }{2(1+|q|)^2}
\left[1+\frac{1-\exp\{-2z_0|q|\}}{|q|}\right] 
\ee  
and finally fitting the extrapolation length $z_0$ and 
diffusion-constant multiplicator $\alpha$ such that height
and slope are equal to the exact values at the origin. Doing this, we find
$\alpha^*\approx 0.86$ within the expected interval $[0.822,1]$ and
$z_0^*\approx0.81$. Figure \ref{CBSplot2.fig} shows the exact CBS
profile (with the single-scattering contribution subtracted as
required)
together with the double scattering contribution plus the full
approximated diffusive CBS profile that turns out to be in
excellent agreement, both for small and large angles.  

\begin{figure}
\begin{center}
\includegraphics[angle=270,width=7cm]{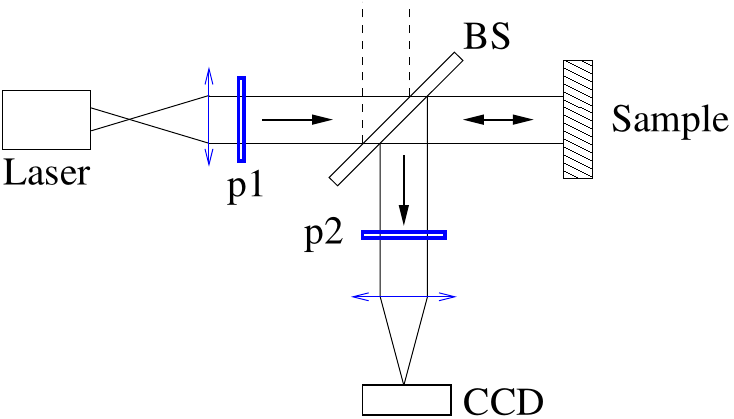}
\caption{\small%
Schematic view of a table-top CBS experiment. Light from a wide laser
beam with small angular dispersion is directed onto a disordered sample. The diffuse retro-reflected
intensity is sent by a beam-splitter (BS) 
in the focal plane of a lens and recorded by a CCD camera, thus imaging the angular
distribution. Polarization elements
(p1 and p2) select a suitable polarization channel;
generally, the helicity-preserving channel of opposite circular
polarization is recommended.} 
\label{cbsexp.fig}
\end{center}
\end{figure}

The full width at half height of the CBS profile is $\Delta
q \approx 0.73 kl \approx 4.59 l/\lambda$, and observing
the CBS peak can be used to measure the transport-mean free path quite
accurately. The explicit occurrence of the wavelength $\lambda$ emphasizes  that CBS is a
genuine interference effect. In many circumstances, the mean-free path is much longer than the
wavelength, such that $kl \sim 10^2...10^3$, and $\Delta\theta$ is at most a
couple of mrad. This makes CBS difficult to observe with the naked
eye, together with the constraint that one has to look exactly toward the
backscattering direction, but it can be easily imaged using standard
optics, as schematically shown in Fig.~\ref{cbsexp.fig}. 

\subsection{Live experiment}

Because the CBS cone is typically very narrow and its maximum height at best equal
to the average background, a source with large angular dispersion will
broaden the signal too much and reduce enhanced
backscattering. Thus, it is highly desirable to use a quasi-parallel beam obtained from a laser source,
with angular divergence smaller than a fraction of mrad. This is turn
requires a large spot, with a diameter larger than the mean free
path, which is easily obtained by expanding
the output beam of a commercial diode laser with a telescope. 
The scattering medium should scatter efficiently
and must not absorb the light: a bright white object is thus chosen. A piece of ordinary paper
turns out to give the best results. A sheet of paper is about 100\,$\mu$m thick and obviously scatters most of
the incoming beam, meaning that the mean free path does not exceed a few
tens of $\mu$m.
A piece of teflon could also be used, but the mean free path is significantly larger, meaning
a narrower CBS cone, much harder to detect. White paint or milk make
also good samples, with the advantage
that the concentration and thus the mean free path can be varied and
that the thermal motion of scatterers inside
the solvent provides us with configuration averaging for free;
however, these samples must be put inside
some transparent container whose surface can produce specular
reflection that is easily confounded with the CBS signal. 

\begin{figure}
\begin{center}
\includegraphics[width=6cm]{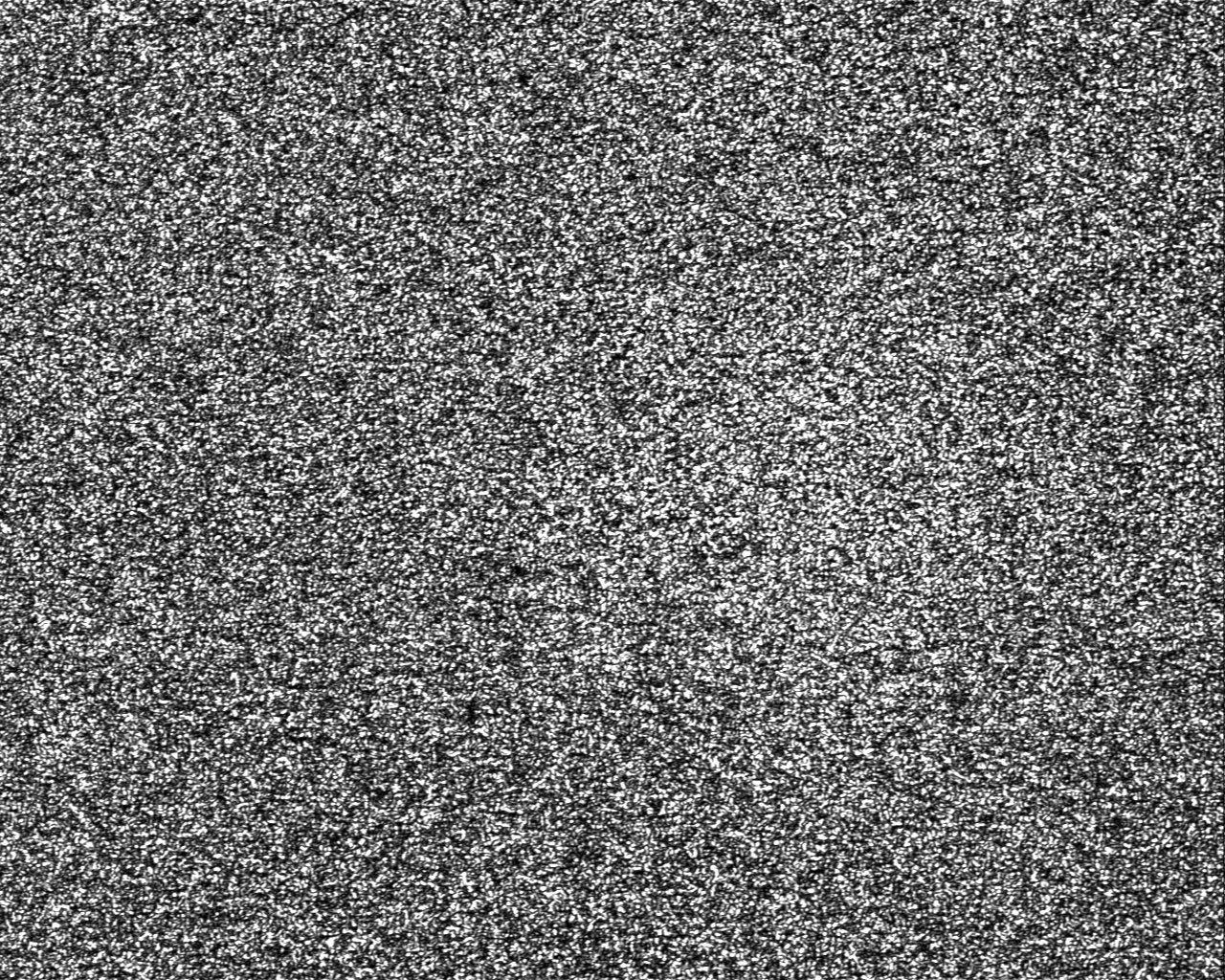}
\includegraphics[width=6cm]{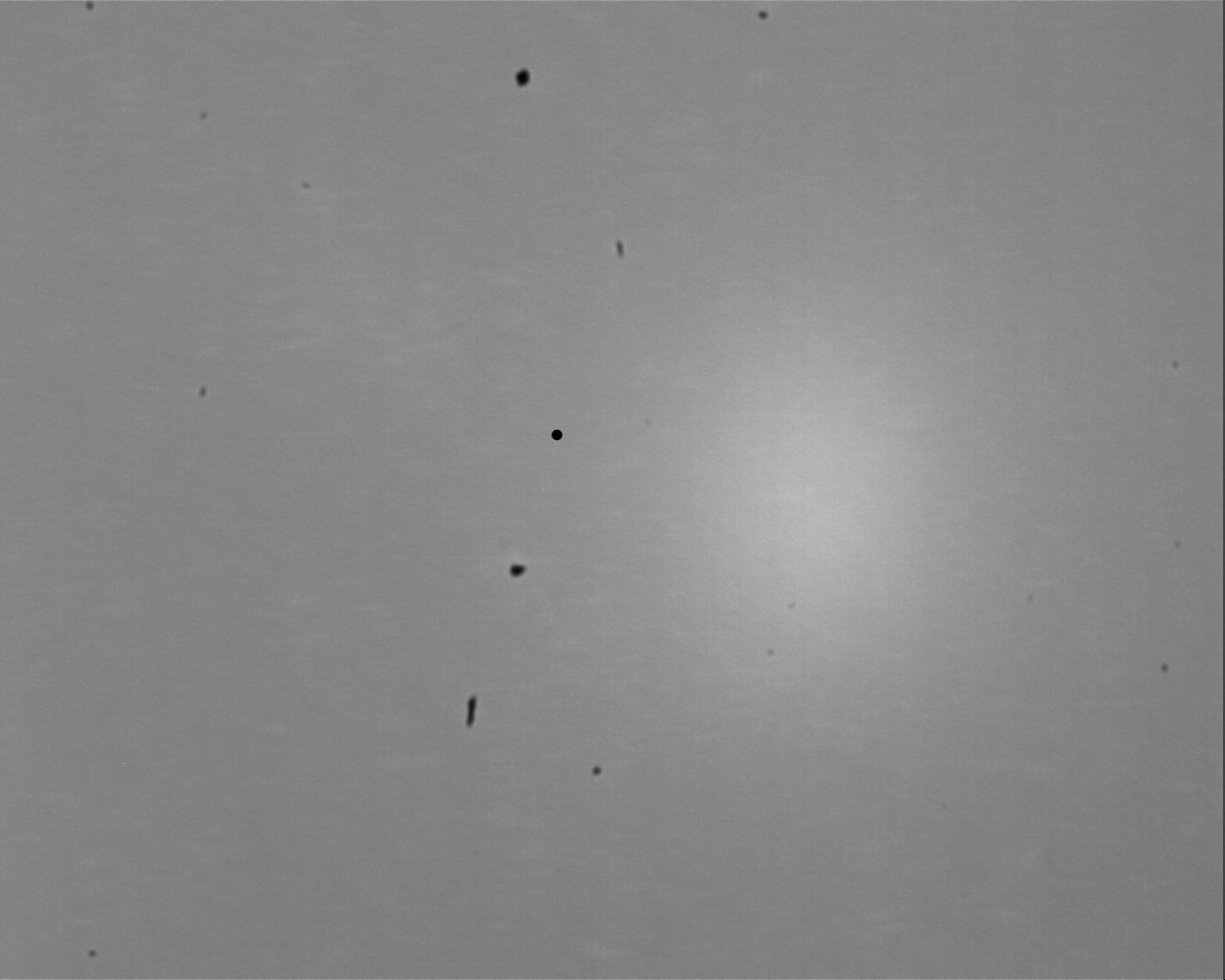}
\caption{\small%
Intensity around the back-scattered direction for a piece
of paper exposed to a parallel laser beam. For a fixed paper (left figure), one observes a characteristic
speckle pattern, due to random interference of phase-coherent
 light scattered by a single configuration of disorder.  
By averaging over various parts of the paper, small-scale random variations are averaged out, 
but an enhanced intensity is clearly visible around exact backscattering. Original data from an 
experiment performed during the Les Houches Summer School in Singapore
on July, 15th, 2009. Special thanks
to David Wilkowski, Kyle Arnold, and Lu Yin from the Center for
Quantum Technologies, National University of Singapore, for generous
support and invaluable help in setting up the experiment.
} 
\label{live.fig}
\end{center}
\end{figure} 

\begin{figure}
\begin{center}
\includegraphics[width=9cm]{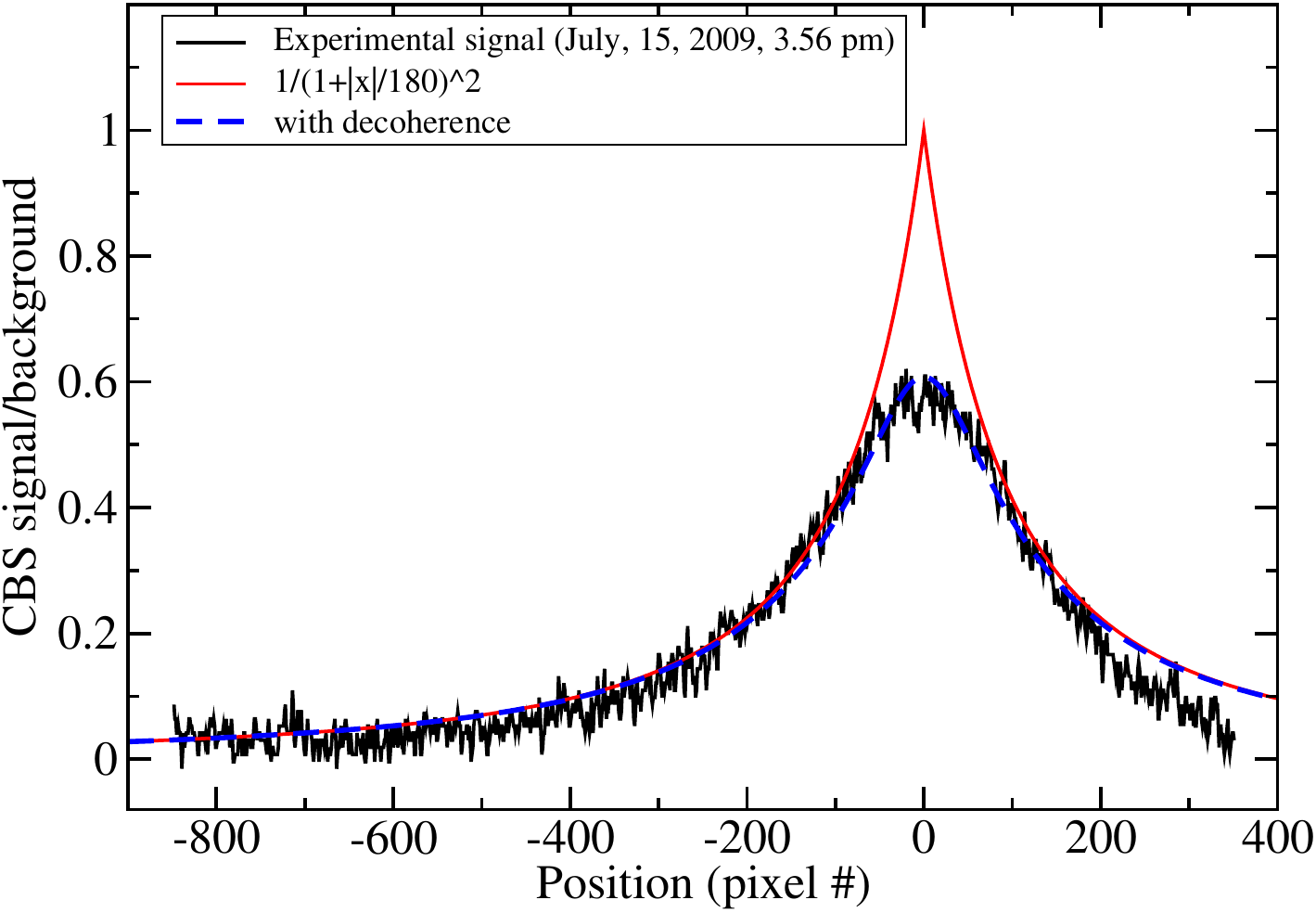}
\caption{\small%
Experimentally measured coherent backscattering signal, together with
a fit to the simplest theoretical formula, \eqref{ICdiff.eq} (full red line), and
to \eqref{ICdiff_decoherence.eq} with phenomenological decoherence 
included (dashed blue). From the angular width of the CBS signal,
one can extract the transport mean free path inside the sheet of paper, here 25\,$\mu$m.
Experimental imperfections limit the coherence of the phenomenon and are responsible for
the deviation from the peaked shape at the center of the cone. When properly taken into account (dashed curve),
the agreement is very good.
} 
\label{fit_live_cone.fig}
\end{center}
\end{figure} 

A semitransparent plate (beamsplitter) 
can be used to send the back-reflected light into a 1280$\times$1024 pixel CCD camera with pixel size
around 5\,$\mu$m, located about 20\,cm from the scattering medium, thus
ensuring a 0.025\,mrad angular resolution. 
Fig.~\ref{live.fig} (left) shows the image recorded from a 
fixed piece of paper. This situation corresponds to a single
realization of the disorder. 
The electric field on each pixel is the coherent sum of the field
amplitudes radiated by each point of the sheet of paper. Because of
its disordered nature, each contribution picks a random
amplitude and phase, resulting in a characteristic speckle pattern on
the CCD camera, the ``optical fingerprint'' of the paper.  
The angular size of the speckle grain is of the order of $1/kL$ where $L$ is the size of the
illuminated spot on the sheet of paper. For our case, it is about
0.1\,mrad, i.e.\ slightly larger than the pixel size,
in agreement with the experimental observation. The attentive reader may notice that the bright
spots look slightly brighter in a roughly circular area on the right
side of the figure. 
In order to \emph{see} the CBS cone, one should perform configuration averaging. This is easily
done by mounting the piece of paper on a rotating device (in our case
a battery-powered computer fan). On the time-averaged 
intensity, shown in Fig.~\ref{live.fig} (right), the  
fluctuating speckle pattern has been washed out, leaving
a uniform background, on top of which appears a smooth bright spot of
approximate width 10\,mrad due to coherent backscattering. The effect
is perhaps not dramatic, as the enhancement factor cannot be larger than 2 (it is 1.6 in this live experiment),
but clearly present and visible with the naked eye.

A cut across the spot center is presented in Fig.~\ref{fit_live_cone.fig} together with a fit
to the simplest theoretical formula, eq.~\eqref{ICdiff.eq}. The fit is
quite good in the wings and allows us to extract the mean free path inside the piece of paper, in our case 25\,$\mu$m.
The fact that the top of the CBS peak is rounded can be attributed to various experimental
imperfections such as the finite angular resolution, geometrical aberrations,
finite thickness and residual absorption of the piece of paper, but
could in principle also highlight the presence of a decoherence
mechanism. 

\subsection{Dephasing/decoherence}
\label{dephasing.sec}

The CBS phenomenon presented so far relies on perfect phase coherence of the multiply scattered wave. 
What happens if some external agent---such as some degree
of freedom inside the paper coupled to the wave---affects
the scattered amplitude in an uncontrolled way? Qualitatively, it is
clear that amplitudes of long scattering paths are more fragile than
those of shorter paths. As very long paths are responsible for the
characteristic triangular shape of the CBS cone around the exact
backscattering direction, it is important to understand the effect of
decoherence on the CBS signal. In return, the CBS enhancement factor
can serve as a sensitive measure for phase coherence. 

There is no universal way of breaking phase coherence, and the effect
on CBS can be different depending on the specific mechanism at work. 
Nevertheless, a simple phenomenological approximation may often be used, and we will see
below that several physical processes are well described by this
approximation. This assumption is that phase coherence
is lost at a constant rate, characterized by a phase-coherence time
$\tau_{\phi}$, also called dephasing time. 
Then, interference terms associated with paths who are visited in
a time $t$  have to be multiplied by a factor $\exp(-t/\tau_{\phi}).$
An example of such a situation is provided
by a Michelson interferometer operated with a classical light source,
where the interference disappears once the optical path length
difference exceeds 
 $c\tau_{\phi},$ with $\tau_{\phi}$ the longitudinal coherence time of the source.
Note that these phenomena are typically called  ``dephasing'' in the context of classical waves, 
and ``decoherence'' for quantum mechanical matter waves. The bottom line is simply
that interference is lost by coupling to some external degree of freedom.

The exponential attenuation of interference as $\exp(-t/\tau_{\phi})$
applies especially often to the Cooperon contribution. In Fourier
space, the effect is simply tantamount to the
replacement 
\belab{tauphi.eq}
\omega \mapsto \omega + \frac{i}{\tau_{\phi}}
\ee
or, in the diffusive propagator \eqref{Phizero.eq}:
\belab{diffphi.eq}
\frac{1}{-i\omega+\DB q^2} \mapsto \frac{1}{-i\omega+\frac{1}{\tau_{\phi}}+\DB q^2} = \frac{1}{-i\omega+\DB \left(q^2
+\frac{1}{\DB \tau_{\phi}}\right)}
\ee
which can be also be obtained via the replacement 
\be
q^2 \mapsto q^2 + \frac{1}{L_{\phi}^2}.  
\ee
The phase coherence length, 
\be
L_{\phi} = \sqrt{\DB \tau_{\phi}}, 
\ee
is the average distance over which the wave propagates \emph{diffusively} before losing its
phase coherence.

This simple replacement can be used to calculate the shape of the CBS cone
in the presence of decoherence effects. Indeed, Sec~\ref{CBStheo.sec}
discusses several approximate
expressions for the shape, all expressed as a function of the transverse momentum
$k_{\perp} \approx k |\theta|,$ which is nothing but the sum of the incoming and outgoing momenta (in the limit of
small angles $\theta \ll 1).$
The substitution $k_{\perp}^2 \mapsto k_{\perp}^2+ 1/\DB \tau_{\phi} $ 
in eq.~\eqref{ICdiff.eq} yields 
\belab{ICdiff_decoherence.eq}
\frac{I_\text{C}(\theta)}{I_\text{L}} \approx
\frac{1}{\left[1+\sqrt{(kl\theta)^2+l^2/L_{\phi}^2}\right]^2}. 
\ee
This expression is now a smooth function of $\theta$ (no cusp at
$\theta=0$ anymore).
In the limiting case $l\ll L_{\phi},$ one recovers the previous expression, only slightly
perturbed near the tip. In the opposite limit $L_{\phi} \ll l$, the
CBS cone  disappears completely,
which is quite natural as interference effects are washed out before the wave travels a single mean free path.
The relative height of the CBS peak, compared to the background at $kl|\theta| \gg 1,$ is
\be
\frac{I_\text{C}(0)}{I_\text{L}} = \frac{1}{\left[1+ l/L_{\phi}\right]^2} \approx 1 - 2 \frac{l}{L_{\phi}} = 1 - 2 \sqrt{\frac{\tau_l}{\tau_{\phi}}}
\ee
where the last two expressions are valid in the limit of weak decoherence $l\ll L_{\phi}.$ Here, $\tau_l=\vthree{\DB/l^2}$
is \vthree{of the order of} the mean free time separating two consecutive scattering events.
This expression emphasizes the sensitivity of the CBS cone to dephasing effects. Indeed, if the dephasing time is say 10 times
larger than the mean free time, its effect on the CBS cone is still very noticeable, reducing its height by almost 50\%.
For example, the experimentally observed CBS cone in the live experiments is well fitted by eq.~\eqref{ICdiff_decoherence.eq},
with a decoherence time $\tau_{\phi}=12.5\tau_l.$

Several other types of decoherence have been studied in great detail
in connection with light and cold atoms. In the following, we present
a few of them qualitatively, referring to the literature for more
details.

\subsubsection{Polarization} 
\label{polar.sec}

\begin{figure}
\begin{center}
\includegraphics[width=5cm,angle=-90]{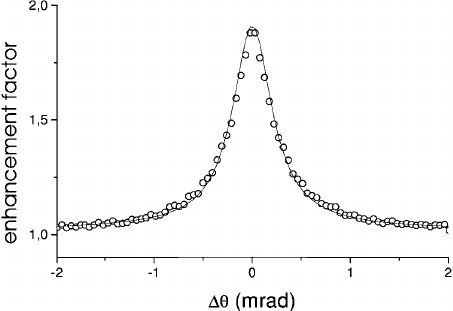}
\caption{\small%
CBS of light by cold Strontium atoms 
\cite{Bidel2002}. The backscattering enhancement
is close to 2, which indicates full phase coherence. Signal observed
in the helicity preserving channel (incoming light circularly polarized, detection in the
opposite circular polarization). Solid line: the result of a calculation taking into account
the finite geometry and inhomogeneous density of the atomic cloud.
} 
\label{strontium_cbs.fig}
\end{center}
\end{figure}

For simplicity, we have up to now considered a scalar complex wave,
describing 
e.g.\ a spinless atomic matter wave. Many real atoms, all electrons
and also electromagnetic waves 
are more complicated because of
their spin/polarization. Especially light scattering 
does not preserve polarization, as is obvious from the requirement of transversality. 
Thus the light back-scattered along the direct and reverse
paths generically emerges from the medium with different
polarization. But orthogonal polarizations
do not interfere, and thus one may expect a reduced enhancement
factor. 

Technically, one has to dress the multiple-scattering Cooperon
contribution with the polarization structure. Two independent 
polarizations of the propagating intensity  
must be taken into account (one in the retarded, one in the advanced
Green function), leading to a tensorial structure for diffuson and
Cooperon alike. 
The complete calculation of this effect is possible by decomposing the
intensity kernel into irreducible tensor moments~\cite{Mueller02}. 
To make a long story short, it is enough to say that each contribution
has a kernel of the  
type~\eqref{diffphi.eq}, with its own $\tau_{\phi}$ of the order of  $\tau_l.$
The physical interpretation is clear: because
scattering will on average lead to depolarization, all channels
associated with specific polarization correlations must decay
during propagation. Only the one channel measuring the
total intensity is protected by conservation of energy,
with $1/\tau_{\phi}=0$, and propagates diffusively.  

If the system is
additionally time-reversal invariant, the same conserved intensity
channel also exists for the Cooperon. The population of the various contributions depends on the
specific choices for the incoming and outgoing polarizations used for recording the CBS signal. 
Using the same linear
polarization for excitation and analysis 
populates the conserved mode and ensures an optimal interference
contrast for long scattering paths, at least for 
classical point-like objects (such as dye molecules) acting as
Rayleigh scatterers (we will discuss in Sec.\ \ref{spinflip.sec} the more general case).
The same is true if the incident field has circular polarization and the
opposite circular polarization is used for 
detection (helicity-preserving channel), with the additional advantage that the single scattering
background of the diffuson is filtered out, allowing in principle the
observation of a a perfect CBS enhancement by a factor of 2 
\cite{Wiersma95,Bidel2002}.

\subsubsection{Residual velocity of the scatterers} 
\label{velocity.sec}

The previous derivation assumed quenched disorder, i.e.\ scatterers
at fixed positions. Moving scatterers are a cause of decoherence: as light travels along two reciprocal paths,
it visits the same scatterers, but in opposite order, i.e.\ at
different times. If, during the time delay separating the 
scattering events on the direct and reversed path, the atom 
has moved by at last one wavelength, the phase coherence between the two paths will be
lost. This phenomenon can alternatively be interpreted in the frequency
domain,  where moving scatterers induce a Doppler shift of the scattered photon
which is different along the direct and reversed path.
Although this phenomenon does not lead to a strict exponential
decay of the phase coherence~\cite{Golubentsev84}, it
reduces the enhancement factor, which has been notably 
observed with cold atoms \cite{Labeyrie2006}. 

In general, interference of waves is suppressed once the environment
has acquired knowledge of the path taken by the scattered object. This can be most
simply seen in experiments of the Young's double-slit type 
\cite{Itano1998}, but applies equally to the CBS by light from moving
atoms, where moreover the storage of which-path information in the atomic
recoil has been studied \cite{Wickles2006}.

\subsubsection{Non-linear atom-light interaction} 
\label{nl_atom_photon.sec}

Because atoms have extremely narrow resonance lines, they have large polarisabilities
and already quite low laser intensities can saturate an atomic transition, in which case
the atom scatters photons inelastically. 
It is easy to understand that such a non-linear inelastic process will reduce the phase coherence of the 
scattered light and the enhancement factor. This indeed has been
observed~\cite{Chaneliere2004}. A full 
quantitative understanding of multiple inelastic scattering is still
not available. For a model system of two atoms driven by a powerful
laser field, a rather complete understanding of the CBS signal has
been achieved (see \cite{Shatokhin2007,Shatokhin2007b} and references therein).  

In the context of matter waves, one may study coherent backscattering of
interacting matter waves, obeying a non-linear equation such as the Gross-Pitaevskii
equation, evolving in a random optical potential. 
It has been shown that already a moderate non-linearity 
induces a phase-shift between the direct and reversed paths and
thus a decrease of the height of the CBS peak, and may in some 
cases even create a negative contribution in the backward direction 
\cite{Hartung2008}. These theoretical predictions still await
experimental realization.

\subsubsection{Internal atomic structure/spin-flip} 
\label{spinflip.sec}

\begin{figure}
\begin{center}
(a) 
\includegraphics{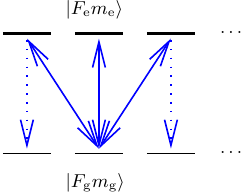}
(b) 
\raisebox{2em}{\includegraphics{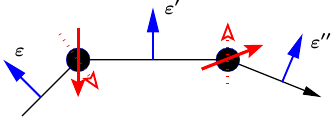}}
\caption{\small%
(a) Light scattering by a degenerate atomic dipole transition
$F_\text{g} \leftrightarrow F_\text{e}$ can
either preserve spin (Rayleigh transition, full arrows) or change the spin
(degenerate Raman transition, dotted arrows).  
(b) Multiple light scattering by randomly placed atoms with internal
spin states involves depolarization/decoherence from both spin-orbit
(transversality) and spin-flip effects.   
} 
\label{degenerate.fig}
\end{center}
\end{figure} 

The preceding description treats atoms as 
Rayleigh point scatterers that radiate a purely dipolar electromagnetic
field, with an induced dipole directly proportional 
to the incoming electric field. 
This is an excellent approximation for atoms with a non-degenerate
electronic ground state, such as Strontium.
The situation is radically different if the atomic ground state is degenerate: indeed, when scattering a photon,
the atom may stay in the same atomic state (Rayleigh
transition) or change to another state 
with the same energy (degenerate Raman
transition), see Fig.~\ref{degenerate.fig}(a). 

The basic rules of quantum mechanics imply that 
orthogonal final state cannot interfere. In other words, two multiple
scattered paths will interfere only if they 
are associated with \emph{the same initial and final states of all atoms}.%
\footnote{Not taking this into account may lead to incorrect results, see for
example~\cite{Assaf2007}, corrected in \cite{Gremaud2008}.} 
Note that there is no need
for the initial and final states to be identical, so that Raman
transitions can very well contribute to interference terms, 
if and only if the same Raman transitions occur 
along the interfering paths.%
\footnote{Thus, one must take statements
like ``Raman scattered light is incoherent'', often made by quantum
opticians, with great care. It is true 
that Raman scattered light does not interfere with the incoming
reference beam---because the final states of the atom are
different---but 
a single Raman-scattered photon along two different paths does
very well interfere with itself.}
A commonly encountered situation is that the degeneracy
of the atomic ground state is due to its non-zero total angular
momentum, see Fig.~\ref{degenerate.fig}(a): Raman transitions then
involve different Zeeman sub-states. 
The detailed calculation of the scattering vertex requires to incorporate also the angular
momentum, i.e.\ the polarization,
of the light. 
Then, the whole structure of the diffuson and the Cooperon boils down to
various kernels of type~\eqref{diffphi.eq},
where the various depolarization/decoherence rates are rotational
invariants that depend only on the
angular momenta $F_\text{g}, F_\text{e}$ \cite{Mueller2001,Mueller2005a}. 

\begin{figure}
\begin{center}
\includegraphics[width=10cm]{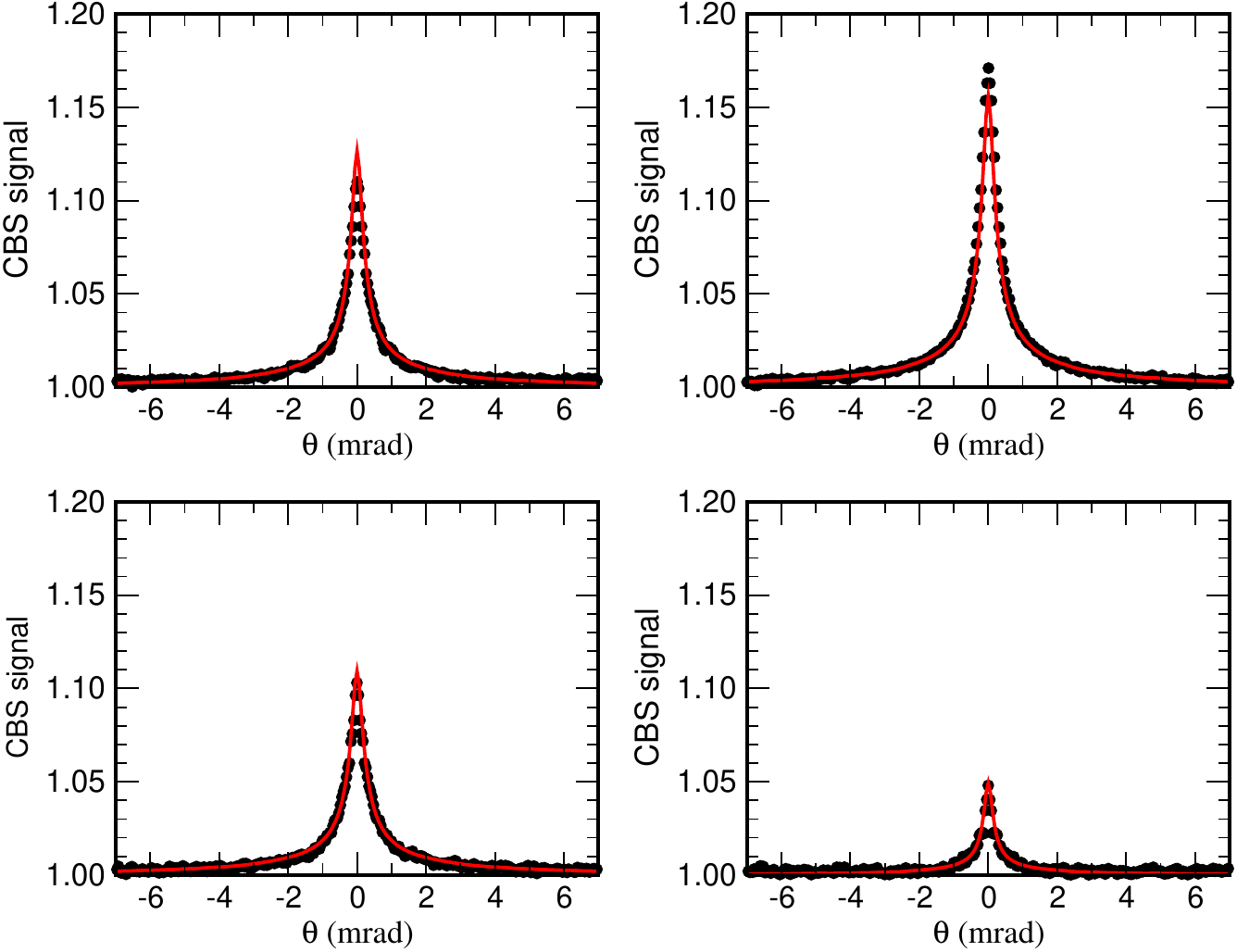}
\caption{\small%
CBS by a cloud of cold
Rubidium atoms, in four polarization channels (upper/lower left:
parallel/perpendicular linear polarization; upper/lower right:
circular polarization 
with non-preserving/preserving helicity). Solid red line: 
calculation taking into account the geometry of the atomic cloud. 
\cite{Labeyrie2003c}.
The enhancement factor is strongly reduced compared to ideal
Rayleigh scatterers such as 
Strontium, Fig.~\ref{strontium_cbs.fig}. This is due to the internal
Zeeman structure of the Rubidium atom, Fig.~\ref{degenerate.fig}a). Note in particular that the
helicity preserving channel---where the largest enhancement is found
for point scatterers such as Strontium---gives 
here the smallest enhancement.}
\label{rubidium_cbs.fig}
\end{center}
\end{figure} 

For a typical alkali atom like $^{85}$Rb with a $F_\text{g}=3 \to F_\text{e}=4$ resonance line, the longest decoherence time
for the Cooperon is $\tau_{\phi}=\frac{19}{21} \tau_l$~\cite{Mueller02}, 
meaning that the CBS interference is quite efficiently killed already
by very few scattering events.
An immediate consequence is that the CBS cone observed on a cold Rb
gas has a much reduced enhancement factor \cite{Jonckheere2000}.
A less trivial feature is that the best Rb CBS signal is not
observed in the same channels than with Sr.
Detailed calculations can be performed and an excellent agreement between the measured and the
calculated CBS signals is observed~\cite{Labeyrie2003c}, see Fig.~\ref{rubidium_cbs.fig}.

The internal atomic degrees of freedom are here responsible for the loss of coherence. Information flows from the light
to the atoms; as long as we do not precisely measure the internal state of each atom, this information is lost
and the interference contrast is reduced. This information-theoretic argument can
be made quantitative by investigating how much which-path information
is stored within the atomic internal degrees of freedom. A
quantitative measure of this wave-particle-duality, developed
originally in the context of Mach-Zehnder-type interferometers 
\cite{Englert1996}, can be investigated analytically in the simplest
cases and highlights the r\^ole of which-path information in the loss
of CBS interference visibility 
\cite{Miniatura2007}.  

A simple way to restore phase coherence is to lift the atomic degeneracy by applying an external
magnetic field. Fields as small as a few Gauss are enough to  detune some of the atomic transitions
far from resonance, thus reducing the effect of Raman transitions. It has been experimentally observed and
theoretically explained how this can increase the enhancement
factor~\cite{Sigwarth2004}. We here face a seemingly paradoxical
situation (in view of the negative magneto-resistance discussed in
Sec.~\ref{wl.sec}), where adding an external magnetic field, which should break the
time-reversal symmetry, has the effect of increasing the interference
between time-reversed paths! Similarly, strong magnetic fields in
electronic samples have been used to align free magnetic impurities,
reduced thus spin-flip effects and
restore Aharonov-Bohm interference \cite{Washburn1986,Pierre2002}.

\section{Weak localization (WL)} 
\label{wl.sec}

As discussed in the preceding section, the Cooperon is responsible for
enhanced 
backscattering, which implies an increased probability to return to the starting point. 
In the bulk of a disordered system,  diffusive transport is thus
hindered. This phenomenon, known as weak localization, 
is quantitatively expressed by a reduction of the diffusion constant
(or dimensionless conductance/conductivity)
with respect to the classical diffusion constant expected for
phase-incoherent transport. 

The weak-localization effect of the
Cooperon is expressed by eq.~\eqref{Dwl}. 
For the sake of concreteness, we will take in the following
quantitative estimates the example of atomic matter
waves with a quadratic dispersion relation $\eps = \hbar^2k^2/2m.$
The free density of states \eqref{dos.def} is 
\belab{N_0.eq}
N_0(\eps) = \frac{S_d}{(2\pi)^d} \frac{m k^{d-2}}{\hbar^2}
\ee
where  $S_d=2\pi^{d/2}/\Gamma(d/2)$ is the area
of the unit sphere in dimension $d$: $S_1=2$, $S_2=2\pi$, $S_3=4\pi.$ 
The Boltzmann diffusion constant $\DB=\hbar k l/dm$ is directly
proportional to the transport mean free path $l$. 
Because the Cooperon is isotropic, the $d$-dimensional integral in
\eqref{Dwl} can be reduced to a trivial $(d-1)$-dimensional angular
integral and a radial integral over momentum $q$, such that 
\belab{weak_loc_int}
\frac{1}{D} = \frac{1}{\DB} \left( 1 + \frac{\hbar}{\pi m k^{d-2}\DB }
\int_0^{\infty}{ \frac{q^{d-1} \rmd q}{q^2  - i0}} \right).
\ee
The result of the $q$-integral depends crucially on the dimensionality of the
system. This is a consequence of the fact, well known from classical
random walks, that the return probability to the origin is the higher,
the lower the spatial dimension $d$. Therefore, weak---and consequently
also strong---localization are immediately seen to have the largest impact in
low dimensional systems. 

\subsection{\texorpdfstring{$d=1$}{d=1}} 
\label{wl1.sec}

In dimension $d=1$, the integral \eqref{weak_loc_int} diverges 
for small $q$. However, for a system of size $L,$ the
momentum $q$ cannot take arbitrary small values, and a lower cutoff
of the order of $1/L$ must be used. A simple way of implementing it---following the recipe of Sec.~\ref{dephasing.sec} for including
decoherence effects---consists in replacing $q^2$ by $q^2+1/L^2.$ One then gets:
\be
\frac{1}{D} = \frac{1}{\DB} \left( 1 + \frac{L}{2l} \right).
\ee 
This expression is valid only if the weak localization contribution is
a small correction, i.e.\ for $L\ll l$. 
To lowest order, we recover $D\approx\DB(1-L/2l)$, which is the exact
result \eqref{betag1asymp1.eq}
already derived for 1d systems in Sec.~\ref{scaling_1d.sec}. 
The interest of the present approach is that a full microscopic theory provides us with the
weak localization correction and thus puts the scaling theory of localization on firm grounds.

\subsection{\texorpdfstring{$d=2$}{d=2}} 
\label{wl2.sec}

In dimension 2, the integral diverges both for small and large $q$. A suitable cutoff at small $q$ is again $1/L,$ the inverse
of the system size. Diffusive transport is a long time, large distance behavior. It is not expected to give an accurate
description on a scale shorter than the mean free path.  Performing
the integral with a natural cutoff $1/l$ at large $q$ thus leads to 
\belab{D_weak_loc_2d}
D \approx \DB \left[ 1 - \frac{2}{\pi k l} \ln\left(\frac{L}{l}\right)
\right]. 
\ee
In terms of the dimensionless conductance $g=2mD/\hbar,$ this implies the following
scaling relation:
\be
\beta(g) = \frac{\rmd \ln g}{\rmd \ln L}  = - \frac{2}{\pi g}. 
\ee
Our microscopic calculation thus gives an explicit prediction that can be 
readily incorporated into scaling theory, as anticipated in Sec.~\ref{scaling_2d.sec}.

\bigskip
\begin{figure}
\begin{center}
\includegraphics[width=10cm]{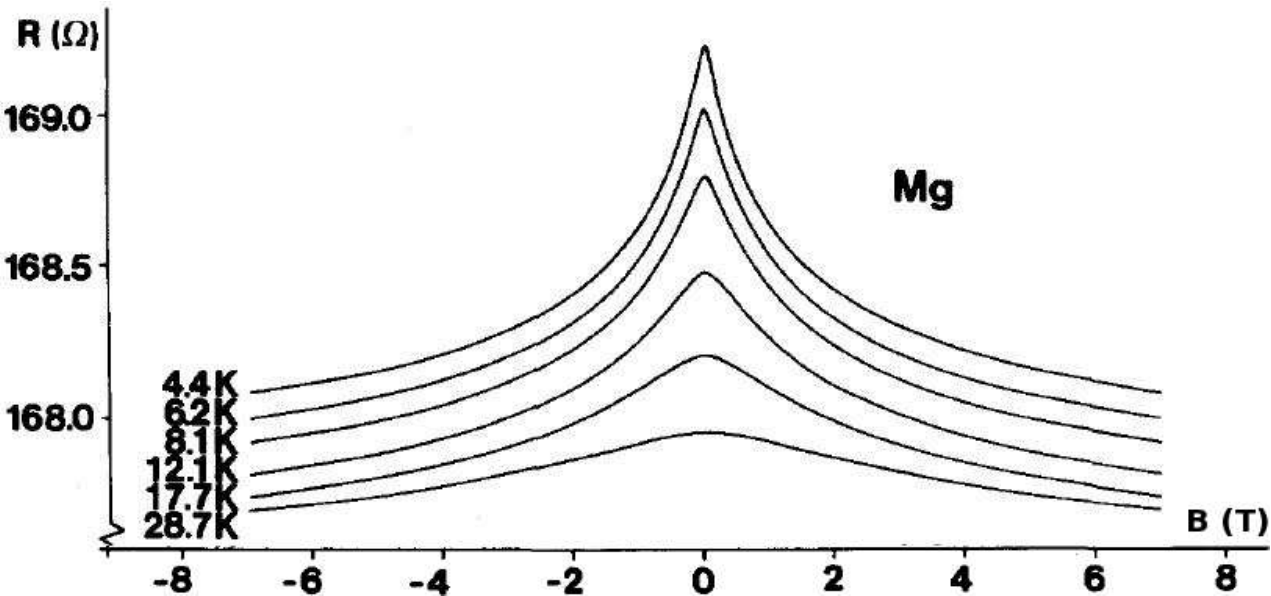}
\caption{\small%
Experimentally measured resistance of a thin Mg film exposed to a perpendicular magnetic field.
The magnetic field breaks the constructive interference of waves
counter-propagation along closed loops, reduces the weak localization
effect, and thus results in \emph{negative magneto-resistance}. 
Similarly, larger temperatures reduce the phase coherence of the electronic wavefunction,
and also reduce weak localization corrections. Adapted from~\cite{Bergmann84} (courtesy of G. Bergmann).} 
\label{weak_loc_exp.fig}
\end{center}
\end{figure} 

How can weak localization be observed experimentally? 
A priori, any measured diffusion constant incorporates already all
interference corrections to the classically expected value.  
Fortunately, the Cooperon contribution to weak localization is due to
the constructive interference between a multiply scattered path and its time-reversal. If one breaks time-reversal
symmetry on purpose, then the delicate interference is likely to
disappear, and an enhancement of diffusive transport should be observed. 

For charged particles---such as electrons in solid state samples---the simplest way is to add
a magnetic field perpendicular to the sample. In the presence of a
vector potential $\vec{A}$, a charged particle picks an additional
phase $\int e\vec{A}\cdot\rmd \vec{l}/\hbar$ along a closed loop.  This is nothing
but $e/\hbar$ times the enclosed magnetic flux. Along each closed
loop, this additional phase
appears in the Cooperon contribution. If this phase fluctuates largely from one loop to the other,
the resulting interferential contribution will vanish. As the smallest
area enclosed by a diffusive loop
is $l^2,$ the weak localization correction is expected to vanish above $B\approx \hbar/el^2.$
For a typical mean free path of a fraction of $\mu$m, this is in the Tesla range.
Figure~\ref{weak_loc_exp.fig} shows the measured resistance of a 2D Mg film vs.\ magnetic field at various temperatures~\cite{Bergmann84}.
At the lowest temperature, propagation is almost fully phase coherent and one observes a decreasing resistance,
i.e.\ a increasing conductance, when a magnetic field is applied. This
\emph{negative magneto-resistance} was a mystery when first observed
and only later explained as a manifestation of weak localization.
When temperature increases, the phase coherence of the electrons diminishes, and the weak localization correction
gets smaller. This is, in a different context, analogous to
decoherence phenomena discussed for coherent backscattering in
Sec.~\ref{dephasing.sec}.
Note that, from such experimental data, it is possible to measure the transport mean free path
(via the width of the weak localization peak) as well as the
temperature-dependence of the decoherence time. 
In recent times, weak localization measurements have been used as very
sensitive detectors for minute concentrations of magnetic
impurities, which induce spin-flip decoherence and are responsible for
finite decoherence times even at zero temperature \cite{MuellerLN2009,Pierre2002,Pierre03}.

\subsection{\texorpdfstring{$d=3$}{d=3}} 
\label{wl3.sec}

In dimension 3, the integral in eqn~\eqref{weak_loc_int} requires
only a cutoff at large $q,$ which we take again as $1/l$ and obtain
\belab{D_3d}
D \approx \DB \left( 1 - \frac{3}{\pi (kl)^2}\right). 
\ee
This diffusive Cooperon contribution to weak localization is found to
scale as $1/(kl)^2.$ However, this is
not the whole story, because other diagrams, not included in the
simple diffusive Cooperon, give contributions that are actually more
important for small disorder $kl\gg 1$. Just as for the CBS cone
discussed in Sec.~\ref{CBStheo.sec}, also here the double-scattering
diagrams appearing in eqn~\eqref{Udiag} contribute to leading order
$1/kl$. In $d=3$, the \emph{static} electronic conductivity was found to be given
by \cite{Belitz1994}
\belab{sigmastaticWL.eq}
\frac{\sigma(\omega=0)}{\class{\sigma}} = 1 - \frac{2\pi}{3kl} -
\frac{\pi^2-4}{(kl)^2} \ln(kl) + O((kl)^{-2}). 
\ee
As long as $kl\gg 1,$ the weak localization is only a small correction, again providing us with
a macroscopic ground for the scaling theory of localization. 
It also gives an approximate criterion for the onset of Anderson
localization, which should set in approximately when the right hand
sides of eqns~\eqref{D_3d} or \eqref{sigmastaticWL.eq} vanish, i.e.\ 
$(kl)_\text{c}  = O(1)$. This is Ioffe-Regel
criterion, eq.~\eqref{ioffe-regel.eq}. However, the precise
calculation of the critical point is a delicate endeavor.   
What precisely happens at the $1/l$ scale is
not universal. The same is true 
for the Ioffe-Regel criterion, but the latter nonetheless yields a first
estimate on where to expect the Anderson transition.

\subsection{Self-consistent theory of localization} 
\label{sc.sec}

Weak localization describes how 
diffusive transport is affected by interference.  
In essence, however, weak localization is a
perturbative result: first, because the Cooperon contribution is
evaluated using a  diffusive kernel valid in the absence of interference; second, because
this simple approach takes into account only a specific type of diagrams. 
The first assumption is especially questionable in 1d, where diffusive transport actually never occurs,
because localization appears at the very same scale (the localization
length) than diffusion (the mean free path). 
Concerning the second point, the dominant
r\^ole of the Cooperon in large systems was recognized 
already by Gorkov et al.\ \cite{Gorkov1979} and Abrahams et
al.~\cite{Abrahams1979}. However, a weak-disorder perturbation theory
in powers of $1/kl$ alone would never be able to describe the Anderson transition (in 3d) for
strong disorder, nor the crossover from weak to strong localization in
1d and 2d systems. 

The self-consistent theory of localization, developed by Vollhardt and
W\"olfle in the 1980s \cite{Vollhardt1980,Vollhardt1982,Vollhardt1992}, 
is an attempt to escape this seemingly hopeless situation by
applying a suitable 
self-consistency scheme, as often employed with success to describe phase
transitions in statistical physics. Rather than
a theory with rigorously controlled approximations, 
it must thought of as a guess, albeit highly educated, about the most important 
contributions of diagrams to all orders. 
The basic observation is that the diffusive contribution of large closed
loops in eq.~\eqref{weak_loc_int} must itself be modified by weak localization: inside
a large loop, the wave explores smaller loops, leading to a decreased diffusion constant for propagation along
the large loop. This argument can of course be repeated: one should
take into account loops within loops within loops..., all the way down
to the smallest loops, stopping at the scale of the 
transport mean free path. 

The whole description  must now be self-consistent, describing what happens
at every scale from the mean free path up to the size of the system---or toward infinity in the bulk.
The simplest idea would be to replace the static 
Boltzmann diffusion constant $\DB$ in the integral of \eqref{weak_loc_int}
by the renormalized diffusion constant $D$ itself, thus providing us with an implicit
equation for $D.$ It turns out that this is not enough: indeed, a
single number---the static diffusion constant $D$---cannot 
describe the full dynamics both for short times, where it is
diffusive, and for long times where 
localization may eventually set in. So we require a scale-dependent
diffusion constant, and it turns out that it is 
simpler to consider
various time scales rather than various spatial scales. 
We thus consider a diffusion constant $D(\omega)$ which 
depends on frequency $\omega.$  
The self-consistent expression for $D(\omega)$ just derives from
\eqref{weak_loc_int} by re-introducing the $\omega$ dependence and replacing $\DB$ by $D(\omega)$ in the integral:
\belab{D_self}
D(\omega) + \frac{\hbar}{\pi m k^{d-2}} \int{\frac{q^{d-1} \rmd q}{q^2 - (i\omega/D(\omega))}} = \DB.
\ee
In the short-time limit $\omega \to \infty,$ the contribution of the integral vanishes and one gets
back to classical Boltzmann diffusive propagation, as expected.
The most interesting part takes place at long times, i.e.\ in the
limit $\omega \to 0$, whose consequences again depend crucially 
on the dimension.

\subsubsection{\texorpdfstring{$d=1$}{d=1}} 
\label{sc1.sec}
\begin{figure}
\begin{center}
\includegraphics[width=11cm]{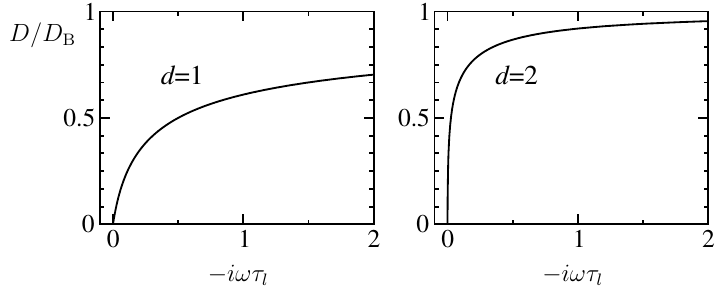}
\caption{\label{D_self_1_2d.fig} 
\small Diffusion constant vs.\ (imaginary) frequency, in units of
Boltzmann diffusion constant and 
mean free time, respectively, as predicted by 
the self-consistent theory of localization, in 1d (left) and 2d (right, for $kl$=1.5).
At large $\omega$ (short time), one recovers the classical Boltzmann diffusive behavior.
At small $\omega,$ the dependence is linear, which implies exponential localization
in configuration space, in agreement with scaling theory, numerical
and experimental observations (in 1d).
In 2d, the localized regime is reached at much smaller frequency
because the 
localization length and time are exponentially large.} 
\end{center}
\end{figure}

At finite $\omega,$ the integral in \eqref{D_self} does not need any
regularization, 
it is simply $\frac{\pi}{2}\sqrt{D(\omega)/(-i \omega)},$ and the implicit equation for $D(\omega)$ is easily solved~\cite{Weaver2005}:
\be
\frac{D(\omega)}{\DB} = \frac{\sqrt{1-16i\omega \tau_l} -1}{\sqrt{1-16i\omega \tau_l} + 1}
\ee
where $\tau_l=l^2/\DB$ is the mean free time between two scattering events.
This function is plotted in the left panel of
Fig.~\ref{D_self_1_2d.fig} as function of $-i\omega$.%
\footnote{Imaginary frequency is only used for convenience, 
as it makes the diffusion constant purely real and thus easier to plot. The most important
transport property is the small-$|\omega|$ behavior, which
is linear in the localized regime, both for real or imaginary $\omega.$} 
In the limit of small $\omega,$
it behaves linearly $D/\DB \approx -4i\omega \tau_l.$ This in turns implies that the propagation kernel 
$1/(-i\omega + D(\omega) q^2)$ is just $1/(-i\omega) \times 1/(1+4l^2q^2).$ When going back from
momentum to configuration space by inverse Fourier transform, it implies that the intensity kernel
is proportional to $\exp(-|z|/2l)$ at long times. It successfully describes
exponential localization with the localization length $\xiloc=2l,$
i.e.\ the exact result for the localization length!
The elementary ingredients used for obtaining this important result are: quantum kinetic theory,
microscopic calculation of the weak localization correction in the perturbative regime
and its self-consistent extension. That the exact result is eventually obtained is a strong hint that
the self-consistent approach catches an important part of the physics
of localization.

But beware! This triumph is somewhat tarnished by the fact that it is the
\emph{typical} intensity that decays with $\xiloc$, whereas the \emph{average}
intensity calculated here should asymptotically decay with
$4\xiloc$, as shown in Sec.~\ref{full_distrib.sec}.
The precise source for this discrepancy escapes
our present understanding. Clearly, the self-consistent theory is built for the average
intensity kernel $\mv{\GR\GA}$ and thus cannot describe the huge fluctuations
in the localized regime. One lacks a diagrammatic expansion
for the typical transmission, which would require to calculate contributions of advanced and
retarded Green functions to all orders. 

Decoherence effects can be easily included in the self-consistent
approach by the replacement $-i \omega \mapsto -i \omega +
1/\tau_{\phi}$ explained in Sec.~\ref{dephasing.sec}. In
Figure~\ref{D_self_1_2d.fig}, this replacement simply translates the
curve horizontally to the left. 
One immediately finds that the diffusion constant no longer
vanishes at $\omega=0,$ but takes a finite value, implying diffusive motion at long times.
In the limit of weak decoherence $\tau_l \ll \tau_{\phi}$, 
the residual diffusion constant is $D\approx 4 \tau_l \DB/\tau_{\phi} = \xiloc^2/\tau_{\phi}.$
It is much smaller than the Boltzmann diffusion constant and allows
for a simple physical interpretation: 
a phase-breaking event, occurring on average every $\tau_\phi$, destroys
the delicate interference responsible for localization. This implies a restart of diffusion
during time $\tau_l$ after which localization sets in again, until the
next phase breaking event, etc. 
 
\subsubsection{\texorpdfstring{$d=2$}{d=2}} 
\label{sc2.sec}

In 2d, the integral in \eqref{D_self} diverges in the large-$q$ limit, requiring a regularization.
The natural short-distance cut-off is the mean free path $l.$
Elementary manipulations shows that $D(\omega)$ is implicitly determined  by  
\belab{D_sc_2d.eq}
\frac{D(\omega)}{\DB} = 1 - \frac{1}{\pi k l} \ln \left( 1 -
\frac{D(\omega)}{\DB} \frac{1} {2i\omega \tau_l}\right). 
\ee
In contrast with the 1d case, $D/\DB$ is not a universal function, it depends on the parameter $kl.$
The right panel of Fig.~\ref{D_self_1_2d.fig} plots it for $kl=1.5.$ 
It displays the classical diffusive behavior $D\approx \DB$
at large $\omega$ (short times), and localization at long
times. Indeed, for $\omega \to 0$, one finds  
$D(\omega) \approx -i \omega \xiloc^2,$ i.e.\ exponential localization
with the localization
length
\be
\xiloc = l \sqrt{\exp{\left(\pi k l\right) -1 }} \approx l
\exp{\left(\frac{\pi k l}{2}\right)}. 
\ee
This provides us with a microscopic derivation of the result of scaling theory, eq.~\eqref{xiloc_2d.eq}. 
The self-consistent approach describes correctly the exponentially
large localization length in 2d.%
\footnote{The same caveats than in 1d exist, concerning average versus
typical quantities.}
Note that, even for strong disorder with a rather small value
$kl=1.5,$
 the linear regime in Fig.~\ref{D_self_1_2d.fig} is observed only at
very small $\omega,$ i.e.\ for very long times.

\begin{figure}
\begin{center}
\includegraphics[width=8cm]{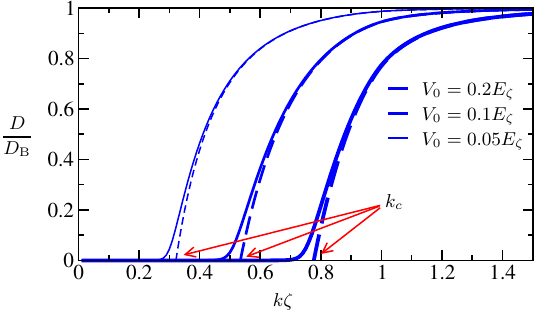}
\caption{\small%
Diffusion constant (normalized to the Boltzmann diffusion constant) computed from
the self-consistent theory of localization, for atomic matter waves with wave vector $k$ exposed to
a 2d speckle potential with correlation length $\lcor$ and different
amplitudes $V_0$. A stronger potential means a smaller value of $kl$. 
Dashed lines are the prediction of the simple
perturbative weak localization correction, eqn~\eqref{D_weak_loc_2d}, solid lines the 
result of the self consistent approach, eqn~\eqref{D_sc_2d.eq}, including residual decoherence due
to spontaneous emission implemented via eqn~\eqref{tauphi.eq}. 
A rather sharp cross-over
between the Boltzmann diffusive behavior at high energy and the quasi-localized behavior
at low energy is observed around a critical value $k_c$ \cite{Miniatura2009}.} 
\label{D_over_DB.fig}
\end{center}
\end{figure} 

Decoherence can be taken into account exactly like in 1d. Instead of a true metal-insulator transition,
one observes a cross-over from classical diffusion at large $kl$
towards a residual diffusion (triggered by decoherence) at small $kl.$ Explicit calculations have been
carried out in \cite{Miniatura2009} for the case of atomic matter waves in a speckle
potential, where residual spontaneous emission is one source of
decoherence that can be experimentally tuned. 
Figure~\ref{D_over_DB.fig} shows
typical results. Because of the exponential dependence in 2d, the cross-over from quasi-localized
behavior at small $k$ to diffusive behavior at large $k$ is rather rapid. In any case,
a crucial requirement is to have very cold atoms, with de Broglie
wavelength shorter than the speckle correlation length.
 
Considering an expanding BEC wave packet released from a harmonic
trap, one can calculate the expected
stationary (for negligible decoherence) density distribution along the
lines of \eqref{wave_packet_final}. Just as in 1d, the
asymptotic decay is governed by the wave vector $k_\text{max}$ of the
fastest atoms, and the density is predicted to be $\mv{|\psi(r)|^2}\approx C
r^{-5/2}\exp\{-r/\xiloc(k_\text{max})\}$ \cite{Miniatura2009}. 

\subsubsection{\texorpdfstring{$d=3$}{d=3}} 
\label{sc3.sec}

\begin{figure}
\begin{center}
\includegraphics[width=8cm]{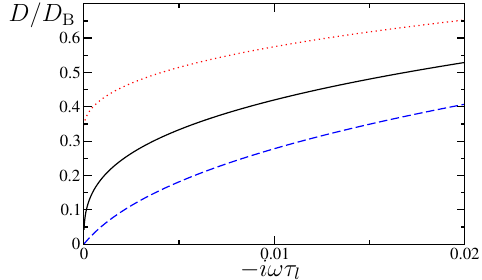}
\caption{\small Diffusion constant vs.\ (imaginary) frequency, in units
of Boltzmann diffusion constant and mean free time, respectively, computed from
the self-consistent theory of localization in 3d,
eqn~\eqref{D_implicit_3d}. Dotted red line:  metallic
regime $kl=1.2$ with finite diffusion constant at long times. Dashed
blue line: insulating regime $kl=0.8$ with a finite localization
length. Solid black line: critical point $kl=(kl)_\text{c}=\sqrt{3/\pi}$ 
of the metal-insulator Anderson transition, where the diffusion constant scales like $(-i\omega)^{1/3}$,
implying an anomalous diffusion $\langle r^2(t)\rangle \propto t^{2/3}$ at long times.} 
\label{D_self_3d.fig}
\end{center}
\end{figure} 

In 3d, the same short-distance regularization than in 2d is necessary,
leading to the following implicit equation, valid in the limit $\omega
\tau_l \ll 1$:%
\footnote{This formula features only the static Cooperon contribution, which
suffices for qualitative predictions, but
should be extended to include the full interference terms appearing in
\eqref{sigmastaticWL.eq} when quantitative precision is necessary.}
\belab{D_implicit_3d}
\frac{D(\omega)}{\DB} + \frac{3}{\pi (kl)^2} \left( 1 -\frac{\pi}{2}
\sqrt{\frac{-3i\omega \tau_l}{D(\omega)/\DB}} \right) =1. 
\ee
The behavior of the solution, shown in Fig.~\ref{D_self_3d.fig},
depends on the Ioffe-Regel parameter $kl$ and defines three distinct regimes: 

\paragraph{Diffusive regime:} For $kl> (kl)_\text{c} = \sqrt{3/\pi},$
$D/\DB$ tends to a constant value in the limit $\omega \to 0,$ 
which means that the system behaves always diffusively, albeit with a diffusion constant smaller
than the Boltzmann diffusion constant. This is the regime of weak localization. 

\paragraph{Localized regime:} For $kl< (kl)_\text{c},$ it is easy to
see that $D/\DB \to 0$ in the limit $\omega \to 0.$ 
More precisely, one has exponential localization:
\be
D(\omega) \approx -i\omega \xiloc^2 \qquad \text{with} \qquad \xiloc
\sim \frac{1}{(kl)_\text{c}-kl}
\ee
immediately below the Anderson transition, on the insulating side.
This means, see eq.~\eqref{xilocnu.eq}, that the critical exponent
deduced from the self-consistent approach is $\nu=1$, quite far from
the true value $\nu=1.58$ known from numerical simulations \cite{Slevin}.
The reason lies in the approximate character of the self-consistent
approach, which disregards the huge fluctuations in the vicinity of
the critical point. Field-theoretic approaches can in
principle capture the effect of fluctuation and have been
quantitatively tested in $d=2+\epsilon$ dimensions~\cite{Evers2008}. For $d=3$,
however, to our knowledge no analytical theory is available that makes
a better quantitative prediction than the self-consistent theory.

\paragraph{Critical regime:} 
At the critical point  $kl= (kl)_\text{c},$ it is easy to see that the solution of eq.~\eqref{D_implicit_3d}
scales like $D(\omega) \sim (-i\omega)^{1/3}.$ Consequently, the critical behavior is anomalous
diffusion where the squared extension increases subdiffusively at long
times: $\langle r^2(t)\rangle \propto t^{2/3}.$
This anomalous diffusion has been experimentally observed with the 
quasi-periodically kicked rotor, see Sec.~\ref{kicked.sec}.

\section{Kicked rotor} 
\label{kicked.sec}

The physics of Anderson localization is, as amply discussed in the preceding sections,
highly dependent on the dimension of the system. While the 1d
situation is fairly well understood---localization is the generic behavior, the localization length
is comparable to the mean free path, and the fluctuation properties
in the localized regime are essentially well understood---the physics of
higher dimensions is much richer still.
Dimension 3 is especially interesting, as one expects a so-called
mobility edge, separating, in the continuum case, 
localized states at low energy/strong disorder from extended states at
high energy/weak disorder. 

As explained in Sec.~\ref{NumAnderson_d3.sec}, it is very difficult to
find a  clean experimental system
to observe this metal-insulator Anderson transition 
unambiguously. Cold atomic matter waves are very attractive 
because they can be directly observed, and because most experimental
imperfections as well as atom-atom interactions can be precisely
controlled, if not reduced
to a minimum. The main difficulty consists
in reaching the Ioffe-Regel threshold $kl=O(1)$, eq.~\eqref{ioffe-regel.eq},
i.e.\ preparing a sufficiently
small $k$ (low energy, large de Broglie wavelength) and short mean free path $l.$
Indeed, the latter cannot be shorter than the correlation length of the disordered
potential, i.e.\ of the order of 1\,$\mu$m for optical speckle.

This limitation on the mean free path can be overcome using a different approach, where disorder
is not provided by an external potential in configuration space, but
by classically chaotic dynamics 
in momentum space. 
This idea has been realized experimentally with the atomic kicked
rotor, and Anderson localization in 1d has been observed as early as
1994~\cite{Raizen1994}, 14 years prior to the widely noticed 
Anderson localization in configuration space~\cite{Billy2008}! 
A key advantage
of the kicked rotor is that is does not require ultra-cold atoms from
a Bose-Einstein condensate: a standard magneto-optical trap suffices 
to prepare the initial state. 
The kicked rotor also has permitted the clean observation of the metal-insulator Anderson transition
in 3D, and the first experimental measurement of the critical exponent, with non-interacting
matter waves~\cite{Chabe2008}.

\subsection{The classical kicked rotor}

We consider a one-dimensional rotor whose
position can be described by the angle $x$ (defined modulo $2\pi$) 
and the associated momentum $p$, and kick it periodically  
with a position-dependent amplitude. In properly scaled units,
the Hamiltonian function can be written as 
\begin{equation}
H = \frac{p^2}{2} \ \vthree{+}\ k \cos x\ \sum_{n=-\infty}^{+\infty}\ \delta(t-nT)
\label{hkr}
\end{equation}
where $T$ and $k$ are period and strength of the kicks, respectively.

Because of the time-dependence, energy is not conserved, but thanks to
the time-periodicity, we can analyze the motion stroboscopically  and build a
Poincar\'e map picturing the evolution once every period.
This map relates the phase space coordinates just before kick $n+1$
to the coordinates just before kick $n$: 
\begin{equation}
\left\{
\begin{array}{l}
\displaystyle I_{n+1} = I_n + K \sin x_n\\
\displaystyle x_{n+1} = x_n + I_{n+1}
\end{array}
\right.
\end{equation}
where $K=kT$ and $I_n=Tp_n.$ 
This is nothing but the celebrated
standard map (also known
as the Chirikov map) that has been widely studied \cite{licht,kr:casati}: 
it is almost fully chaotic and ergodic around $K=10$ and above.

When the stochasticity parameter  $K$ is very large, each kick is so strong that the positions of
the consecutive kicks can be taken statistically uncorrelated. By 
averaging, one thus gets:
\begin{equation}
\langle p_{n+1}^2 \rangle \simeq \langle p_n^2\rangle +  k^2 \langle \sin^2{x_n}\rangle  \simeq  \langle p_n^2\rangle  + \frac{k^2}{2}
\end{equation}
It follows that the motion in momentum space is diffusive
($\langle p^2\rangle $ increases linearly with time) with 
diffusion constant 
\begin{equation}
D = \frac{k^2}{2T}. 
\label{diffconst}
\end{equation}
Numerical experiments~\cite{licht} show that this expression works
well for $K\geq 10.$ 
Note that the kicked rotor is a perfectly deterministic system, without any randomness.
It is the chaotic nature of the classical motion, and thus its extreme sensitivity to 
perturbations, which renders the deterministic classical motion diffusive \emph{on average}.

\subsection{The Quantum Kicked Rotor}
\label{qkr}

The quantum Hamiltonian is obtained from the classical one, eq.~(\ref{hkr}),
through the canonical replacement of $p$ by 
$-i\hbar\partial_x.$
The evolution operator over one period is the product of the free evolution
operator and the instantaneous kick operator:
\begin{equation}
U = U(T,0) = \exp\left(-\frac{i}{\hbar}\frac{p^2T}{2}\right)
\exp\left( \vthree{-} \frac{i}{\hbar}\ k\cos x\right)
\label{ut}
\end{equation}
The long-time dynamics is generated by successive iterations of $U.$
Thus, one can use the eigenstates of $U$ as a basis set. $U$ being
unitary, its eigenvalues are complex numbers with unit modulus:
\begin{equation}
U|\phi_i\rangle = \exp\left( -\frac{iE_iT}{\hbar}\right)\ |\phi_i\rangle
\label{qe}
\end{equation}
with $0 < E_i \leq 2\pi\hbar/T$ are defined modulo $2\pi\hbar/T$. They are not exactly the energy
levels of the system---the $|\phi_i\rangle$ are not stationary states
of the time evolution, but are only periodic---and are
called quasi-energy levels, the $|\phi_i\rangle$ being the Floquet
eigenstates. This Floquet description is the time-analog of the Bloch
theorem that applies to spatially periodic potentials.

\subsection{Dynamical Localization} 

The quantum dynamics of the kicked rotor can  be quite simply
studied numerically 
by repeated application of the one-period evolution operator $U$ to
the initial state, alternating free propagation phases with
instantaneous scattering events in momentum space induced by the kicks.
The free evolution between kicks, $\exp\left( - ip^2T/2\hbar\right)$, is
diagonal in momentum representation, such
that each momentum eigenstate, characterized by its momentum $m\hbar$ with
integer $m$, picks up a different phase shift.
The kick operator $\exp\left(ik\cos \theta/\hbar\right)$, in contrast, is diagonal
in position representation and couples different momenta. Being unitary, it plays the role
of a scattering matrix in momentum space and contains the quantum
amplitude for changing an incoming
initial momentum in an outgoing one, under the influence of one kick; $k$ is the parameter controlling
the scattering strength. The dynamics of the kicked rotor can be seen as a sequence of scattering events
interleaved with free propagation phases.

For sufficiently large $K=kT$, the classical dynamics is diffusive in
momentum space, but it should come as no surprise
to the reader now familiar with 1d Anderson localization, that the quantum dynamics
may be localized at long times. 
This localization was baptized ``dynamical localization" when it was observed
in numerical simulations~\cite{Casati1979}. Only later, people  realized that it is nothing
but the Anderson scenario of 1d localization, as explained below.

Dynamical localization has been experimentally observed in the dynamics of a Rydberg electron
exposed to an external microwave field~\cite{abu}. Arguably the simplest
observation uses a cold atomic gas, prepared in a standard
magneto-optical trap with a typical velocity spread
of few recoil velocities \cite{Raizen1994,nz,Chabe2008}. 
After the trap is switched off, a 
periodic train of laser pulses is applied to the atoms.
Each pulse is composed of two far-detuned 
counter-propagating laser beams
producing a spatially modulated optical potential. Each laser pulse
thus produces a kick on the atom velocity, whose amplitude is 
proportional to the gradient of the optical potential.

If the kicks are infinitely short, we recover exactly the kicked
rotor, eq.~(\ref{hkr}), where the position of the atom in the standing wave 
plays the role of the $x$ variable and its velocity is the $p$ variable. 
The kick strength $k$ is proportional to the laser intensity divided 
by the detuning.
The spatial dimensions perpendicular to the laser beams do not play
any role in the problem, so that we have an effectively one-dimensional
time-dependent problem.
The mapping of the dimensionfull Hamiltonian for cold atoms to the kicked rotor
Hamiltonian, eq.~(\ref{hkr}), shows that the effective Planck's constant of the problem~\cite{Lemarie2009} is 
$\hbar_\text{eff}= 4\hbar k_{L}^{2}T/M=8\omega_r T$ where $k_{L}$ is the laser wavenumber and $M$ the atomic mass.
Up to a numerical factor, it is the ratio of the atomic recoil frequency $\omega_r$ to the pulse frequency,
and can be easily varied in the experiment, from the semiclassical regime $\hbar_\text{eff}\ll 1$
to the quantum regime  $\hbar_\text{eff}\sim 1.$

\begin{figure}
\begin{center}
\includegraphics[width=7cm]{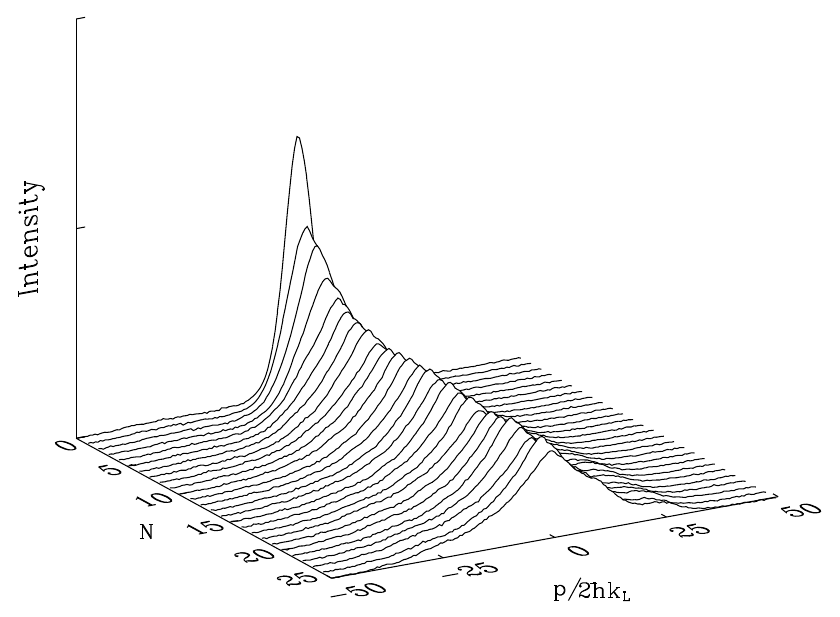}
\end{center}
\caption{\small%
Experimental time evolution of the momentum distribution
of the atomic kicked rotor~\cite{Raizen1994},
from the initial Gaussian distribution until the exponentially
localized distribution at long time; $N$ is the number of kicks (courtesy of M. Raizen).}
\label{distp_raizen}
\end{figure}

After the series of pulses is applied, the momentum distribution is measured either by
a time of flight technique~\cite{Raizen1994} or velocity selective Raman transitions~\cite{Chabe2008}.
Figure \ref{distp_raizen} 
shows the momentum distribution as a function of time.
While, at short time, the distribution is Gaussian---as expected for
a classical diffusion---, its shape changes around the localization time
and evolves toward an exponential shape $\exp (-|p|/\xi_\text{loc})$ at long time, a clear-cut
manifestation of Anderson/dynamical localization. 

Adding decoherence on the system---either by adding spontaneous emission~\cite{nz} or by
weakly breaking the temporal periodicity~\cite{raizen:noise}---induces some residual diffusion at long time,
in accordance with the discussion in Sec.~\ref{dephasing.sec} and \ref{sc.sec}. This is another proof
that dynamical localization is based on delicate destructive interference. 

\subsection{Link between dynamical and Anderson localizations}
 
So far, we have only made plausible that dynamical localization with
the quantum kicked rotor is similar to Anderson localization 
in a spatially disordered medium. We now demonstrate the connection
between the two phenomena, following \cite{fishman}.
Consider the evolution operator, eq.~(\ref{ut}), and 
the associated eigenstate $|\phi\rangle$ with quasi-energy $E.$ 
The part of the evolution operator associated
with the kick can be written as:
\begin{equation}
\exp\left( \vthree{-} \frac{i}{\hbar}\ k\cos x\right) = \frac{1+iW(x)}{1-iW(x)}
\end{equation}
where $W(x)$ is a periodic Hermitean operator
which can be Fourier-expanded:
\begin{equation}
W(x) = \sum_{r=-\infty}^{\infty}{W_r\ \exp{(irx)}}. 
\end{equation}
Similarly, the kinetic part can be written as:
\begin{equation}
\exp\left[ -\frac{i}{\hbar}\left(\frac{p^2}{2}-E\right)\ T\right] = 
\frac{1+iV}{1-iV}
\end{equation}
The operator $V$ is diagonal is the eigenbasis of $p$, labeled by the integer $m$ (see above).
If one performs the following expansion in this basis set, 
\begin{equation}
\frac{1}{1-iW(x)} |\phi\rangle = \sum_m{\chi_m\ |m\rangle},
\end{equation} 
it is straightforward to show that the eigenvalue-equation \eqref{qe} can be
rewritten as
\begin{equation}
\epsilon_m\chi_m + \sum_{r\neq 0}{W_r\chi_{m-r}} = - W_0\chi_m
\label{andersoneq}
\end{equation}
where
\belab{epsilon_m}
\epsilon_m = \tan \left[\left(E - \tfrac{1}{2}m^2\hbar^2\right) T/
{2 \hbar} \right]. 
\ee

Equation \eqref{andersoneq} is the time-independent Schr\"odinger
equation for a one-dimensional Anderson model,
cf.~\eqref{Anderson_discrete_1d.eq}, with 
site index $m$, on-site energy $\epsilon_m$,
coupling $W_r$ to the nearest sites and  total energy $W_0$.
Compared to \eqref{Anderson_discrete_1d.eq}, there are two new ingredients: 
firstly, there are additional hopping amplitudes to other neighbors.
But since they decrease sufficiently fast at large distance, they do not
play a major role. Secondly, the $\epsilon_m$ values, determined 
deterministically by \eqref{epsilon_m}, 
are not really random variables, but only pseudo-random%
\footnote{As is well known, ``random-number generators''
implemented in computers also generate deterministic, merely pseudo-random
sequences; most of them use formulae analogous to \eqref{epsilon_m}.}
with
a Lorentzian distribution.%
\footnote{The non-random character appears for example, when the
product $\hbar T/2$ is chosen as an integer multiple of $2\pi$. Then all $\epsilon_m$ are equal,
the motion is ballistic and localization is absent. Similarly, when $\hbar T$
is commensurate with $\pi$ (the so-called quantum resonances), the
sequence $\epsilon_m$ becomes periodic, and Anderson localization
does not take place, giving way to Bloch-band transport.}
Still, localization is expected and indeed observed.
The computation of the localization length follows the general lines explained in section~\ref{microscopic.sec},
and is in good agreement with experimental observations.

It should be emphasized that space and time play different roles in the
Anderson model and in dynamical localization.
What plays the role of the sites of the Anderson model are the
momentum states. This is why dynamical localization is not
observed in configuration space, but in momentum space.

\subsection{The quasi-periodically kicked rotor}

How can the kicked rotor be used to study Anderson localization in more than one dimension?
The first idea is to use a higher-dimensional rotor with a classically chaotic dynamics
and to kick it periodically. It turns out that this is not easily realized experimentally, as it requires
to build a specially crafted spatial dependence\cite{Garcia2009}. 
Yet, remember that time and space have switched roles,  
and so a simpler idea
is to use additional temporal dimensions rather than spatial dimensions. Instead of kicking the system periodically
with kicks of constant strength, one may use a temporally quasi-periodic excitation. Various schemes
have been used~\cite{bicolor}, but the one allowing to map on a multi-dimensional Anderson model
uses a quasi-periodic modulation of the kick strength, 
the kicks being applied at fixed time interval~\cite{Shepelyansky1989}.

We will be interested in a 3d Anderson model, obtained by adding two quasi-periods to the system:%
\footnote{In this section, we take the kicking period $T$ as unit of time}
\begin{equation}
\mathcal{H}_{\mathrm{qp}}=\frac{p^{2}}{2}+\mathcal{K}(t)\cos x\sum_{n}\delta(t-n)\;,
\label{eq:KRquasiper}
\end{equation}
with
\begin{equation}
\mathcal{K}(t)=K\left[1+\varepsilon\cos\left(\omega_{2}t+\varphi_{2}\right)\cos\left(\omega_{3}t+\varphi_{3}\right)\right]\;.\label{eq:Kdet}
\end{equation}
Now where is the three dimensional aspect in
the latter Hamiltonian? The answer lies in a formal
analogy between this quasi-periodic kicked rotor and a 3d kicked rotor
with the special initial condition of a  ``plane source'', as
follows. 

Take the Hamiltonian of a 3d, periodically kicked rotor: 
\belab{eqKR3DquasiperH}
\mathcal{H}=\frac{p_{1}^{2}}{2}+\omega_{2}p_{2}+\omega_{3}p_{3}+K\cos
x_{1}\left[1+\varepsilon\cos x_{2}\cos
x_{3}\right]\sum_{n}\delta(t-n), 
\ee
and consider the evolution of a wavefunction $\Psi$ with the initial
condition 
\begin{equation}
\label{eqPsi3}
 \Psi({x}_{1},{x}_{2},{x}_{3},t=0)\equiv\psi({x}_{1},t=0)\delta({x}_{2}-\varphi_{2})\delta({x}_{3}-\varphi_{3}). 
\end{equation}
This initial state, perfectly localized in $x_{2}$ and $x_{3}$ 
and therefore entirely delocalized in the conjugate momenta $p_{2}$
and $p_{3}$, is a  ``plane source" in  momentum space  \cite{Weaver2005}. 
A simple calculation shows that the stroboscopic evolution of $\Psi$
under (\ref{eqKR3DquasiperH}) 
 coincides exactly with the evolution of the initial state
$\psi(x=x_1,t=0)$ under the Hamiltonian \eqref{eq:KRquasiper} of the 
quasi-periodically kicked rotor (for details, see~\cite{Lemarie2009}). 
An experiment with
the quasi-periodic kicked rotor can thus be seen as a localization
experiment in a 3d disordered system, where localization is
actually observed in the direction perpendicular
to the plane source. In other words, the situation is comparable to a transmission experiment
where the sample is illuminated by a plane wave and the exponential localization is only measured along the 
wave vector direction. Therefore, the behavior of the quasi-periodic
kicked rotor (\ref{eq:KRquasiper}) matches  \emph{all} dynamic properties  of the quantum 3d kicked rotor.

The classical dynamics has been shown to be a chaotic diffusion,
provided the parameter $\varepsilon$ is sufficiently
large to ensure efficient coupling between the 3 degrees of freedom~\cite{Lemarie2010}.
As for the
standard 3d kicked rotor (\ref{eqKR3DquasiperH}), its  quantum
dynamics can be studied using the Floquet states via mapping to a 3d Anderson-like model: 
\begin{equation}
\epsilon_{\mathbf{m}}\Phi_{\mathbf{m}}+\sum_{\mathbf{r}\neq0}W_{\mathbf{r}}\Phi_{\mathbf{m}-\mathbf{r}}=-W_{\mathbf{0}}\Phi_{\mathbf{m}}\;,
\label{eqAndersonmodelKRquasiper}
\end{equation}
 where $\mathbf{m}\equiv(m_{1},m_{2},m_{3})$ labels
sites in a 3d cubic lattice, the on-site energy $\epsilon_{\mathbf{m}}$
is 
\begin{equation}
\epsilon_{\mathbf{m}}=\tan\left\lbrace \frac{1}{2}\left[\omega-\left(\hbar\frac{{m_{1}}^{2}}{2}+\omega_{2}m_{2}+\omega_{3}m_{3}\right)\right]\right\rbrace \;,
\label{eq:pseudo-random-disorder-quasiperKR}
\end{equation}
 and the hopping amplitudes $W_{\mathbf{r}}$ are the Fourier expansion coefficients of
\begin{equation}
W({x}_{1},{x}_{2},{x}_{3})= \vthree{-} \tan\left[K\cos{x}_{1}(1+\varepsilon\cos{x}_{2}\cos{x}_{3})\right/2\hbar]\;.
\label{eq:hoppingamplitudes3DQKR}
\end{equation}

A necessary condition for localization is obviously that
$\epsilon_{\mathbf{m}}$ not be periodic. 
This is achieved if $(\hbar,\omega_{2},\omega_{3},\pi)$
are incommensurate. When these conditions are verified, localization effects as predicted
for the 3d Anderson model are expected, namely either a diffusive
or a localized regime. Localized states would be observed if the disorder
strength is large compared to the hopping. In the case of the 
model~(\ref{eqAndersonmodelKRquasiper}), the amplitude of the disorder
is fixed, but the hopping amplitudes can be controlled by changing
the stochasticity parameter $K$ (and/or the modulation amplitude
$\varepsilon$): $W_{\mathbf{r}}$ is easily seen to increase with
$K$. In other words, the larger $K$, the smaller the disorder. One
thus expects to observe diffusion for large stochasticity $K$ and/or
modulation amplitude $\varepsilon$ (small disorder) and localization for small
$K$ and/or $\varepsilon$ (large disorder). It should be emphasized
that \emph{stricto sensu} there is no mobility edge in our system that
would separate localized from delocalized eigenstates. 
Depending on the parameters $K,\hbar,\varepsilon,\omega_{2},\omega_{3},$
either \emph{all} Floquet states are localized or all are delocalized. The boundary
of the metal-insulator transition is in the
$(K,\hbar,\varepsilon,\omega_{2},\omega_{3})$-parameter space. 
As seen below, $K$ and $\varepsilon$ are the primarily important parameters.

\begin{figure}
\begin{center}
\includegraphics[width=7cm]{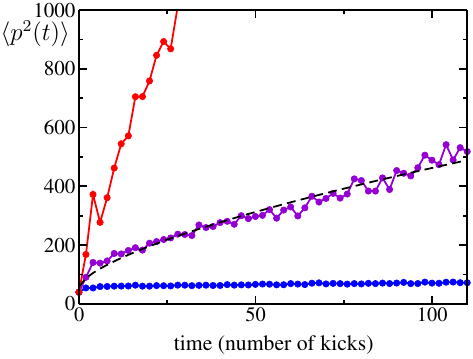} 
\end{center}
\caption{\label{p2_exp.fig}\small Experimentally measured temporal dynamics of
the quasi-periodically kicked rotor, for increasing values of the kick strength.
The average kinetic energy $\langle p^2(t)\rangle$ tends to a constant in the localized regime (lower
curve, $K=4,$ $\varepsilon=0.1$), increases
linearly with time in the diffusive regime (upper curve, $K=9,$ $\varepsilon=0.8$). 
At the critical point $K=K_{c}\approx6.4$ 
(middle curve), anomalous diffusion $\langle p^2(t)\rangle \sim t^{2/3}$ (dashed curve)
is clearly observed.}
\end{figure}

In the experiment performed at the University of Lille
\cite{Chabe2008}, kicks are applied to atoms with an initially narrow momentum distribution,
and the final momentum distribution is measured using
velocity-selective Raman transitions.%
\footnote{Measuring the average $\langle p^2(t)\rangle$ is tedious and
very sensitive to noise in the wings of the momentum distribution. It is much
easier to measure the atomic population $\Pi_0(t)$ at zero momentum. Because of atom number conservation,
 $\langle p^2(t)\rangle$ is roughly proportional to $1/\Pi_0^2(t).$ The proportionality factor
 depends on the shape of the distribution, but does not show large changes. As we are interested
 in scaling properties, $1/\Pi_0^2(t)$ or $\langle p^2(t)\rangle$ are essentially equivalent.} 
Figure~\ref{p2_exp.fig} shows the experimental data. For large
disorder, one clearly
sees the initial diffusive phase and the freezing of the quantum
dynamics in the localized regime (lower curve). 
In the diffusive regime (upper curve),
$\langle p^2(t)\rangle$ is seen to increase linearly with time.
The intermediate curve displays an anomalous diffusion $\langle p^2(t)\rangle \sim t^{2/3}.$
The anomalous exponent 2/3 is exactly the prediction of the
self-consistent theory of localization, section~\ref{sc3.sec}, which
also fully agrees with the scaling theory of localization. 
Time here plays the role of the system size $L$ in the scaling theory: going to longer times
means following the renormalization flow in fig.~\ref{betad.fig}. Only exactly at the unstable critical point
will the anomalous diffusion subsist for arbitrarily long times. At
slightly larger (resp.\ smaller) $K,$ the motion
will eventually turn diffusive (resp.\ localized) at long time.  
Experimental constraints prevent the observation beyond 150-200 kicks. Numerical simulations may extend much beyond:
it has been checked that the anomalous diffusion with exponent 2/3 is
followed for at least $10^8$ kicks~\cite{Lemarie2009}. 
 
Since in numerical or experimental practice one always works in finite-size systems, 
we should emphasize that there is an important difference between a
true metal-insulator transition and a cross-over between two limiting
behaviors. For example, consider the simplest 1d situation
 where the dynamics eventually localizes for sure, with a 
localization time depending on the kick strength $K$.
 Over a finite experimental time, one may observe an apparently diffusive behavior if the localization time
 is longer than the duration of the experiment.%
\footnote{This also occurs for 1d Anderson localization in a speckle
potential~\cite{Billy2008}, as already pointed out in
Sec.~\ref{loc_length_1d.sec}: 
Because the localization length and time 
 vary rapidly with energy, one observes localization at low energy and apparently diffusive behavior
 at high energy. In between, an apparent mobility edge
appears~\cite{Lugan2009}, which should not be confounded with the true
Anderson metal-insulator transition taking place in the thermodynamic
limit, although the experimental signatures may be similar.}
 An intermediate situation with the localization time 
comparable to the duration of the experiment could produce data looking like anomalous diffusion.  
However, this could be only a transient behavior and a longer
measurement  will eventually show localization. In contrast, the
$t^{2/3}$ behavior at the critical point of the Anderson
 transition is not a transient behavior, it extends to infinity,
highlighting the scale-free behavior with fluctuations of all sizes
present right at the critical point. 
 
The unavoidable experimental limitation by finite size can also be
turned into a powerful tool of analysis.  It is known
as finite-size scaling~\cite{finite-size-scaling} and has its roots in the scaling properties observed in the vicinity of the transition.
The idea is that all results, obtained for various values of parameters and time, are described by a universal
scaling law depending on a single parameter, namely, the distance to the
critical manifold. Close to the transition, there is only one characteristic
length (which diverges at the critical point) and all details below this scale are irrelevant.
Such an approach has been extremely successful to extract critical parameters from numerical simulations
of the Anderson model for various system sizes. The approach has been transposed to the
kicked rotor---see~\cite{Lemarie2009,Lemarie_thesis} for details---and
makes it possible to extract the localization length (in momentum space) from
numerical or experimental data acquired over a restricted time interval. 

\begin{figure}
\begin{centering}
\includegraphics[width=6cm]{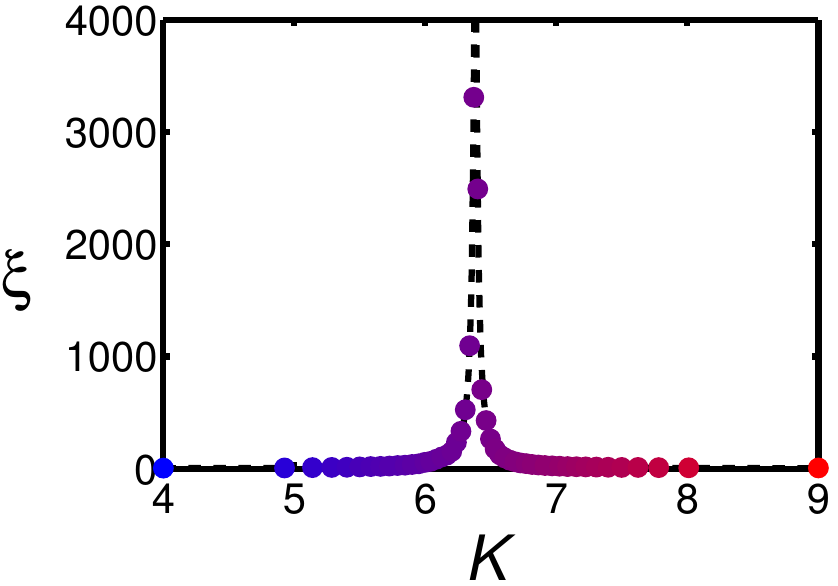} 
\includegraphics[width=6cm]{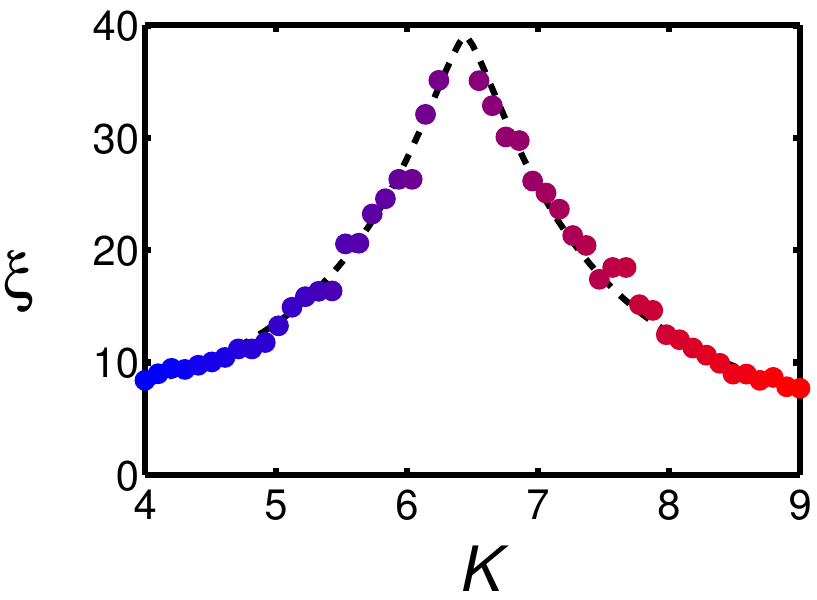} 
\end{centering}
\caption{\small%
Characteristic length (for localization in momentum space) extracted
from numerical (left) and real (right) experiments on the quasi-periodically
kicked rotor, in the vicinity of the metal-insulator Anderson transition~\cite{Chabe2008}. Finite-size
scaling is used. The characteristic length is proportional to the localization length
on the insulating side, and to the inverse of the diffusion constant on the metallic side.
It has an algebraic divergence $1/|K-K_\text{c}|^{\nu}$ at the transition, smoothed
by finite size and decoherence, which are more important
in the real experiment (limited to 110 kicks) than in the numerical
calculations (up to one million kicks). In both cases, it is possible
to extract a rather precise estimate of the
critical exponent $\nu \approx 1.5.$} 
\label{xiloc_rotor.fig} 
\end{figure}

The results are shown in Fig.~\ref{xiloc_rotor.fig} for both numerical simulations and experimental observations.
One clearly sees the divergence of the characteristic length (the localization length on the insulator side)
in the vicinity of the transition. The divergence is smoothed by experimental imperfections
and the finite duration of numerical and real experiments. The smoothing is much more important
in the latter case than in the former one, because the duration of the real experiment (110 kicks maximum)
is about 4 orders of magnitude shorter than in numerical experiments.
It is nevertheless possible to extract the critical exponent of the transition. For the numerical
experiments, one finds $\nu=1.58\pm0.01$ in perfect agreement with the best determination
on the Anderson model. Moreover, it has been checked that this
exponent is universal, i.e.\ independent
of the microscopic details such as the choice of the parameters
$\hbar,\omega_2,\omega_3$~\cite{Lemarie2009b}. This is an additional 
confirmation that the transition observed is actually the metal-insulator Anderson transition.

The critical exponent can also be determined---albeit with reduced
accuracy---from the experimental data~\cite{Chabe2008}. For the data
of Fig.~\ref{xiloc_rotor.fig}, one obtains: 
\be
\nu_\text{exp} = 1.4 \pm 0.3
\ee
These values
are in excellent agreement with the numerical results. The key point is that the
exponent significantly differs from unity, which is the value deduced from
solid state measurements, see Fig.~\ref{katsumoto.fig}, and the
prediction of the self-consistent approach.

Since the 
atom-atom contact interaction in a cold dilute gas is much smaller than the 
electron-electron Coulomb interaction in a solid sample, 
and since atoms are less easily lost than photons, 
cold atoms appear particularly suitable for precise measurements of the Anderson transition. 
Moreover, the possibility to picture wave functions directly opens the way to 
studies of fluctuations in the vicinity of the critical point~\cite{Lemarie2010},
and may even permit to observe multifractal behavior~\cite{Page2009} with matter waves.
The flexibility of the kicked rotor could also be used to study
the Anderson transition in lower dimensions (by reducing the number of quasi-periods)
or, why not, even higher dimensions (by increasing it beyond 3). 
In any case, it is an attractive alternative to experiments on
spatially disordered systems.


\phantomsection

\end{document}